\documentclass[final,3p,times]{elsarticle}

\usepackage{zmy_elsevier}

\begin{document}
\begin{frontmatter}



\title{Parameter estimation of structural dynamics with neural operators enabled surrogate modeling}

\author[1,2]{Mingyuan Zhou}
\author[3]{Haoze Song}
\author[2]{Wenjing Ye}
\author[3,5]{Wei Wang}
\author[1,4]{Zhilu Lai~\corref{c1}}
\cortext[c1]{Corresponding author. E-mail address: zhilulai@ust.hk}

\affiliation[1]{organization={Internet of Things Thrust, The Hong Kong University of Science and Technology (Guangzhou)}, 
            state={Guangzhou},
            country={China}}

\affiliation[2]{organization={Department of Mechanical and Aerospace Engineering, The Hong Kong University of Science and Technology},
            state={Hong Kong},
            country={China}}

\affiliation[3]{organization={Data Science and Analytics Thrust, The Hong Kong University of Science and Technology (Guangzhou)}, 
            state={Guangzhou},
            country={China}}

\affiliation[4]{organization={Department of Civil and Environmental Engineering, The Hong Kong University of Science and Technology},
            state={Hong Kong},
            country={China}} 

\affiliation[5]{organization={Department of Computer Science and Engineering, The Hong Kong University of Science and Technology},
            state={Hong Kong},
            country={China}}

\begin{abstract}

Parameter estimation in structural dynamics generally involves inferring the values of physical, geometric, or even customized parameters based on first principles or expert knowledge, which is challenging for complex structural systems.
In this work, we present a unified deep learning-based framework for \textit{parameterization}, \textit{forward modeling}, and \textit{inverse modeling} of structural dynamics.
The parameterization is flexible and can be user-defined, including physical and/or non-physical (customized) parameters.
In the forward modeling, we train a neural operator for response prediction -- forming a surrogate model, which leverages the defined system parameters and excitation forces as inputs to the model.
The inverse modeling focuses on estimating system parameters.
In particular, the learned forward surrogate model (which is differentiable) is utilized for preliminary parameter estimation via gradient-based optimization;
to further boost the parameter estimation, we introduce a neural refinement method to mitigate ill-posed problems, which often occur in the former.
The framework's effectiveness is verified numerically and experimentally, in both interpolation and extrapolation cases,
indicating its capability to capture intrinsic dynamics of structural systems from both forward and inverse perspectives.
Moreover, the framework's flexibility is expected to support a wide range of applications, including surrogate modeling, structural identification, damage detection, and inverse design of structural systems.

\end{abstract}



\end{frontmatter}

\section{Introduction}
The study of physical systems generally comprises three fundamental steps~\cite{tarantola2005inverse}: parameterization, forward modeling, and inverse modeling.
The parameterization aims to discover a minimal set of parameters that fully characterize the investigated system. 
Subsequently, the forward modeling aims to make predictions on the system's state, given values of system parameters and system inputs.
Conversely, the inverse modeling involves estimating actual parameter values or other inputs of the system using some measurements of its states.
These methodological steps have been foundational in addressing diverse problems in science and engineering. 
In this work, we draw inspiration from these steps and endeavor to integrate them with deep learning scheme for structural dynamics applications. 

In structural dynamics applications, 
many sub-fields and tasks rely on effective modeling of structural systems~\cite{gallet2022structural}.
For instance, forward modeling and inverse modeling are essential for \textit{system identification}~\cite{noel2017nonlinear,kerschen2006past} and \textit{structural health monitoring}~\cite{zhang2022vibration,bao2019state},
which are discussed in Section~\ref{sec:related_1}.
Specifically, structural \textit{response prediction}~\cite{eshkevari2021dynnet} focuses on predicting how structures will respond to external loading and initial/boundary conditions, which is viewed as forward modeling. 
Conversely, physical \textit{parameter estimation}~\cite{naets2015online,abdessalem2018model} is inverse modeling, which seeks to determine the values of structural parameters, such as stiffness and damping coefficients. 
Similarly, \textit{force identification}~\cite{liu2022dynamic,yang2023dynamic} in mechanical or civil structures belongs to inverse modeling, which aims to infer the external forces acting on the structures from the dynamic response measurements.
Overall, building accurate mappings between excitation force, system parameters, and dynamic responses is vitally important for structural dynamics applications.

The approaches of modeling structural dynamics generally fall into two streams: \textit{solving-based} (model-based, first principles) approaches, and \textit{learning-based} (data-driven) approaches.
In the solving-based methods~\cite{moore2012model,mottershead1993model}, mathematical models of structural systems are typically derived from first principles.
Particularly, continuous structures are generally represented by partial differential equations (PDEs).
The structural state is then determined by solving these PDEs analytically or numerically.
Nevertheless, model-based methods can be challenging and computationally expensive for complex or realistic structural systems with unknown mechanisms.

In contrast, learning-based methods aim to model structural dynamics by closely capturing major dynamics directly from measured data, 
without necessarily requiring prior knowledge of structures.
A comprehensive review of learning-based methods in structural dynamics can be found in~\cite{worden2011natural}.
Among various methods, machine learning models~\cite{cunha2023review} such as multi-layer perceptron~\cite{xu2004direct}, autoencoder~\cite{lei2023displacement}, convolutional neural networks~\cite{wu2019deep}, and physics-informed neural networks~\cite{lai2022neural} have been extensively applied in structural dynamics.
More recently, neural operators have emerged as powerful tools for solving PDEs in diverse fields~\cite{lu2021learning,azizzadenesheli2024neural,cao2024laplace,li2020fourier,he2023mgno}, as introduced in Section~\ref{sec:related_no}.
In structural dynamics, neural operators~\cite{lu2023deep,lin2023learning,najera2023structure} have demonstrated improved response prediction performance compared to standard neural networks.
However, most works focus on the forward modeling of structural dynamics, while the jointly forward and inverse modeling of structural dynamics is less explored in learning-based methods.

The two streams of approaches are not mutually exclusive but rather represent the different \textit{prior} information used to address structural dynamics problems.
In this work, we focus on the \textit{learning-based} approaches to model structural dynamics. 
As many existing models focus on forward and inverse modeling separately, 
we hypothesize that they are strongly related and can be effectively addressed within a unified framework -- a well trained forward surrogate model that captures more intrinsic dynamics often leads to more accurate inverse parameter estimation.
Additionally, deep learning models often learn structural system parameters implicitly with trainable weights.
Developing learning-based methods that can estimate system parameters via neural networks remains an open, challenging problem in structural dynamics.
 
Recent progress in deep learning provides potential solutions for inverse modeling in structural dynamics.
Researchers have explored neural networks in solving inverse problems,
such as inverse design~\cite{wu2024compositional,allen2022inverse}, parameter estimation~\cite{gaskin2023neural,kutz2023machine}, initial/boundary conditions estimation~\cite{takamoto2022pdebench,mackinlay2021model}, control problem~\cite{khimin2024optimal}, etc.
A typical workflow involves first learning a neural surrogate model to approximate forward physical dynamics~\cite{takamoto2022pdebench},
and then performing back-propagation to optimize the input parameters of interest. 
As highlighted in~\cite{azizzadenesheli2024neural,kutz2023machine}, deep learning models offer a promising technique for solving inverse problems.

Inspired by these deep learning approaches, 
to the best of our knowledge, this work pioneers to integrate the steps of parameterization, forward modeling, and inverse modeling with deep learning architectures and optimization techniques for structural dynamics applications.
More specially, the contributions of this work are as follows:
\begin{itemize}
    \item We propose a unified deep learning-based framework for modeling structural dynamics (Section~\ref{sec:struc_dym}), beginning with the parameterization of structural systems. 
    The parameterization is flexible, that can include physical (Section~\ref{sec:param_case_1}) and/or non-physical (customized) (Section~\ref{sec:param_case_2}) parameters.
    The system parameters are subsequently utilized as input to the forward surrogate model for supervised training. 
    \item Within this framework, we train a neural network as the surrogate model for forward modeling (Section~\ref{sec:forward}). 
    Specifically, we introduce a variant of the deep operator networks (Parametric DeepONet),
    with its sub-modules specially designed for structural dynamics application, which merges excitation forces and system parameters to predict dynamic responses at one or multiple locations. 
    \item In the inverse modeling (Section~\ref{sec:inverse}), we perform parameter estimation using \textit{gradient-based initialization} on the learned forward surrogate model.
    To address potential inaccuracies in this initialization, we employ \textit{neural refinement} to further boost the performance of parameter estimation. 
    Ultimately, the inverse modeling offers a more complete data-driven solution for modeling structural dynamics.
\end{itemize}

The effectiveness of the framework is verified through a numerical Duffing oscillator and an experimental wind turbine blade. 
Notably, both interpolation and extrapolation test cases are considered to comprehensively evaluate the proposed framework.

\section{Related works}
\subsection{Neural operators} 
\label{sec:related_no}

Recently, considerable efforts have been dedicated to developing neural networks to learn operators from data ~\cite{lu2021learning,azizzadenesheli2024neural,cao2024laplace,li2020fourier,he2023mgno}, which map input function spaces to output function spaces.
Lu et al.~\cite{lu2021learning} propose the deep operator networks (DeepONet), which employ a branch-trunk architecture that is based on the universal approximation theorem of operators~\cite{chen1995universal}.
Li et al.~\cite{li2020fourier} introduce the Fourier neural operator, which parameterizes a general convolutional operator to approximate nonlinear operators using the fast Fourier transform.
Additionally, many other neural architectures such as graph neural operators~\cite{li2020neural}, multi-grid neural operators~\cite{he2023mgno}, Laplacian neural operators~\cite{cao2024laplace} have also been proposed,
exhibiting improved generalization performance for solving partial differential equations than standard neural networks.

Neural operators are more natural solutions to learn structural dynamics from data, considering that the underlying mechanism of solving structural dynamics is applying certain operators.
For instance,
Garg et al.~\cite{garg2022assessment} use DeepONet to predict the displacement of dynamical systems under stochastic excitation force.
Cao et al.~\cite{cao2024deep} employ different neural operators to predict the structural responses of the floating structures under irregularly driven waves, which can be viewed as varying excitations.
Additional studies~\cite{lu2023deep,lin2023learning,najera2023structure} have explored the use of DeepONet to predict the response of dynamical systems with different initial conditions and input functions.

Previous research mainly investigates the response prediction of dynamical systems under varying excitations, and our work considers structural response prediction given varying system parameters -- learning a family of operators other than a specific one. More importantly, we integrate the trained operators to serve for the parameter estimation in the framework -- an open and challenging problem that is less addressed for neural network-based methods.

\subsection{Structural system identification and health monitoring} 
\label{sec:related_1}
\textit{Structural system identification} is essential in understanding structural systems, 
which aims to develop models based on the input and output (or output only) measurements from the investigated system.
Extensive research has been conducted for nonlinear systems~\cite{kerschen2006past,noel2017nonlinear},
which typically involve three main steps: detection, characterization, and parameter estimation.
Detection aims to confirm the presence of nonlinear effects between the system's input and output~\cite{allemang1998survey}.
Subsequently, characterization identifies the location, type, and functional form of these nonlinearities~\cite{lin1995location}.
According to the formulation in~\cite{kerschen2006past}, the functional form can generally be represented by a solution functional (operator) $\mathcal{M}_{\mathbf{\vect{\mu}}}$ parameterized by $\vect{\mu}$, 
which maps the system's input $f(t)$ to its output $y(t)$, expressed as $y(t) = \mathcal{M}_{\mathbf{\vect{\mu}}}\left(f(t)\right)$.
Building this solution functional relates to forward modeling, and conventional methods heavily rely on physics-based models to have an explicit $\mathcal{M}_{\mathbf{\vect{\mu}}}$, which can be infeasible if there are unknown mechanisms embedded in complex systems.
Also, conventional methods may not work if one opts for customizing the parameters $\vect{\mu}$. 
Lastly, parameter estimation~\cite{poulimenos2006parametric} involves determining the system coefficients ${\mathbf{\vect{\mu}}}$, which belongs to inverse modeling.
Typical methods include structural model updating~\cite{mottershead1993model}, restoring force surface method~\cite{rogers2022latent}, auto-regressive models~\cite{billings2013nonlinear}, neural networks~\cite{lai2021structural}, etc.

Vibration-based \textit{structual health monitoring (SHM)} is another typical application in structural dynamics,
aiming to identify potential damage of structural systems.
Most vibration-based SHM methods can be categorized as either pattern recognition~\cite{farrar2012structural,bao2019state} or inverse problem~\cite{friswell2007damage}.
For pattern recognition-based methods, specific properties (often called damage-sensitive) of response data are extracted and used for damage identification tasks, formulated as novelty detection~\cite{soleimani2022toward}, classification~\cite{zhou2024structural}, and regression~\cite{shahidi2015structural}, etc.
For inverse problem-based methods, which are closely related to parameter estimation,  
involving determining structural parameters and their changes, which can be related to structural integrity and health~\cite{spiridonakos2009parametric,ebrahimian2018bayesian,rubaiyat2024data}.

The proposed unified framework for modeling structural dynamics can be conveniently adapted to certain tasks of system identification and SHM.
Particularly, the forward modeling is equivalent to nonlinearity characterization in system identification, where we employ neural networks to learn a solution operator $\mathcal{M}_{\mathbf{\vect{\mu}}}$.
Additionally, the inverse modeling can estimate the system parameters, which can be user-defined values for describing structural states and health.

\section{Methodology}
\subsection{Problem formulation}
\label{sec:struc_dym}

Consider a \textit{structural dynamic system} defined by the mapping between input and output functions as follows: 
\begin{align}
    \label{eq:forward}
        y(t) = \mathcal{M}_{\boldsymbol{\mu}} \left(f(t) \right),
\end{align}
where $\mathbf{\vect{\mu}} = \{\mu_1, \mu_2, ...\}$ represents a set of parameters that characterize the system; $f(t)$ denotes the excitation force function, and $y(t)$ is the system's dynamic response function; $t$ is the time.   
The operator $\mathcal{M}_{\mathbf{\vect{\mu}}}$ parametrized by $\mathbf{\vect{\mu}}$ maps the excitation force function to the dynamic response function, which typically involves solving the intrinsic dynamics of the system.
For instance, $\mathcal{M}_{\mathbf{\vect{\mu}}}$ can be considered as a Duhamel's integral for a linear single-degree-of-freedom (SDOF) system.

\begin{figure}[!h]
    \centering
    \includegraphics[width=\linewidth]{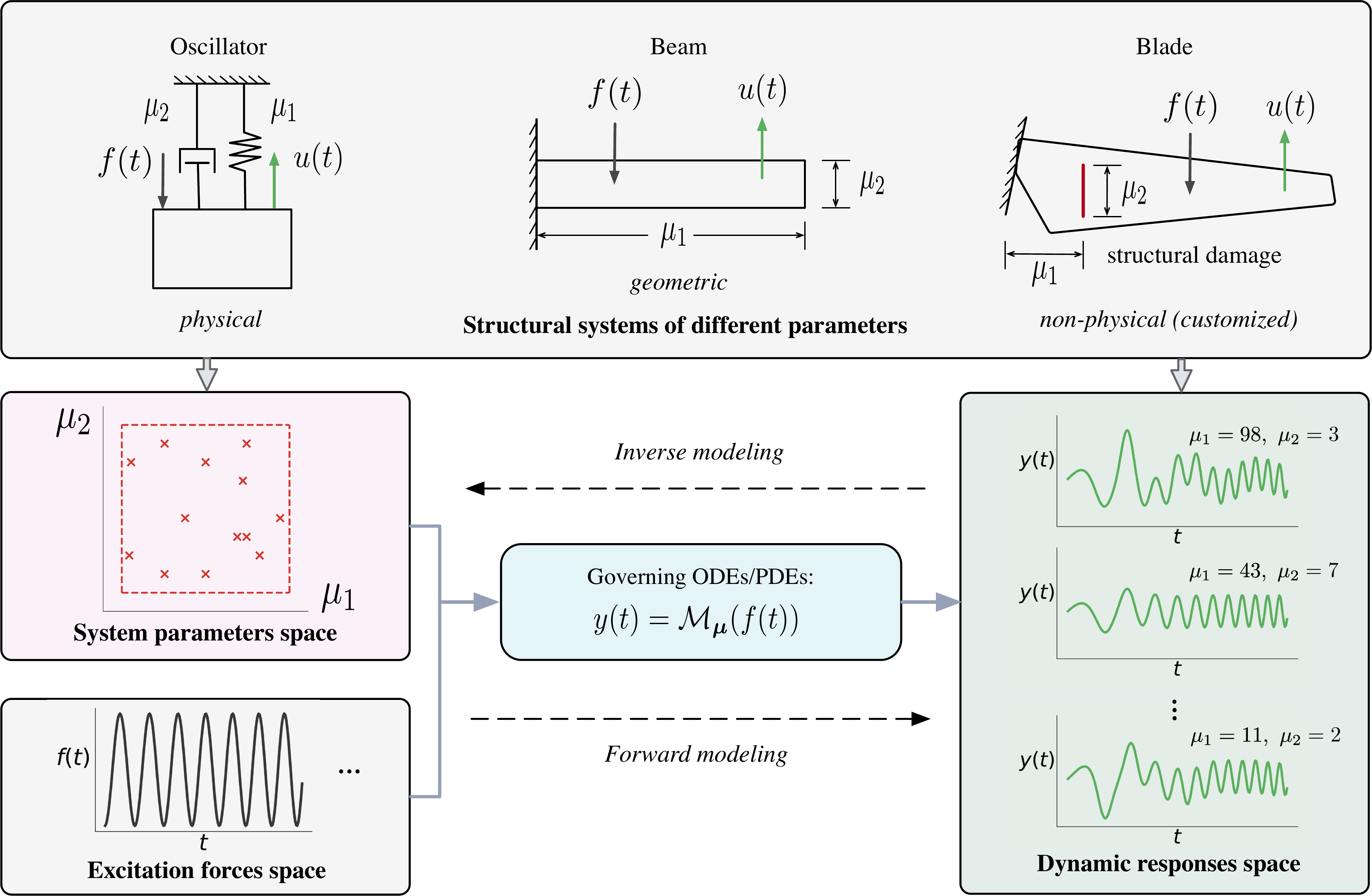}
    \caption{Structural dynamics is defined by the mapping (parameterized by system parameters) from excitation forces to dynamic responses.}
    \label{fig:sd}
\end{figure}

As illustrated in Figure~\ref{fig:sd}, structural dynamics such as oscillators, beams, or blades can be characterized by different classes of parameters, including physical parameters (e.g., stiffness and damping), geometric parameters (e.g., length and width), and non-physical (customized) parameters (e.g., crack damage parameterized via the location and the length of the crack).
For a specific scenario, one may consider a bounded parameter space.
Given data samples from the defined parameter space, this work aims to learn structural dynamics via a data-driven method that generalizes effectively across the defined parameter space.

The proposed framework is comprised of two major tasks as follows:
\begin{itemize}
    \item \textit{Forward modeling}: Predicting dynamic response from system parameters and excitation force. 
    We focus on response prediction when system parameters (either physical or non-physical) differ from those in the training data.
    \item \textit{Inverse modeling}: Estimating system parameters from the pair of dynamic response and excitation force.
    We aim to directly infer the values of system parameters that were \textit{not} included in the training data.
\end{itemize}

Although excitation force estimation (force identification) is another case of inverse modeling, it is not considered in our current attempt.
The primary objective of this research is to develop a unified data-driven framework that addresses both response prediction (forward modeling) and parameter estimation (inverse modeling) in structural dynamics.

\subsection{Neural operators for learning structural dynamics}
\subsubsection{Vanilla DeepONet}
\label{sec:deeponet}
In the setting of vanilla DeepONet~\cite{lu2021learning,zhu2023reliable}, two function spaces are considered.
Let the input function $v$'s variable $x$ be defined on the domain $D \subset \mathbb{R}^d$:
\begin{align}
    v: x \mapsto v(x), x \in D, 
\end{align}
and the output function $u$'s variable $x^\prime$ be defined on a domain $D^{\prime} \subset \mathbb{R}^{d^{\prime}}$:
\begin{align}
    u: x^{\prime} \mapsto u(x^{\prime}), x^{\prime} \in  D^{\prime},
\end{align}
where $x$ and $x^\prime$ can be identical or distinct depending on specific applications~\cite{lu2022comprehensive}. 
Let $\mathcal{V}$ and $\mathcal{U}$ be the function spaces of the input and output functions, respectively.
The mapping between the input function $v$ and the output function $u$ is defined by an operator $\mathcal{G}$:
\begin{equation}
    \mathcal{G}: v \in \mathcal{V} \mapsto u \in \mathcal{U}.
\end{equation}

\begin{figure*}[!h]
    \centering
    \begin{subfigure}{0.5\linewidth}
        \centering
        \includegraphics[height=4.5cm]{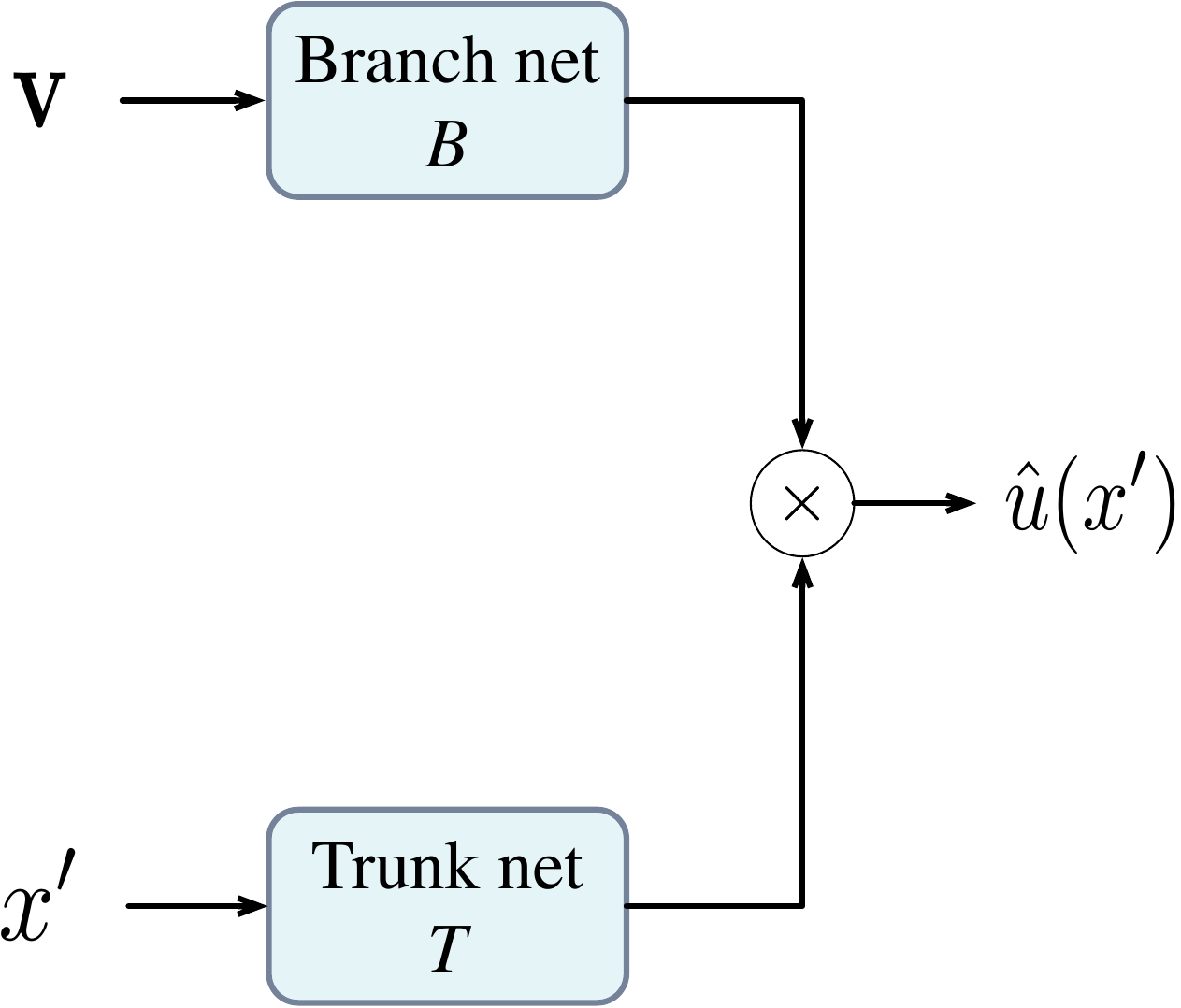}
        \caption{DeepONet}
        \label{fig:deeponet-a}
    \end{subfigure}%
    \begin{subfigure}{0.5\linewidth}
        \centering
        \includegraphics[height=4.5cm]{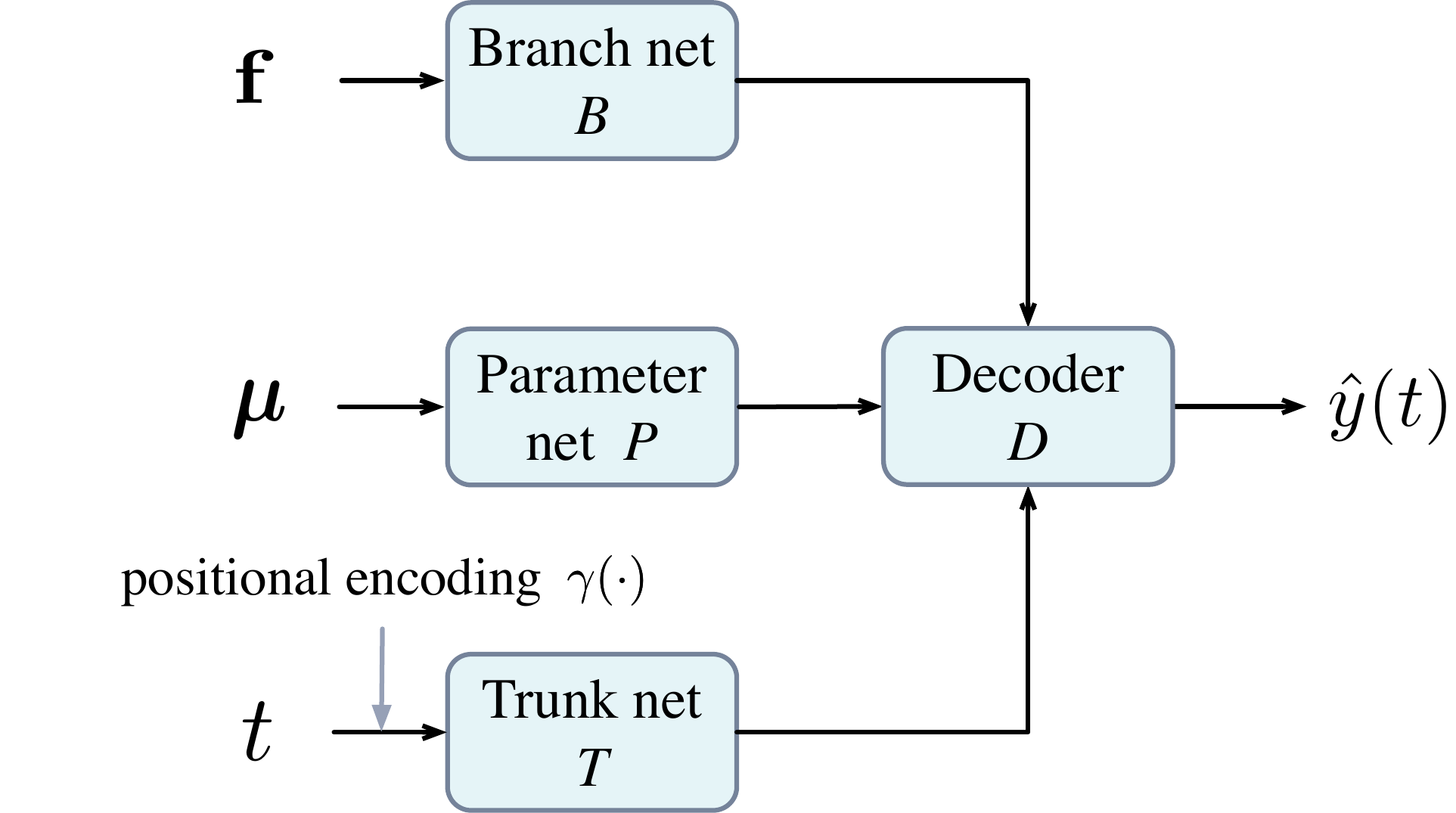}
        \caption{Parametric DeepONet}
        \label{fig:deeponet-b}
    \end{subfigure}
    \caption{Illustrations of vanilla DeepONet and Parametric DeepONet.}
\end{figure*}

Given a finite collection of input-output pairs, the motivation of DeepONet is to approximate the true underlying operator via a specially designed neural network architecture. 
As illustrated in Figure~\ref{fig:deeponet-a}, DeepONet consists of two sub-modules: a branch net $B$ and a trunk net $T$. 
The function $v$ is assumed to be accessed at $m$ evaluation points $\left\{x_1, x_2, \ldots, x_m\right\}$ in $D$, resulting in a discretized input function $ \mathbf{v} = \left[v(x_1), v(x_2), \ldots, v(x_m)\right]$.
The branch net takes the discretized input function $ \mathbf{v}$ as input, generating the $\left[b_1, b_2, \ldots b_n \right]$ as output.
The trunk net takes the evaluation points $x^{\prime}$ in $D^{\prime}$ as input and outputs $\left[\tau_1(x^{\prime}), \tau_2(x^{\prime}), \ldots, \tau_n(x^{\prime})\right]$.
The forward process is defined as: 
\begin{align}
    u(x^{\prime}) & = \mathcal{G}(v)(x^{\prime}) \\
    & \approx 
    \sum_{k=1}^{n} b_{k}(\mathbf{v}) \tau_{k}(x^{\prime}), 
\end{align}
where
$\left[b_1, b_2, ..., b_n \right]$ and $\left[\tau_1, \tau_2, ..., \tau_n \right]$ are the outputs of the branch net $B$ and trunk net $T$, respectively.
 This architecture can be understood as the output function $u(x^\prime)$ is learned as the linear combination of basis function $\tau_k(x^\prime)$ via the associated weight $b_k(\mathbf{v})$.

\subsubsection{Parametric DeepONet for modeling structural dynamics}
\label{sec:para_deeponet}
To apply vanilla DeepONet in modeling structural dynamics, the mapping from excitation force to dynamic response can be considered as:
\begin{equation}
    y(t) = \mathcal{G}(f)(t)
     \approx \sum_{k=1}^{n} b_{k}(\mathbf{f}) \tau_{k}(t),
    \label{eq:forward_deeponet}
\end{equation}
where $\mathbf{f}$ is the discretized excitation force function, $t$ is the time coordinate.
However, applying vanilla DeepONet to model structural dynamics as in Eq.\eqref{eq:forward_deeponet} faces three limitations:
(1) Vanilla DeepONet has a branch net to encode input functions and a trunk net to encode output evaluation points, while it does not have modules for explicitly encoding system parameters $\vect{\mu}$,
making it unsuitable to model parametric structural dynamics formulated in Eq.~\eqref{eq:forward}.
(2) Vanilla DeepONet has difficulty in predicting high-frequency responses due to the spectral bias of trunk net~\cite{tancik2020fourier}, which limits its performance in modeling structural systems with high frequency patterns (see~\ref{sec:pe}).
(3) Vanilla DeepONet generates single-channel output from a single-channel input function, making it unsuitable for modeling multi-degree-of-freedom (MDOF) systems. Therefore, researchers propose to apply multiple independent DeepONets to generate MDOF responses~\cite{garg2022assessment,cai2021deepm}, which leads to the increase of training complexity.

To address these limitations, we introduce the Parametric DeepONet, as illustrated in Figure~\ref{fig:deeponet-b}. This enhanced DeepONet incorporates following modules designed to improve the performance and flexibility in modeling structural dynamics:
\begin{enumerate}
    \item[(1)] \textbf{Parameter net:} The parameter net $P$ aims to encode system parameters $\boldsymbol{\mu}$. Inspired by the multiple-input DeepONet~\cite{jin2022mionet} and conditioning input mechanism~\cite{dumoulin2018feature-wise}, the parameter net operates in parallel with the branch net $B$ and trunk net $T$. Its output is expressed as:
    \begin{equation}
        P(\boldsymbol{\mu}) = [p_1(\boldsymbol{\mu}), \ldots , p_n(\boldsymbol{\mu})] = [p_1, \ldots , p_n].
        \label{eq:param_net}
    \end{equation}
    \item[(2)] \textbf{Positional encoding (PE):} PE enhances the model's performance on high frequency data. It has proven effective in coordinated-based multi-layer perceptrons~\cite{tancik2020fourier} and neural operators~\cite{bunker2024autoencoders,seidman2023variational}. In structural dynamics, responses from both SDOF and MDOF systems could exhibit high or varying frequencies.
    To capture these patterns, we apply a PE called Fourier features mapping $\gamma(t)$ to the time coordinate $t$. For more details on PE, please refer to~\ref{sec:pe}.
    The role of PE is to introduce prior over the frequency spectrum, enabling neural networks to tune the frequencies of output~\cite {tancik2020fourier}. Importantly, PE comes without any additional trainable parameters. In our proposed Parametric DeepONet, PE is applied to the trunk net input, and the forward process becomes: 
    \begin{equation}
        T(\gamma(t)) = [\tau_1, \tau_2, \ldots, \tau_n].
        \label{eq:pdon_trunk}
    \end{equation}
    \item[(3)] \textbf{Decoder:} The decoder $D$ synthesizes output responses from the outputs of the branch net, parameter net, and trunk net. We interpret the dot product operation from the vanilla DeepONet as a linear decoder (LD).
    Alternatively, we introduce a nonlinear decoder (ND) — a neural network that processes the output of LD to further improve the performance of Parametric DeepONet.
    Details of these two decoders are introduced in~\ref{sec:decoder}.
\end{enumerate}

We term the architecture as Parametric DeepONet, denoted as $\hat{\mathcal{M}}$, with its modules designed specifically for the parametric modeling of SDOF and MDOF structural systems.
The overall objective is to approximate the structural dynamics defined in Eq.~\eqref{eq:forward} as follows:
\begin{align}
    y(t) & = \mathcal{M}_{\boldsymbol{\mu}} (f)(t) \\
    & \approx \hat{\mathcal{M}}(\boldsymbol{\mu}, \mathbf{f}, t) 
    = D\left(B(\mathbf{f}), P(\boldsymbol{\mu}), T(\gamma(t))\right),
    \label{eq:parametric_deeponet}
\end{align}
where $\mathbf{f}$ is the discretized excitation force function, $t$ is the time coordinate; $ B(\mathbf{f})$, $ P(\boldsymbol{\mu})$ and $T(\gamma(t))$ are the outputs of the branch net $B$, parameter net $P$ and trunk net $T$, respectively.

\subsection{A unified framework of forward modeling and inverse modeling}
\subsubsection{Forward modeling}
\label{sec:forward}
As introduced in Eq.~\eqref{eq:parametric_deeponet}, the predicted response $\hat{y}(t)=\hat{\mathcal{M}}(\boldsymbol{\mu}, \mathbf{f}, t)$ is a continuous approximation of the true dynamic response function, which can be evaluated at any arbitrary $t$.
However, directly minimizing the functional distance between $\hat{y}(t)$ and $y(t)$ over the entire continuous domain is generally intractable.

In practice, the neural network $\hat{\mathcal{M}}$ is optimized via empirical risk minimization using discretized data.
Given a training dataset $\mathcal{D}_{\text{train}}$, where each $i^{\text{th}}$ sample $ s_i = \left(\mathbf{f}_i, \vect{\mu}_i, \mathbf{y}_i\right)$ is a triple comprised of excitation force $\mathbf{f}$, system parameters $\mathbf{\boldsymbol{\mu}} $, and dynamic response $\mathbf{y}$. 
The resolution of the excitation force and the response could be identical or different according to applications.
In our implementation of model training, we assume the same time resolution $r$ for excitation force and dynamic response, i.e., $\mathbf{f}_i = [f_i(t_1), \ldots, f_i(t_{r})]$ and $\hat{\mathbf{y}}_i = [\hat{y}_i(t_1), \ldots, \hat{y}_i(t_{r})]$.
During training, the network parameters $\theta$ are optimized as:
\begin{align}
   \hat{\mathcal{M}}^*  
    = \operatorname*{arg\,min}_{\theta} \sum_{\mathbf{y}_i \ \in \ \mathcal{D}_{\text{train}}} \mathcal{L}( \hat{\mathbf{y}}_i, \mathbf{y}_i),
 \end{align}
where $\mathcal{L}(\cdot)$ denotes the loss function,  and our implementation uses the normalized root mean squared error (NRMSE)defined in Eq.\eqref{eq:nrmse}.
Notably, once trained, $\hat{y}(t)$ remains a continuous function of time, allowing for flexible evaluation at arbitrary $t$. This property enables applications such as zero-shot super-resolution, presented in~\ref{sec:super-resolution}.
\begin{figure}[!h]
    \centering
    \includegraphics[width=0.8\linewidth]{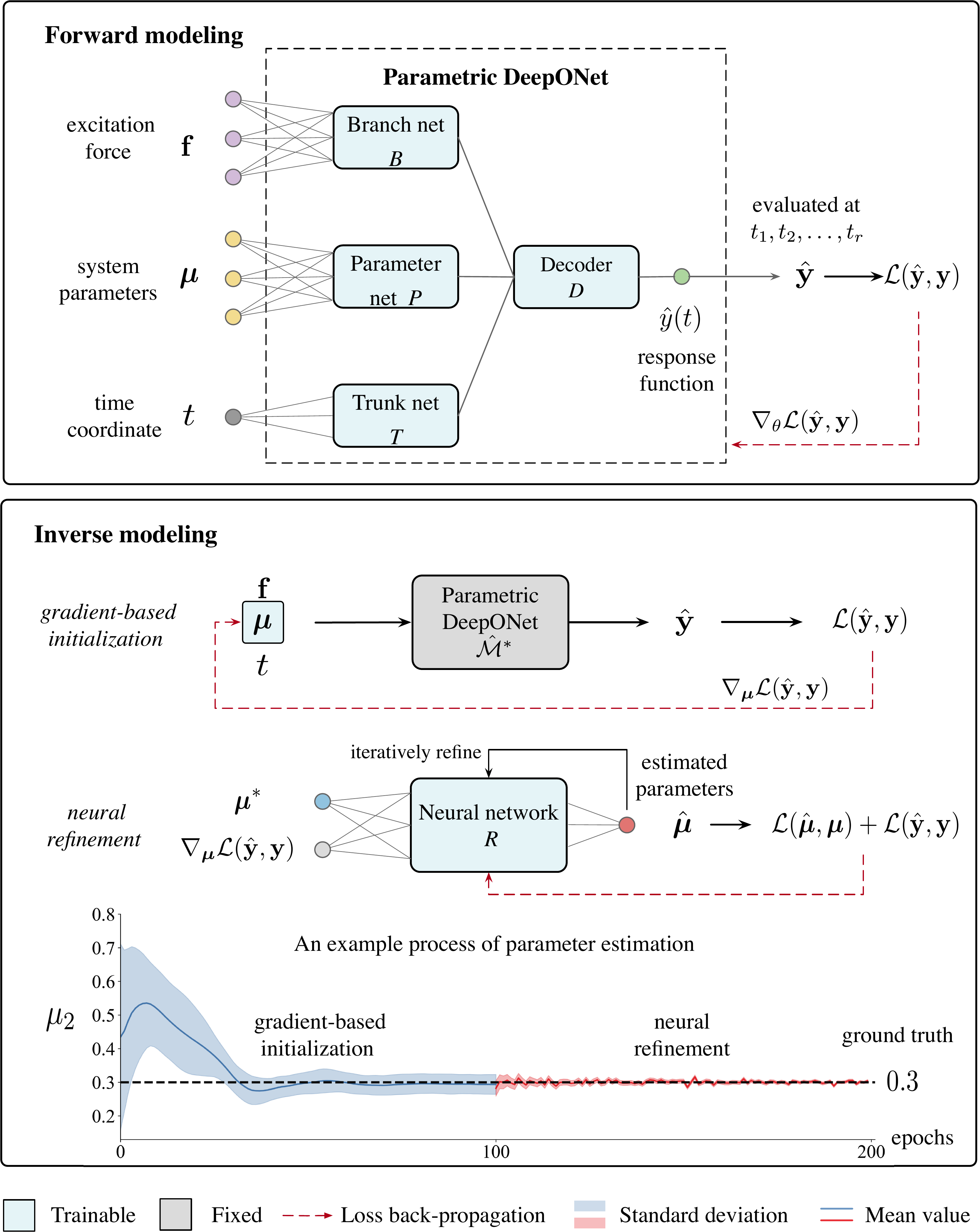}
    \caption{A unified framework for forward and inverse modeling. 
    In forward modeling, a neural network takes excitation force $f$ and system parameter $\vect{\mu}$ as input and predicts dynamic response $y$. 
    In inverse modeling, gradient-based initialization initializes system parameters $\vect{\mu}$ by minimizing the forward prediction loss given the $(f, y)$; subsequent neural refinement employs a neural network to take the initially estimated parameter as input and generates refined parameter estimation results. The standard deviation and mean value are computed based on parameter estimation with different initializations.
    A detailed workflow is further explained in Algorithm~\ref{algo_forward} and~\ref{algo2}.
    }
    \label{fig:framework_nn}
\end{figure}

\subsubsection{Inverse modeling}
\label{sec:inverse}
In inverse modeling, the framework aims to estimate system parameters from pairs of excitation forces and dynamic responses.  
The parameter estimation involves two steps: \textit{gradient-based initialization}  and \textit{neural refinement}. 
The first step provides an initial estimate of system parameters via forward loss minimization, while the second step refines these estimates by another trainable neural network $R$ to improve the performance.

In gradient-based initialization, we hypothesize that a well-trained neural network $\hat{\mathcal{M}}^*$ in forward modeling can serve as a surrogate model that predicts responses $\hat{\mathbf{y}}$. 
To find an optimal system parameter ${\vect{\mu}}^*$, a pragmatic way is to minimize the discrepancy between the response prediction $\hat{\mathbf{y}}$ and its desired dynamic response $\mathbf{y}$. For summing each $i^{\text{}th}$ training sample:
\begin{align}
    \label{eq:inverse_2}
    \vect{\mu}^* = \operatorname*{arg\,min}_{\vect{\mu}} \sum_{ \mathbf{y}_i \ \in \ \mathcal{D}_{\text{train}}} \mathcal{L} \left(\hat{\mathbf{y}}_i,\mathbf{y}_i \right) .
\end{align}
The parameters of the pre-trained neural network $\hat{\mathcal{M}}^*$ are fixed during inverse modeling.
Since the neural network is differentiable with respect to $\vect{\mu}$, the parameter estimation can be achieved via the gradient-based optimization on the forward prediction loss. $\vect{\mu}^*$ denotes the parameter estimation result with minimized forward prediction loss.

However, estimating system parameters $\vect{\mu}$ solely relying on the forward loss gradient with respect to $\vect{\mu}$, i.e., $\nabla_{\vect{\mu}} \mathcal{L} \left(\hat{\mathbf{y}_i}, \mathbf{y}_i \right)$ can encounter ill-posed issues due to the non-uniqueness of the solutions. 
Therefore, the results of gradient-based initialization are often inaccurate, being stuck at local minima.
As shown in Figure~\ref{fig:framework_nn}, for the selected example parameter, although the gradient-based initialization process generates satisfactory mean values, the high standard deviation implies the ill-posed problem -- choosing different initialization of $\vect{\mu}$ leads to very distinct estimation results, i.e., the estimation is very sensitive to the initialization.

To alleviate this issue, we introduce the \textit{neural refinement} method.
A refinement network termed as $R$, is designed to learn the gradient required for updating system parameters.
This approach is inspired by the iterative deep neural networks~\cite{adler2017solving},
which have shown effectiveness in mitigating ill-posed problems in tomographic inversion. 
In the context of parameter estimation, the parameter refinement network takes the initially estimated parameters $\vect{\mu}^*$ from gradient-based initialization, together with the forward loss gradient $\nabla_{\vect{\mu}} \mathcal{L} (\hat{\mathbf{y}},\mathbf{y})$ as inputs, to iteratively generate learned steps $\Delta \vect{\mu}$ for updating initially estimates $\vect{\mu}^*$.
This parameter refinement network is implemented as follows: 
\begin{align}
    \hat{\vect{\mu}} = \textsc{Iterative} \left(R, \vect{\mu}^*, \nabla_{\vect{\mu}} \mathcal{L} (\hat{\mathbf{y}},\mathbf{y}) \right),
\end{align}
where $\hat{\vect{\mu}}$ is the refinement results of parameter estimation, and the \textsc{Iterative}$(\cdot)$ is an iterative process detailed in lines 14-19 of Algorithm~\ref{algo2}. 

Suppose the parameter refinement network $R$ is parameterized by $\beta$. During the training stage for the inverse modeling, not only estimated parameter $\hat{\vect{\mu}}$ should meet its labeled value $\vect{\mu}$, but also we demand that the corresponding predicted dynamic response $y(\hat{\vect{\mu}})$ should match its desired value $y$, evaluated at $t_1,\ldots, t_{r} $.  This is optimized as follows:
\begin{align}
     R^* =
     R_{\beta^*} = 
     \operatorname*{arg\,min}_{\beta} \sum_{(\vect{\mu}_i,\mathbf{y}_i) \ \in \ \mathcal{D}_{\text{train}}} \mathcal{L} (\hat{\vect{\mu}}_i, \vect{\mu}_i) + \mathcal{L}( \hat{\mathbf{y}}_i, \mathbf{y}_i).
\end{align}

As illustrated in Figure~\ref{fig:framework_nn}, the neural refinement steadily improves the parameter estimation, as well as reduces the standard deviation of the estimated parameters. This process effectively mitigates the ill-posed problems, leading to more robust and accurate results in parameter estimation. 

\subsection{Summary}
The training processes of forward modeling and inverse modeling are presented in Algorithms~\ref{algo_forward} and ~\ref{algo2}, respectively.
After training, the test process of forward modeling involves applying the trained forward model $\hat{\mathcal{M}^*}$ to the test data samples, predicting dynamic responses.
The test process of inverse modeling proceeds in two steps as: first, gradient-based initialization is applied to the test data samples to obtain the preliminary estimate of the system parameters; second, the trained refinement network $R^*$ can iteratively generate the update steps, and yield the refined estimates as the final parameter estimation results. 
\begin{algorithm}[H]
    \caption{The Training Process of Forward Modeling}
    \label{algo_forward}
    \begin{algorithmic}[1]
    \State \textbf{Inputs:} 
        Training dataset $\mathcal{D}_{\text{train}} = \{s_i = (\mathbf{f}_i, \vect{\mu}_i ,\mathbf{y}_i)\}_{i=1}^{N_t}$,
        forward net $\hat{\mathcal{M}}$
        \hrule \vspace{0.7em}
        \For{epoch $= 1 : N $}
        \For{data sample $s_i$, $i =  1 : B $ in data batch}
        \State  Predict dynamic response: $\hat{\mathbf{y}}_i = \hat{\mathcal{M}}(\vect{\mu}_i, \mathbf{f}_i)$ 
        \State  Compute forward loss: $\mathcal{L}_{\text{forward}} = \sum_{i=1}^{B} \mathcal{L}(\hat{\mathbf{y}}_i, \mathbf{y}_i) / B$ 
        \State Update $\hat{\mathcal{M}}$ via gradient descent w.r.t $\nabla \mathcal{L}_{\text{forward}}$
        \EndFor
        \EndFor
        \vspace{0.2em}
        \State Save $\hat{{\mathcal{M}}}$ with minimized forward prediction loss as $\hat{\mathcal{M}}^*$
    \end{algorithmic}
    \end{algorithm}
\begin{algorithm}[!h]
    \caption{The Training Process of Inverse Modeling}
    \label{algo2}
    \begin{algorithmic}[1]
    \State \textbf{Inputs:} Training dataset $\mathcal{D}_{\text{train}} = \{s_i = (\mathbf{f}_i, \vect{\mu}_i ,\mathbf{y}_i)\}_{i=1}^{N_t}$, 
    trained forward net $\hat{\mathcal{M}}^*$, parameter refinement net $R$
    \hrule \vspace{0.5em}
    \Procedure{Gradient-based initialization}{} 
    \For{epoch $= 1 : N $}
    \For{data sample $s_i$, $i =  1 : B $ in data batch}
    \State Sample random system parameters: $\tilde{\vect{\mu}}_i \thicksim U(\vect{\mu}_{\text{min}}, \vect{\mu}_{\text{max}})$
    \Comment{ $U(\vect{\mu}_{\text{min}}, \vect{\mu}_{\text{max}})$ is a uniform distribution}
    \State Predict dynamic response: $\hat{\mathbf{y}}_i = \hat{\mathcal{M}}^*(\tilde{\vect{\mu}}_i , \mathbf{f}_i)$
    \State Compute forward loss: $\mathcal{L}_{\text{forward}} = \sum_{i=1}^{B} \mathcal{L}(\hat{\mathbf{y}}_i, \mathbf{y}_i) / B $ 
    \EndFor \State Update $\tilde{\vect{\mu}}_i$ with gradient descent via forward loss minimization w.r.t $\nabla \mathcal{L}_{\text{forward}}$
    \EndFor \State Save $\tilde{\vect{\mu}}$ with minimized forward loss as $\vect{\mu}^*$
    \EndProcedure
    \State ======================================================================
    \vspace{0.2em}
    \Procedure{Neural refinement}{}
    \For{epoch $= 1 : N $}
    \For{data sample $s_i$, $i =  1 : B $ in data batch}
    \Procedure{Iterative}{$R,\vect{\mu}^*, \nabla \mathcal{L} (\hat{\mathbf{y}}, \mathbf{y})$}
    \State Iteration start: $\vect{\mu}_i^0 = \vect{\mu}^*_i$ 
    \For{refine iteration j = 1 : J}
    \State  Generate an update step: $\Delta \vect{\mu}_i^j = R\left(\vect{\mu}_i^{j-1}, \mathcal{L}\left(\hat{\mathbf{y}}_i(\vect{\mu}_i^{j-1}), \mathbf{y}_i\right) \right)$
    \State Update parameters: $\vect{\mu}_i^j = \vect{\mu}_i^{j-1} + \Delta \vect{\mu}_i^j$
    \EndFor \EndProcedure
    \State \textbf{Output:} $\hat{\vect{\mu}}_i = \vect{\mu}_i^{J}$ \Comment{Iterative refinement results}
    \State   Compute inverse estimation loss: $\mathcal{L}_{\text{inverse}} = \sum_{i=1}^{B} \mathcal{L}(\hat{\vect{\mu}}_i, \vect{\mu}_i) / B $
    \State   Compute forward prediction loss: $\mathcal{L}_{\text{forward}} = \sum_{i=1}^{B} \mathcal{L}(\hat{\mathbf{y}}_i(\hat{\vect{\mu}}_i), \mathbf{y}_i) / B$ 
     \State Update $R$ with gradient descent w.r.t $\nabla (\mathcal{L}_{\text{inverse}} + \mathcal{L}_{\text{forward}}) $
    \EndFor \EndFor \State Save $R$ with minimized loss as $R^*$
    \EndProcedure
    \end{algorithmic}
\end{algorithm}

\section{Experiments}
We consider two validation cases to evaluate the effectiveness of the proposed framework: 
Case 1 involves a forced Duffing oscillator, which is an illustrative example of a single-degree-of-freedom (SDOF) system with nonlinearity;
Case 2 involves a laboratorial wind turbine blade, an experimental structure with multiple degrees of freedom (MDOF).

\subsection{General setting}
The proposed framework is a deep learning-based method, which requires simulation or real data for supervised training. 
The normalized root mean squared error (NRMSE) is utilized as loss function and evaluation metric, which is defined as:
\begin{align}
    \text{NRMSE} =\sqrt{\frac{\sum_{i=1}^{N}||\hat{\mathbf{y}}_{i}-\mathbf{y}_{i}||^2}{\sum_{i=1}^{N}{\mathbf{y}}_{i}^2}},
    \label{eq:nrmse}
\end{align}
where $\hat{\mathbf{y}}_{i}$ and $\mathbf{y}_{i}$ are the predicted and true values, respectively; $N$ is the number of data samples. For comparative analysis, the performance of the proposed framework is compared with the following mainstream deep learning models:

\begin{itemize}
    \item Parametric DeepONet: As introduced in Section~\ref{sec:para_deeponet}, the inputs of the branch net, parameter net, and trunk net are the excitation force $\mathbf{f}$, system parameters $\vect{\mu}$, and response evaluation point $t$, respectively. The output is the response acceleration $\ddot{\mathbf{x}} = [x(t_1),\ldots, x(t_r)]$.
    Parametric DeepONet (LD) (with a linear decoder) and Parametric DeepONet (ND) (with a nonlinear decoder) are implemented as two options of Parametric DeepONet. The implementation details of LD and ND is provided in~\ref{sec:decoder}. 
    \item Vanilla DeepONet: It includes one branch net and one trunk net as introduced in Section~\ref{sec:deeponet}.
    The input of the branch net is the concatenation of excitation force $\mathbf{f}$ and system parameters $\vect{\mu}$, and the input of the trunk net is the evaluation points of dynamics response, which is time instances $t$. The output of vanilla DeepONet is the response acceleration $\ddot{\mathbf{x}}$.
    \item Multi-layer perceptron (MLP): It maps the input to the output through multiple fully connected layers with non-linear activation functions.
    The concatenation of excitation force $\mathbf{f}$ and system parameters $\vect{\mu}$ is the input of MLP. 
    The response acceleration $\ddot{\mathbf{x}}$ is the output of MLP.
    \item Convolutional neural network (CNN): It consists of several 1D-convolutional and 1D-deconvolutional layers. The convolutional layers map the 
    concatenation of excitation force $\mathbf{f}$ and system parameters $\vect{\mu}$ to a latent space, followed by  1D deconvolutional layers mapping the latent space to the response acceleration $\ddot{\mathbf{x}}$.
\end{itemize}
The analysis of baseline models' architecture, trainable parameters, and other details are systematically investigated in~\ref{sec:optimal_architecture}.

\subsection{Case 1 - Duffing oscillator}
\label{sec:case1}
We first consider a numerical SDOF system - the Duffing oscillator.
The Duffing oscillator is one of the prototype systems in nonlinear dynamics~\cite{kovacic2011duffing,duffing1918erzwungene},
commonly used for modeling stiffening springs, bulking beams, nonlinear electronic circuits, etc. 
The governing equation is a second-order, nonlinear ordinary differential equation:
\begin{align}
     \ddot{x}(t) = -\mu_1 x(t) - \mu_2 \dot{x}(t) - \mu_3 x^3(t) + f(t) ,
    \label{eq:duffing}
\end{align}
where $\mu_1$ controls the stiffness; $\mu_2$ controls the damping; $\mu_3$ is the nonlinearity parameter; $f(t)$ is the externally driven (excitation) force; $x(t), \dot{x}(t), \ddot{x}(t)$ are the displacement, velocity, and acceleration of the dynamic response, respectively.
In this work, the oscillator is considered to be subjected to zero initial conditions, $x(0) = 0$ and $\dot{x}(0) = 0$. 
We consider a sine sweep excitation force with linearly increasing frequency~\cite{farina2000simultaneous}, 
which is defined as:
\begin{align}
    f(t) = A \sin{\left(2\pi \left[f_{\text{low}}  \cdot t + 
    \frac{ (f_{\text{up}}-f_{\text{low}}) \cdot t^2 }{2T}\right] \right)} ,
    \label{eq:sine-sweep}
\end{align}
where $A$ is the amplitude, $ f_{\text{low}}, f_{\text{up}}$ (in Hz) are the lower and upper limits of the sweep frequency, respectively, and $T$ is the total duration of the excitation.

\subsubsection{Parametrization}
\label{sec:param_case_1}
We parameterize the Duffing oscillator by stiffness $\mu_1$ and damping $\mu_2$ -- only two parameters for the reason of more intuitive visualization to illustrate the framework. 
Thus, the system parameters are two-dimensional physical parameters, denoted as $ \vect{\mu} = (\mu_1, \mu_2)$.
To comprehensively evaluate the generalization performance, we design four cases of increasing complexity (Case 1a - Case 1d), 
where the system parameters for training and testing include \textit{interpolation} and \textit{extrapolation} scenarios. It is noted that most data-driven methods focus on the interpolation cases, while extrapolation cases are less explored.
Figure~\ref{fig:case1_params} shows the parameter ranges for each case, detailed as follows:
\begin{figure*}[!htb]
    \centering
    \begin{subfigure}{\textwidth}
        \centering
        \includegraphics[width=0.6\linewidth]{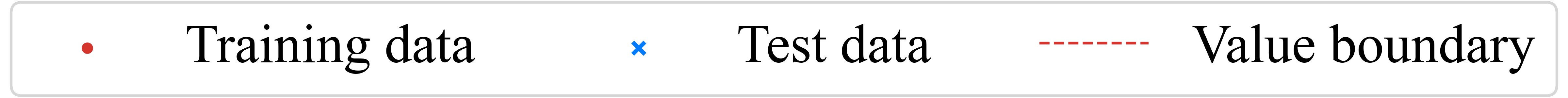}
    \end{subfigure}
    
    \begin{subfigure}{0.4\textwidth}
        \includegraphics[width=\linewidth]{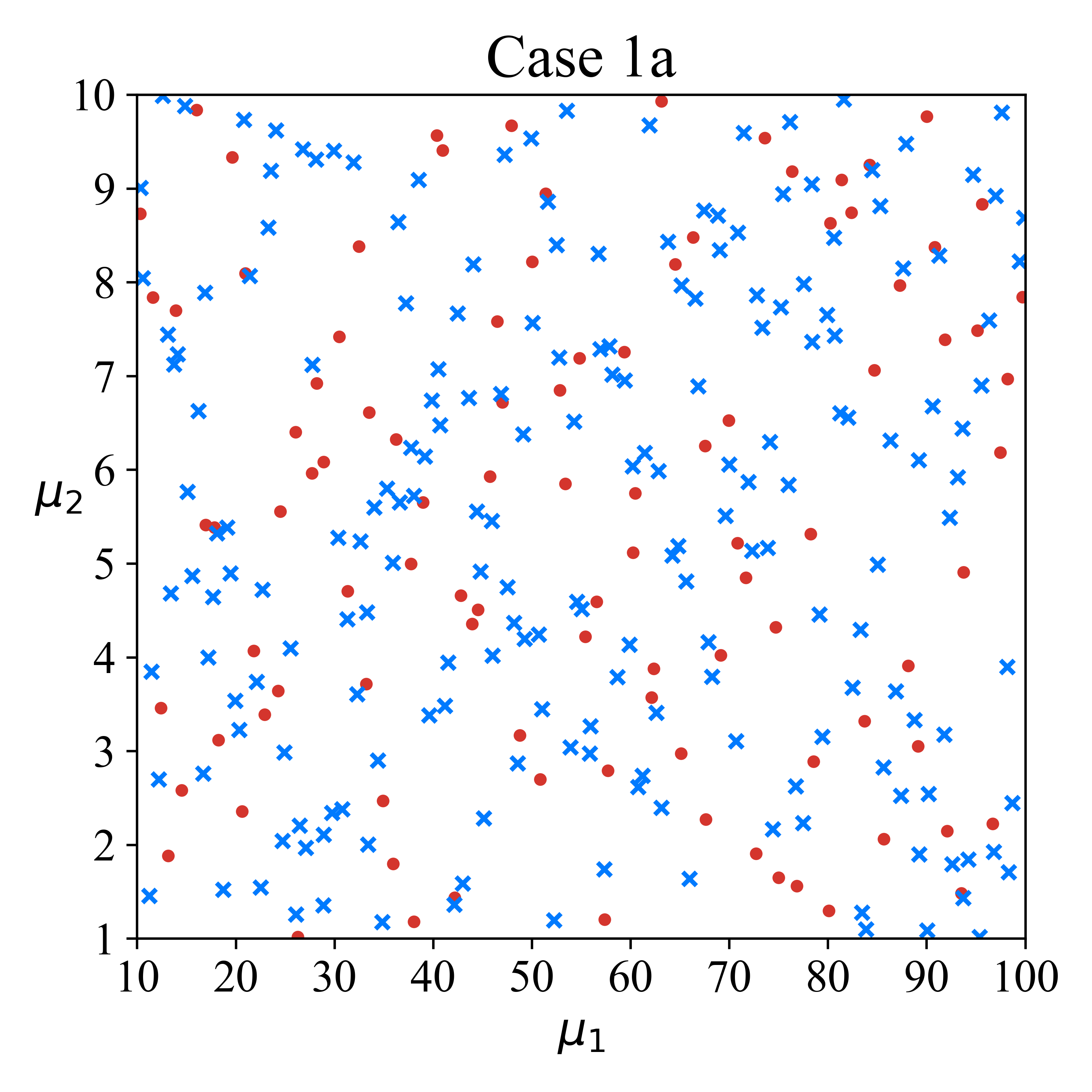}
    \end{subfigure}%
    \begin{subfigure}{0.4\textwidth}
        \includegraphics[width=\linewidth]{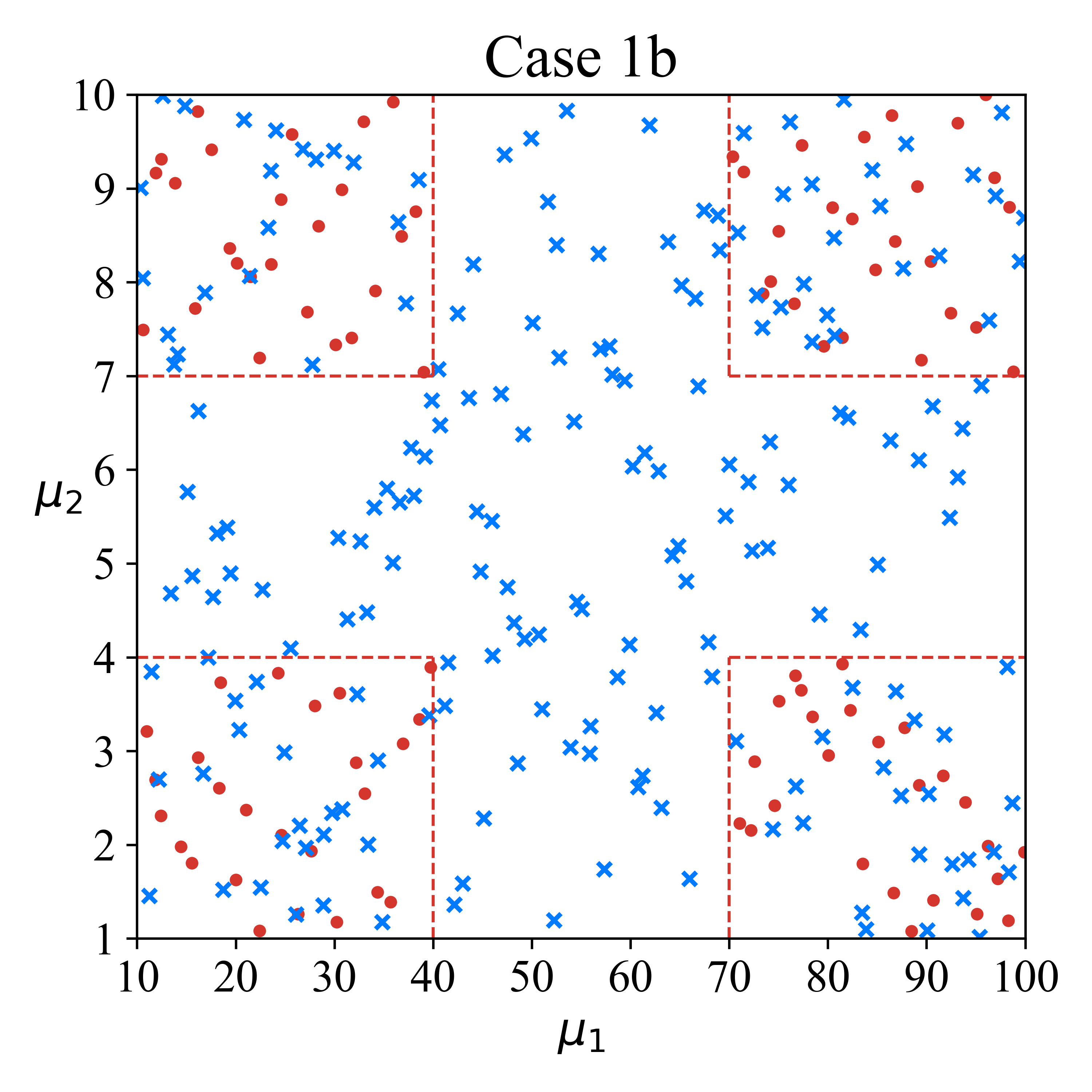}
    \end{subfigure}
    
    \begin{subfigure}{0.4\textwidth}
        \includegraphics[width=\linewidth]{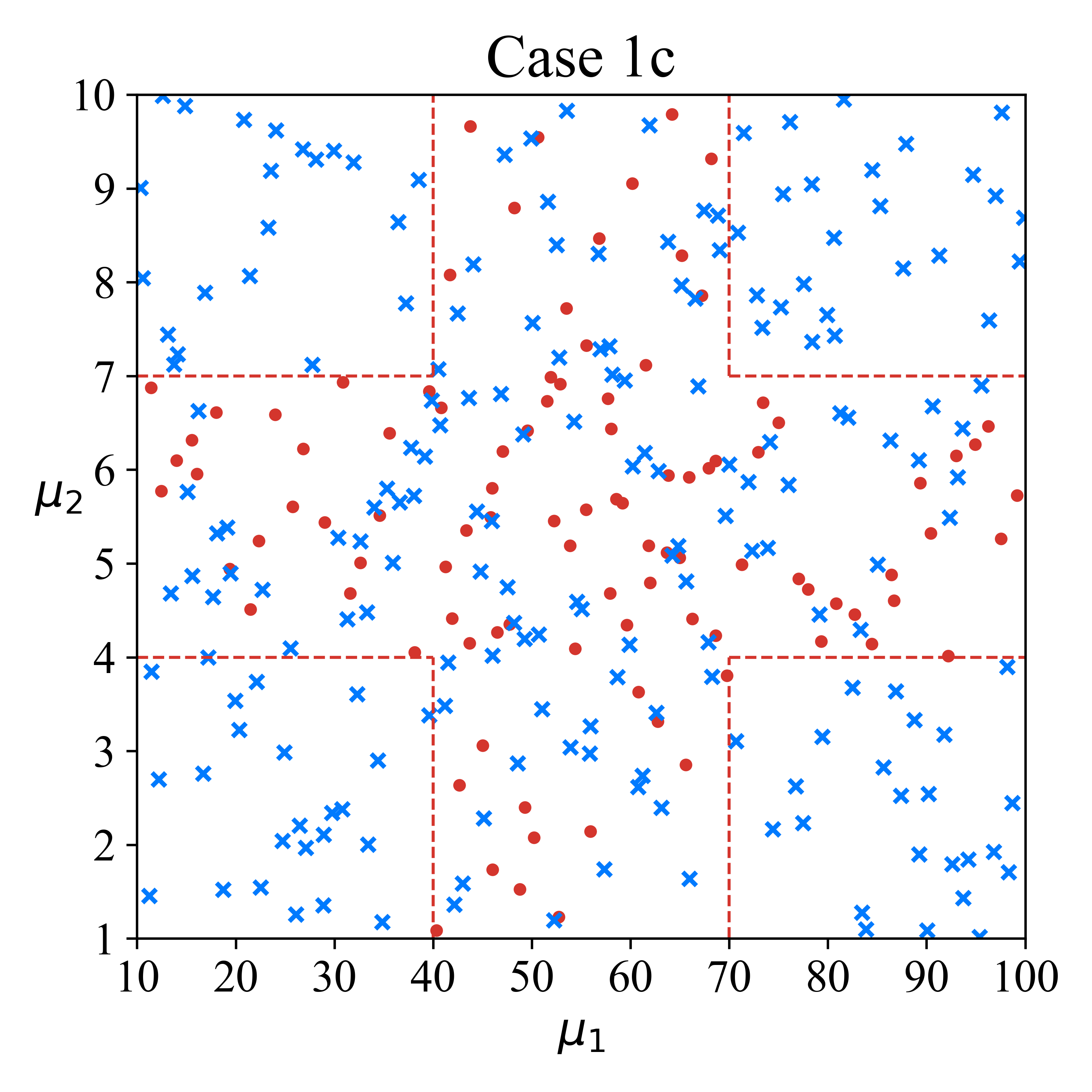}
    \end{subfigure}%
    \begin{subfigure}{0.4\textwidth}
        \includegraphics[width=\linewidth]{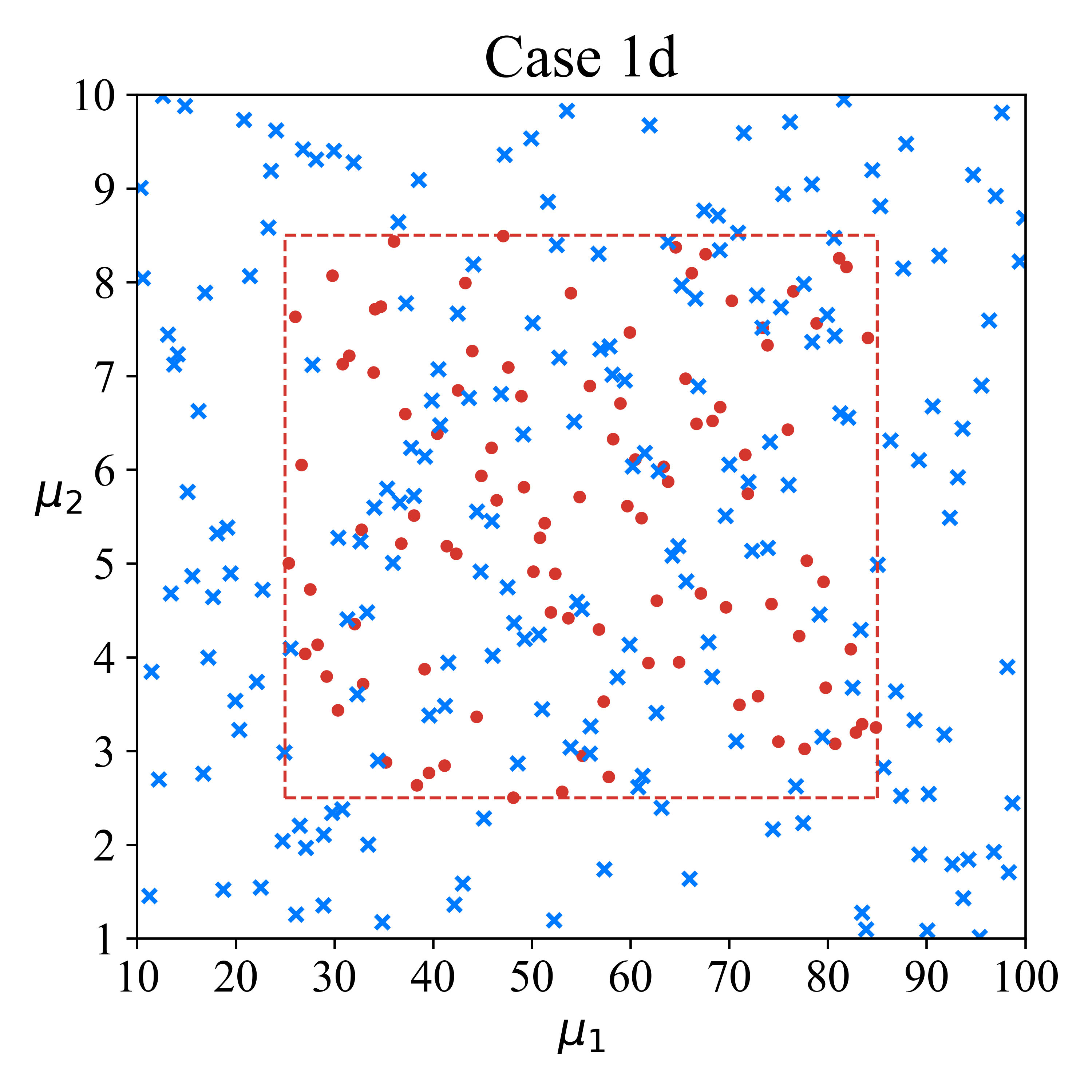}
    \end{subfigure}
    \caption{Ranges of system parameters of training and test data in Case 1 ($\mu_1$ is the stiffness, $\mu_2$ is the damping).}
    \label{fig:case1_params}
\end{figure*}

\begin{figure}[!htb]
    \centering
    \begin{subfigure}{0.45\textwidth}
    \includegraphics[width=\linewidth]{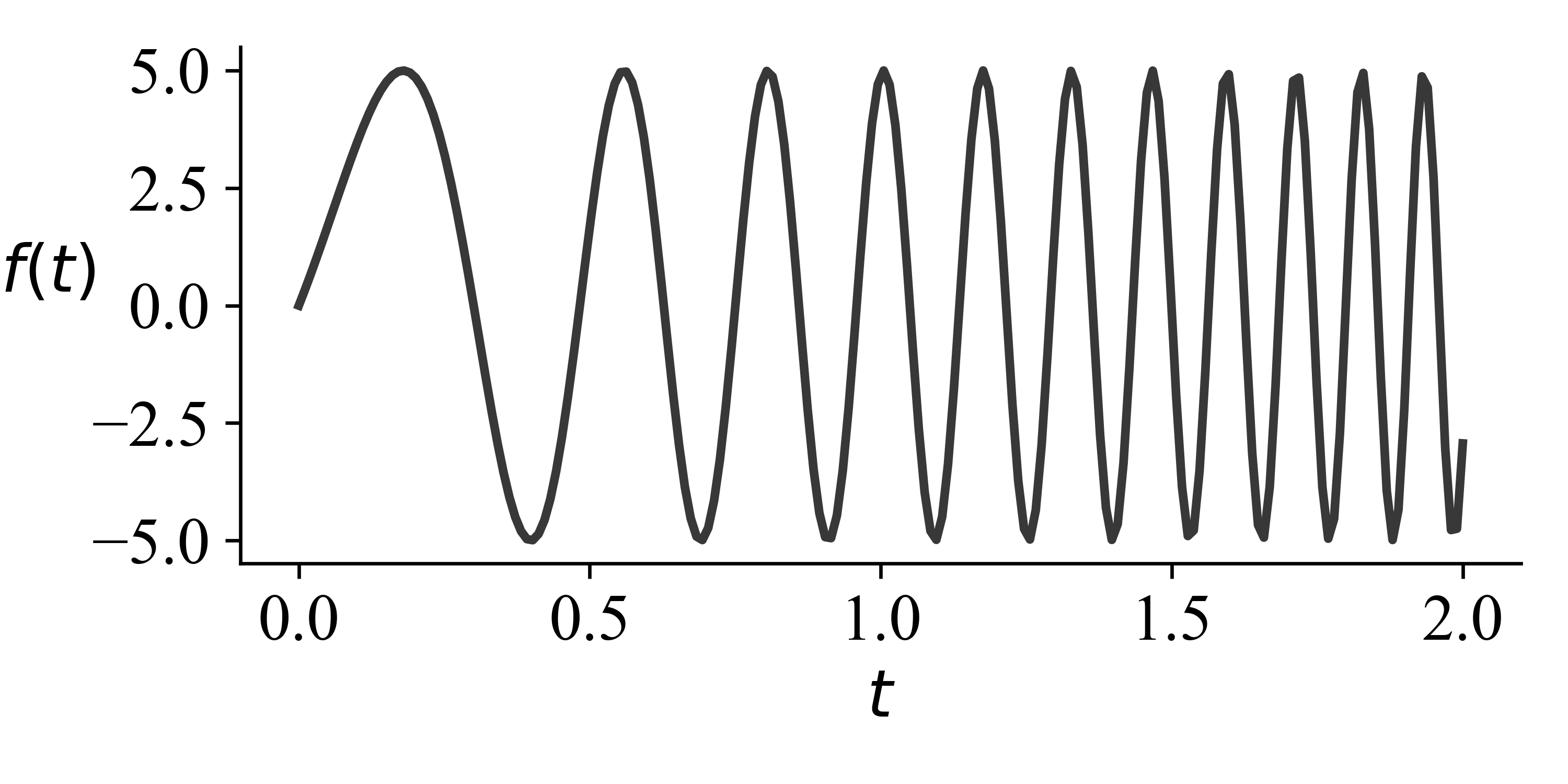}
    \end{subfigure}%
    \begin{subfigure}{0.45\textwidth}
    \includegraphics[width=\linewidth]{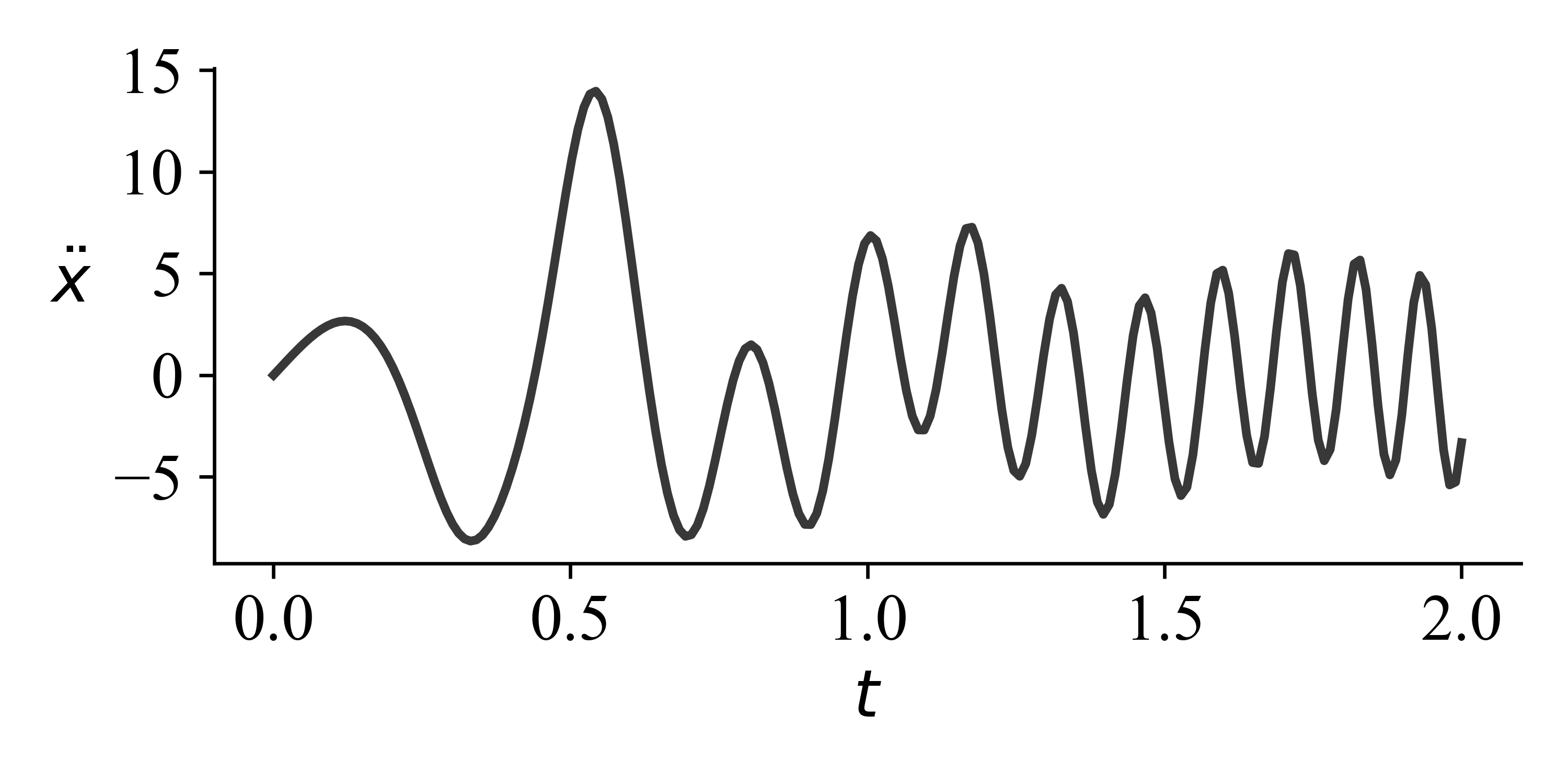}
    \end{subfigure}
    \caption{A data sample in Case 1. Left: the excitation force; Right: the response acceleration ($f(t), \ddot{x} - m/s^2 ,t - s$ ). }
    \label{fig:case1-excitation}
\end{figure}

\begin{itemize}
\item \textit{Case 1a}: This is an \textit{interpolation} case; system parameters are sampled from the bounded domain for training and testing.
The stiffness $\mu_1$ ranges from $[10, 100]$ and the damping $\mu_2$ ranges from $[1,10]$.

\item \textit{Case 1b}: This case involves \textit{extrapolation}, 
where the system parameters in the training dataset cover some limited subdomains within the bounded range. 
The stiffness $\mu_1$ is in range $[10, 40] \cup [70, 100]$ and the damping $\mu_2$ is in range $[1, 4] \cup [7, 10]$ .

\item \textit{Case 1c}: This case involves \textit{extrapolation}, where the system parameters in the training dataset cover the disjoint subdomains (compared to Case 1b) within the bounded range. 
The stiffness $\mu_1$ ranges from $[10, 100]$ and the damping $\mu_2$ ranges from $[1,10]$.

\item \textit{Case 1d}: Another \textit{extrapolation} case, where the system parameters for training are confined to a more central subdomain. 
The stiffness $\mu_1$ ranges from $[25, 85]$ and the damping $\mu_2$ ranges from $[2.5, 8.5]$.
\end{itemize}

For dataset preparation, the system parameters $\vect{\mu} = (\mu_1,\mu_2)$ are generated by Latin hypercube sampling method~\cite{mckay2000comparison}, 
based on the ranges defined in Cases 1a-1d.
The nonlinear parameter $\mu_3$ is set to $1 \times 10^4$ in all cases.
Figure~\ref{fig:case1_params} visualizes the distribution of training and test data, which consist of 100 and 200 data samples, respectively.
In addition, the test data remains the same across Cases 1a-1d to allow for a fair comparison of generalization performances.

For a data sample, the response acceleration is generated through numerical simulation of the ODE defined in Eq.~\eqref{eq:duffing} by the \texttt{scipy} package and the \texttt{odeint} function in Python. 
The excitation force $f(t)$ in Eq.~\eqref{eq:sine-sweep} is with $A = 5$, $ (f_{\text{low}},f_{\text{up}}) = (1, 10) $ Hz. 
The excitation force $f(t)$ and response acceleration $\ddot{x}(t)$ are simulated for a duration of $T = 2$ seconds with the time step $\Delta t = 0.01$ seconds, resulting in a length of 200 for both excitation and response. 
Thus, each data sample is a triple of $(\mathbf{f}, \vect{\mu}, \ddot{\mathbf{x}})$ with $\mathbf{f} \in \mathbb{R}^{200\times1} $, $ \vect{\mu} \in \mathbb{R}^{2}$ and $\ddot{\mathbf{x}} \in \mathbb{R}^{200\times1} $. 
Figure~\ref{fig:case1-excitation} shows the excitation force and the response acceleration of a selected data sample.

\subsubsection{Forward modeling results}
\label{sec:case1-forward}
Table~\ref{tab:case1_forward} presents the quantitative results of response prediction of different models in Case 1.
Parametric DeepONet (ND) consistently achieves the best performance (in terms of NRMSE) across Case 1a-1d, with a slight advantage over Parametric DeepONet (LD) and MLP.
Parametric DeepONet (LD) and MLP achieve second or third-best performance.
In contrast, DeepONet performs noticeably worse in Case 1a and 1b, exhibiting a clear performance gap compared to the top three models. 
The CNN exhibits the highest and nearly constant NRMSEs in all cases, indicating that it struggles to capture the forward dynamics in Case 1.
Concatenating system parameters with excitation force is not an effective encoding method for the CNN — particularly when the dimensionality of system parameters is significantly smaller than that of the input force. Parametric information at the boundary of CNN's input may not be effectively propagated by shallow networks. Moreover, the convolutional layer with stride operation can lead to information loss by discarding values at the boundary.

\begin{table*}[!htb] 
    \centering
    \caption{NRMSE of response prediction on the \textbf{test} dataset in Case 1, for different cases and models. All values are scaled by $10^{-2}$.}
    \label{tab:case1_forward} 
    \begin{tabular}{lcc>{\centering\arraybackslash}m{2cm}>{\centering\arraybackslash}m{2cm}>{\centering\arraybackslash}m{2cm}}
        \hline
       Case & \makecell{Parametric \\ DeepONet (LD)} & \makecell{Parametric \\ DeepONet (ND)} & DeepONet   & MLP  &  CNN \\
       \cline{2-5}
        \hline
        Case 1a  & $ 2.63 $ & $2.50$ & $ 4.68 $ & $ 2.69 $ &$ 30.34 $  \\
        Case 1b  & $ 2.56 $  & $2.45$ &$ 4.36 $ & $ 2.87 $ &$ 30.42 $  \\
        Case 1c  & $ 16.2 $ & $12.2$ &$ 14.3 $ & $ 14.3 $ & $ 30.72 $ \\
        Case 1d  & $ 11.6 $ & $11.4$ &$ 12.8 $ & $ 12.1 $  &$ 30.57 $\\ 
        \hline
    \end{tabular}
\end{table*}

\begin{figure*}[!htb]
    \centering
    \includegraphics[width=\linewidth]{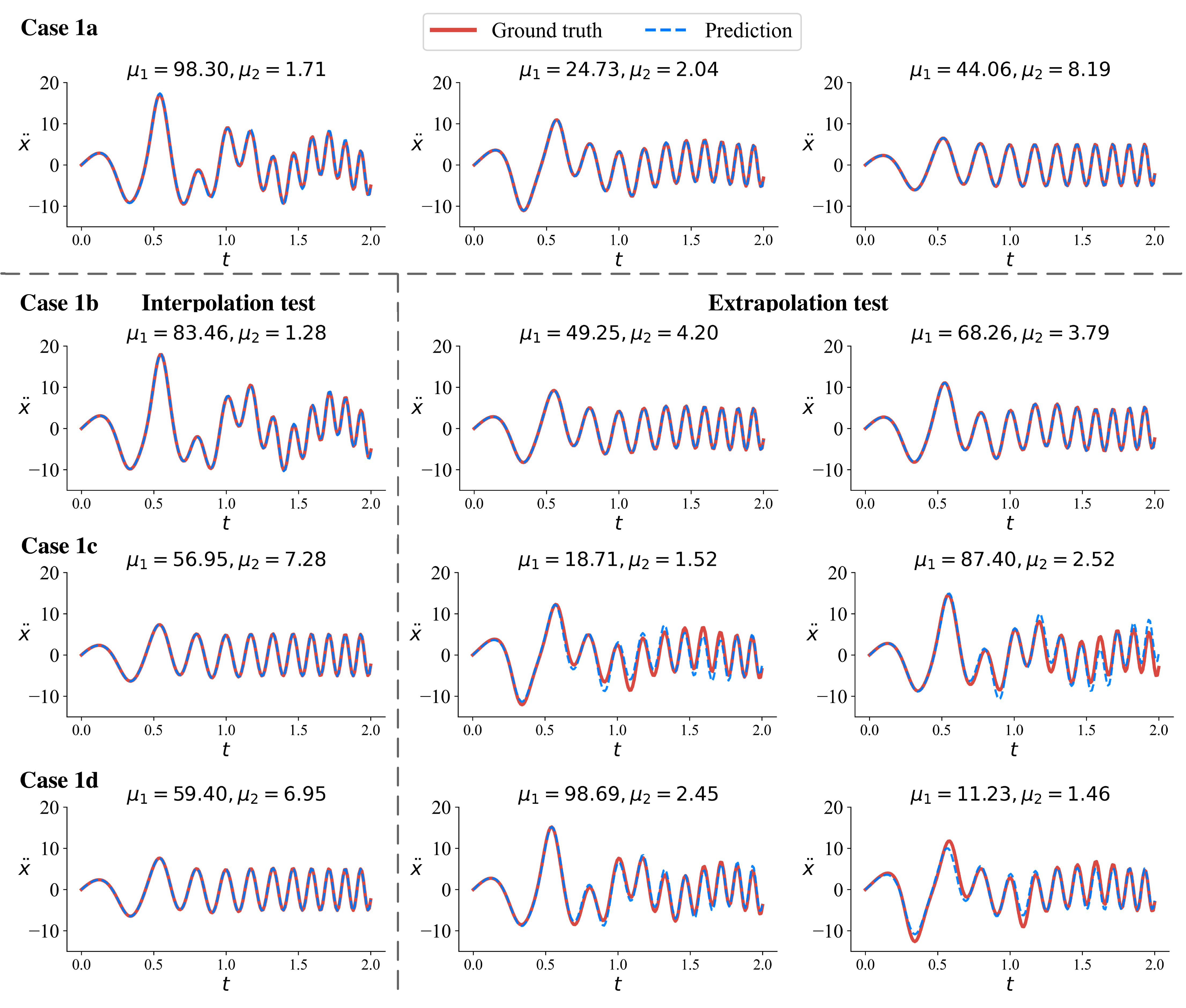}
    \caption{Response prediction results by Parametric DeepONet (ND) of selected \textbf{test} data samples in Case 1 ($\ddot{x}$ - $m^2/s$, $t$ - s). }
    \label{fig:case1_forward}
\end{figure*}

\begin{figure*}[!htb]
    \centering
    \includegraphics[width=0.6\linewidth]{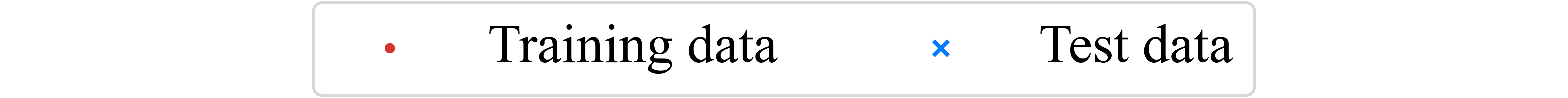}
    \begin{subfigure}{0.4\textwidth}
        \includegraphics[width=\linewidth]{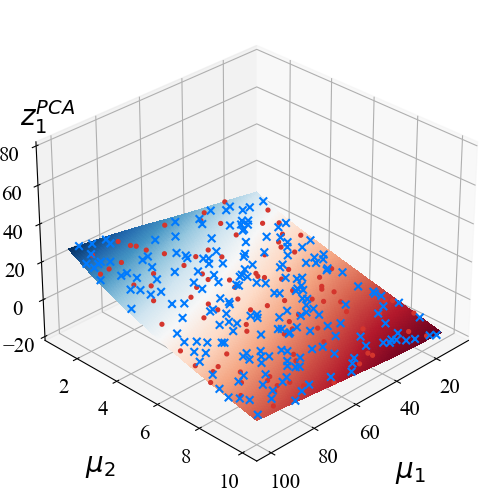}
        \caption{Case 1a}
    \end{subfigure}%
     \begin{subfigure}{0.4\textwidth}
        \includegraphics[width=\linewidth]{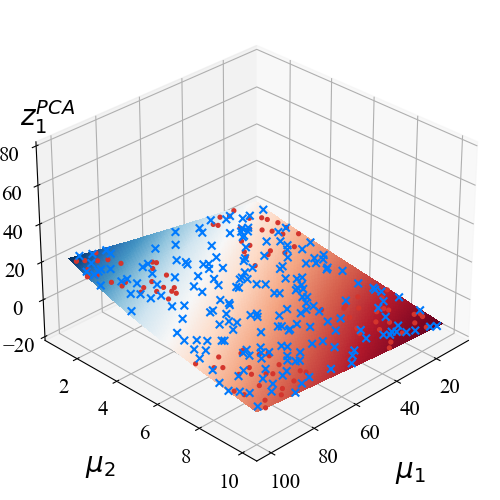}
        \caption{Case 1b}
    \end{subfigure}%
    
    \begin{subfigure}{0.4\textwidth}
        \includegraphics[width=\linewidth]{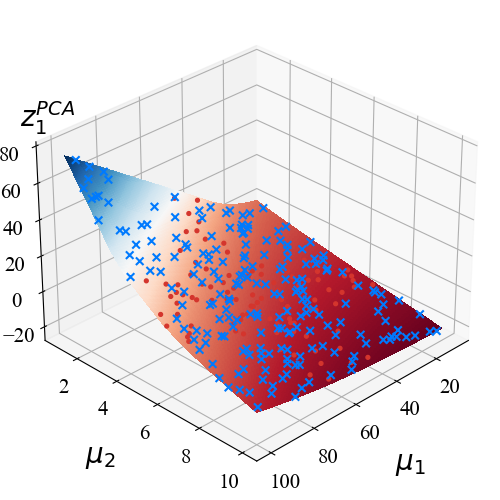}
        \caption{Case 1c}
    \end{subfigure}%
     \begin{subfigure}{0.4\textwidth}
        \includegraphics[width=\linewidth]{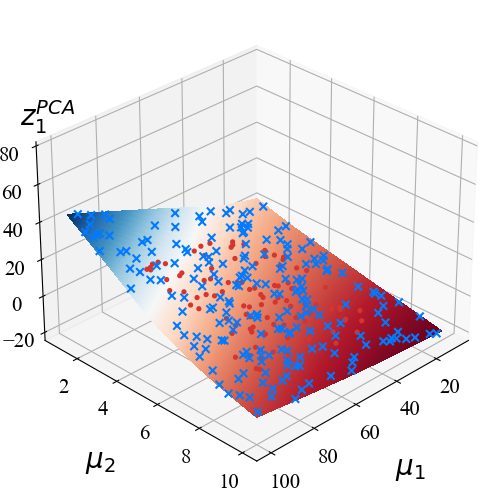}
        \caption{Case 1d}
    \end{subfigure}
    \caption{Visualization of the $z^{\text{PCA}}_1$ in Case 1 ($z^{\text{PCA}}_1$ is the first principle component of latent features from parameter net in Parametric DeepONet (ND)).}
    \label{fig:latent_pca}
\end{figure*}

Figure~\ref{fig:case1_forward} further presents the qualitative results of the response prediction of Parametric DeepONet (ND) in different cases.
In Cases 1a and 1b, the response acceleration of test data samples is very accurately predicted.
In Cases 1c and 1d, the response acceleration of the interpolation test data samples is also accurately predicted. 
The response acceleration of the extrapolation test data samples (the values of the parameter are beyond the bound of training samples) is consistently well predicted, capturing major dynamics albeit with minor errors.

To intuitively explore the framework's effectiveness, we visualize the parameter net latent features of Parametric DeepONet (ND) in Eq.\eqref{eq:param_net}, denoted as $ z = P(\boldsymbol{\mu})$.
Since the latent features are high-dimensional, we employ Principle Component Analysis (PCA) to reduce their dimensionality for visualization and better interpretability.
Figure~\ref{fig:latent_pca} presents the 1st principle component (PC) $z^{\text{PCA}}_1$, plotted against the system parameter $\mu_1$ and $\mu_2$ on the bounded domain.
We focus on the 1st PC because it captures the largest variance in the latent space, making it the most informative element to present how the latent features relate to the system parameters. It is found that $z_1^{\text{PCA}}(\mu_1, \mu_2)$  exhibits a simple and continuous representation, explaining that even if the parameter selection is outside of the training data set, the prediction is reasonable.

In summary, all baseline models except CNN show effectiveness in response prediction in Case 1.
Parametric DeepONet (ND) achieves the best performance.
Notably, CNN has proven to be unable to learn the parametric response prediction in Case 1. 
For strong extrapolation cases (Cases 1c and 1d), all baseline models encounter increased generalization errors.

\subsubsection{Inverse modeling results}
\label{sec:case1_inverse}
Table~\ref{tab:case1_inverse} presents the quantitative results of parameter estimation for different models in Case 1.
For gradient-based initialization,
Parametric DeepONet (ND) achieves the best performance in Case 1a, while MLP outperforms other models in Case 1b.
In Cases 1c and 1d, Parametric DeepONet (LD), Parametric DeepONet (ND), and MLP show comparable performance, achieving similar lowest NRMSEs.
In contrast, CNN shows the highest NRMSE in all cases, indicating that it fails to estimate the input parameters.
After neural refinement, the estimation performance improves for nearly all models — except CNN.
Notably, the results from the CNN suggest that, without proper gradient-based initialization, supervised neural refinement is insufficient to achieve satisfactory performance in parameter estimation.
In some cases, neural refinement may lead to slightly increased NRMSE, likely due to overfitting of the parameter refinement network to the training data.
This issue is observed rarely and can be readily addressed by further introducing the validation dataset to prevent overfitting.

\begin{table*}[!h]
    \centering
    \caption{NRMSE of parameter estimation of \textbf{test} datasets in Case 1, for different cases and models. All values are scaled by $10^{-2}$.}
    \begin{tabular}{ccccccccccc}
        \hline
        \multirow{2}{*}{Case} & \multicolumn{2}{c}{\makecell{Parametric \\ DeepONet (LD)}} & \multicolumn{2}{c}{\makecell{Parametric \\DeepONet (ND)}} & \multicolumn{2}{c}{DeepONet} & \multicolumn{2}{c}{MLP} & \multicolumn{2}{c}{CNN}  \\
        \cline{2-11}
        &  $\mu_1$ & $\mu_2$  &  $\mu_1$ & $\mu_2$  &  $\mu_1$ & $\mu_2$ &  $\mu_1$ & $\mu_2$  &  $\mu_1$ & $\mu_2$\\
        \hline
        \multicolumn{5}{l}{\textit{Gradient-based intialization}} \\
        Case 1a  & $ 2.88 $ & $ 1.88$ & 1.22 & 1.09 & $2.84$ & $4.81$  & $ 3.58 $ & $ 1.79 $ & $ 47.5 $ & $ 48.6 $\\
        Case 1b  & $ 7.16 $ & $ 3.01$ & 2.15 & 2.68 & $7.19$ & $1.42$  & $ 1.81 $ & $ 1.69 $ & $ 52.3 $ & $ 47.9 $\\
        Case 1c  & $ 7.68 $ & $ 7.32$ & 7.16 & 6.33 & $10.1$ & $5.69$ & $ 10.0 $ & $ 4.01 $ & $ 47.0 $ & $ 49.8 $\\
        Case 1d  & $ 6.10 $ & $ 4.10$ & 6.26 & 2.92 & $9.69$ & $5.88$  & $ 6.67 $ & $ 4.14 $ & $ 47.7 $ & $ 49.9 $\\
        \hline
        \multicolumn{5}{l}{\textit{Neural refinement}} \\
        Case 1a & $ 2.51 $ & $ 1.09 $ & 1.19 & 0.99 & $2.07$ & $1.59$  & $ 3.14 $ & $ 1.66 $ & $ 43.0 $ & $ 47.7 $ \\ 
        Case 1b & $ 6.59 $ & $ 2.97 $ & 1.86 & 2.19 & $3.08$ & $1.77$  & $ 1.81 $ & $ 1.56 $ & $ 55.6 $ & $ 44.2 $\\
        Case 1c & $ 7.44 $ & $ 4.41 $ & 6.84 & 5.32 & $10.4$ & $3.90$ & $10.0$ & $4.31$ & $ 44.1 $ & $ 45.7 $  \\
        Case 1d & $ 5.93 $ & $4.62 $ & 6.09 & 2.73 & $7.23$ & $5.76$ & $6.52$ & $4.20$ & $ 43.5 $ & $ 43.9 $ \\
         \hline
    \end{tabular}
    \label{tab:case1_inverse}
\end{table*}

\begin{figure*}[!h]
    \centering
    \begin{subfigure}{0.45\textwidth}
        \includegraphics[width=0.5\linewidth]{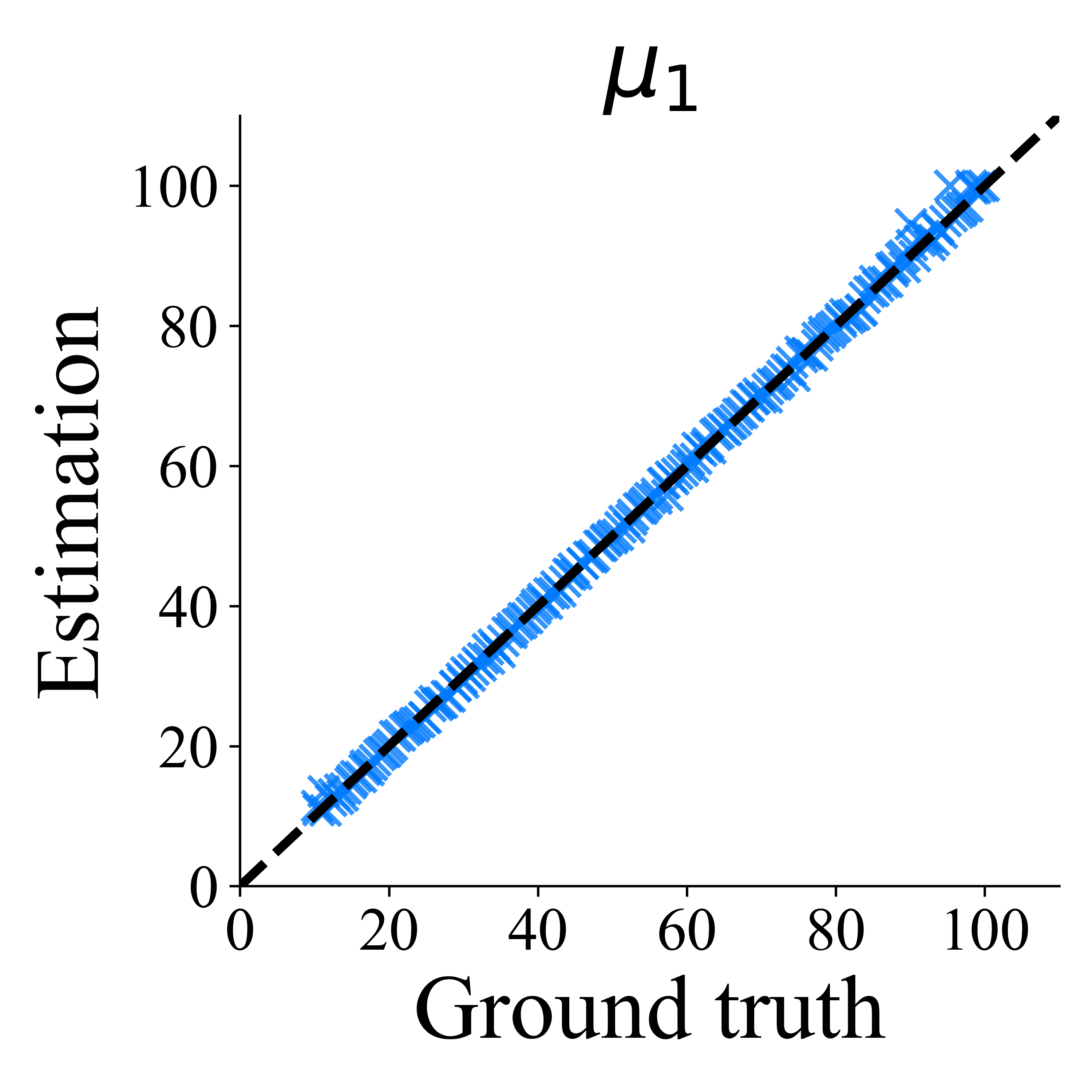}%
        \includegraphics[width=0.5\linewidth]{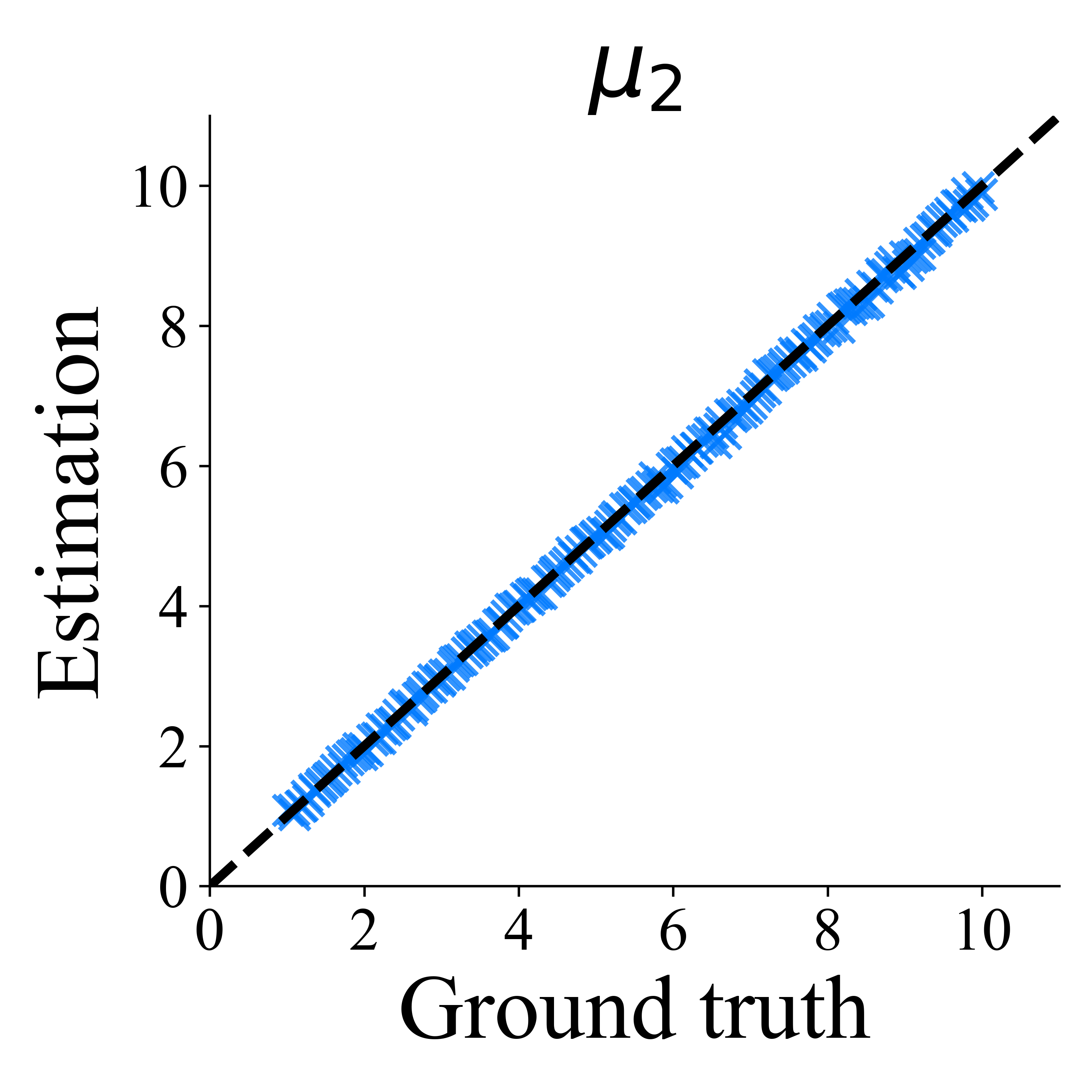}
        \caption{Case 1a, gradient-based initialization}
    \end{subfigure}%
    \begin{subfigure}{0.45\textwidth}
        \includegraphics[width=0.5\linewidth]{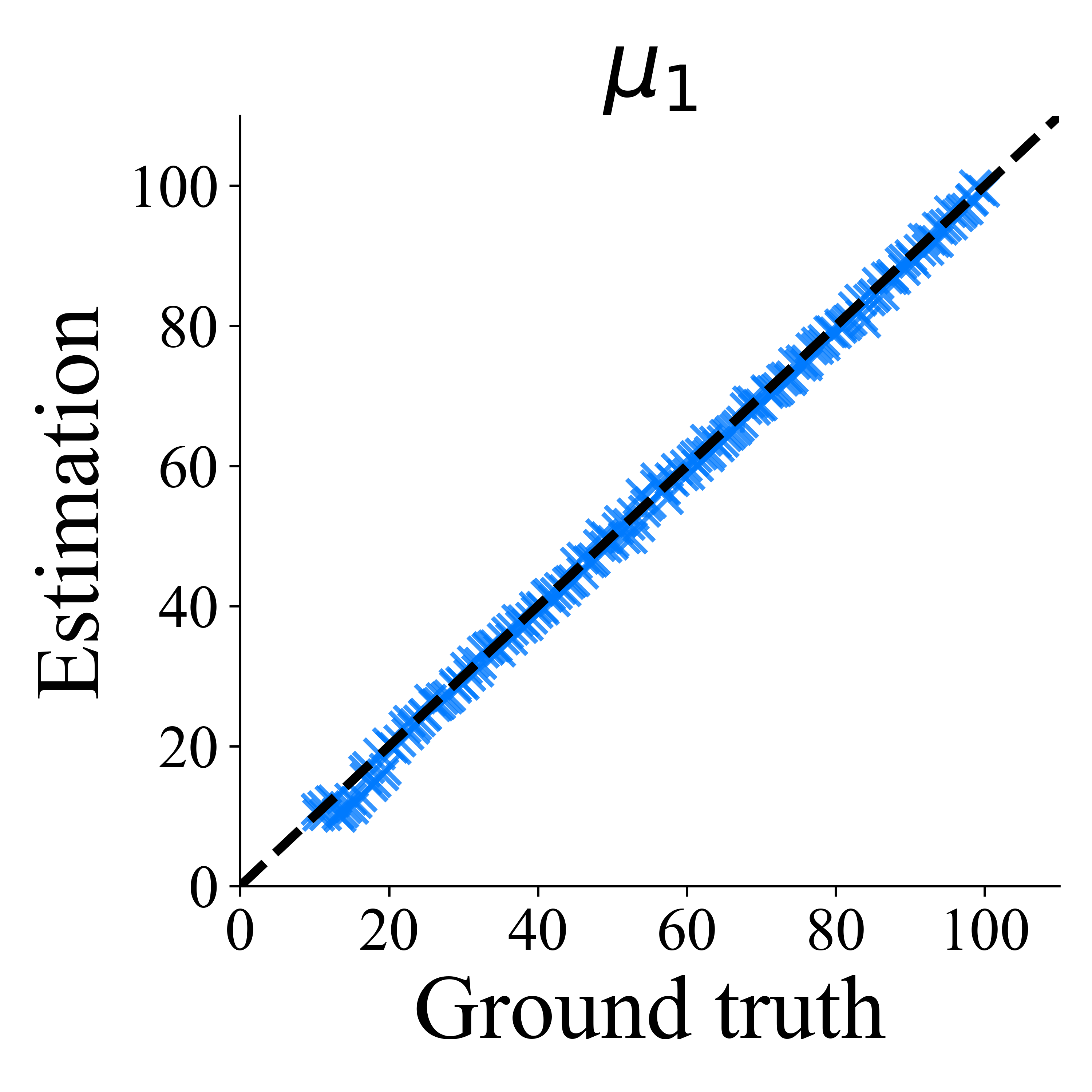}%
        \includegraphics[width=0.5\linewidth]{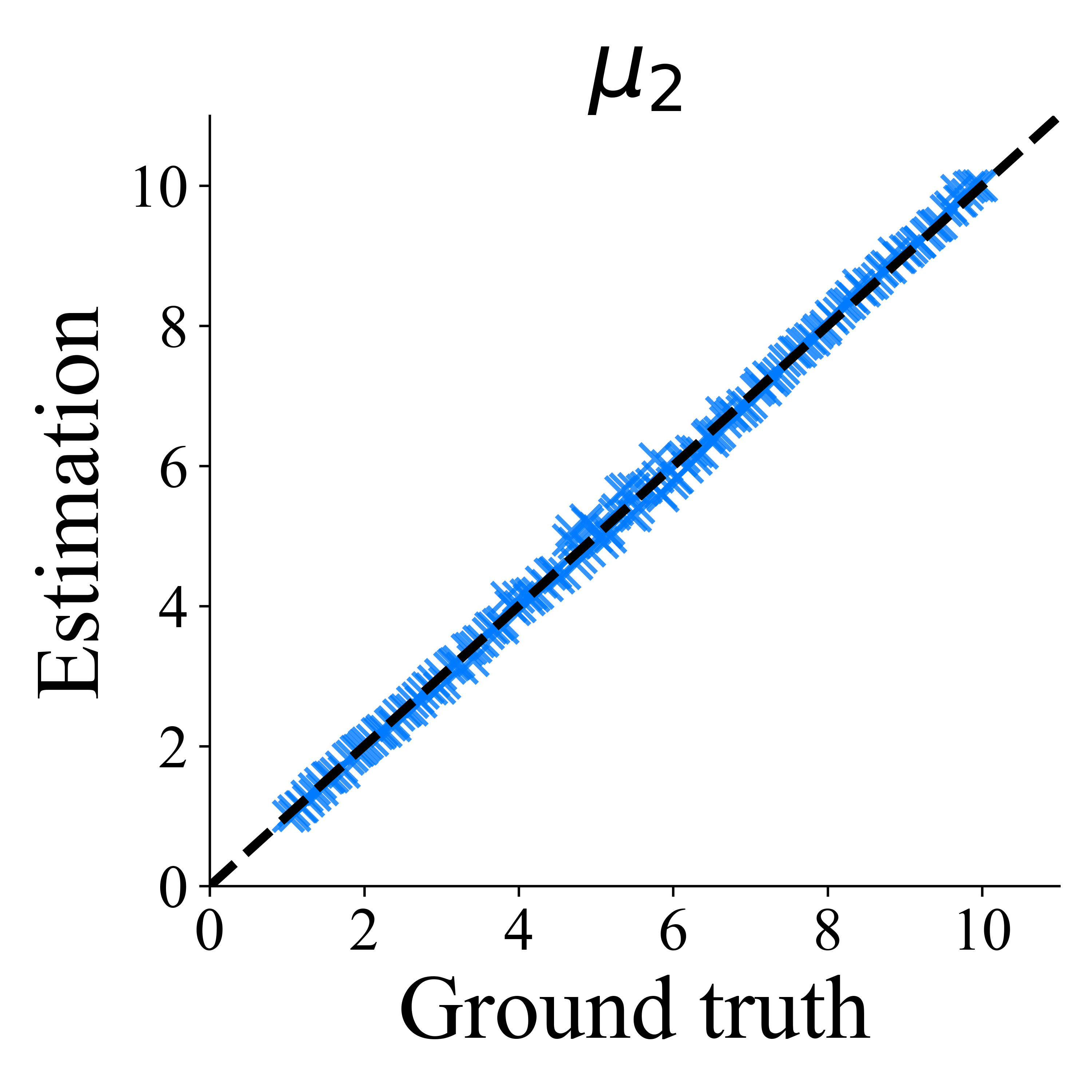}
        \caption{Case 1b, gradient-based initialization}
    \end{subfigure}

    \begin{subfigure}{0.45\textwidth}
        \includegraphics[width=0.5\linewidth]{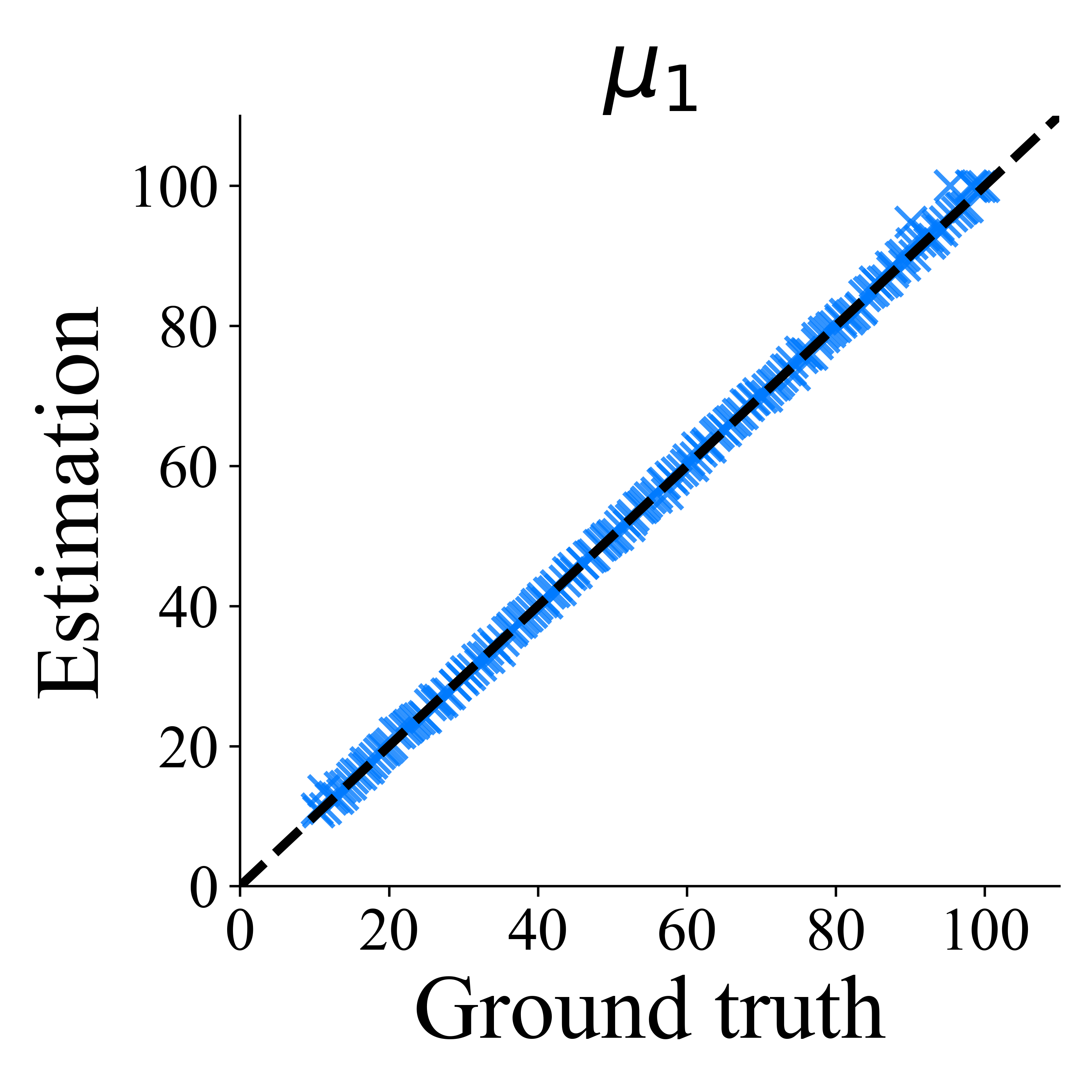}%
        \includegraphics[width=0.5\linewidth]{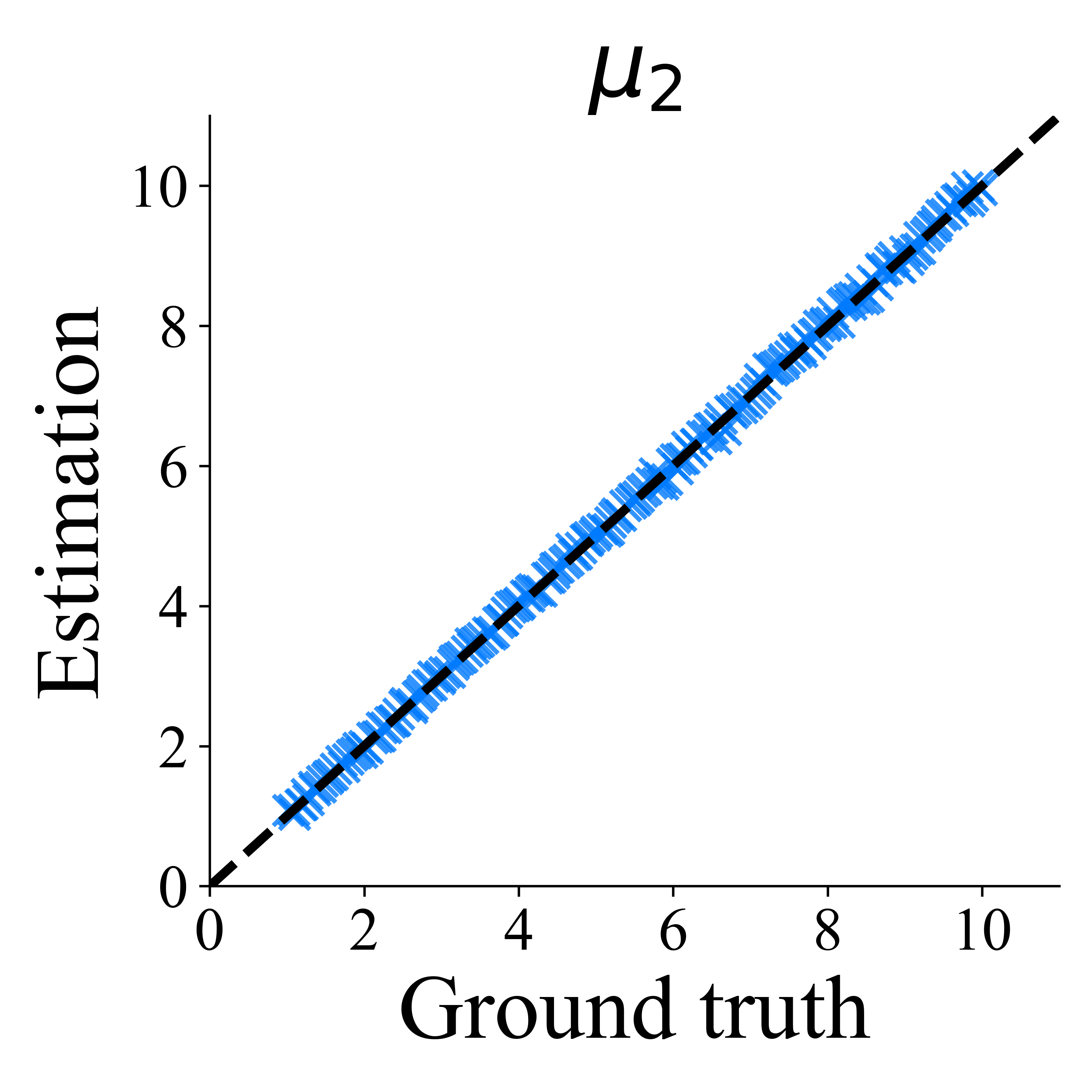}
        \caption{Case 1a, neural refinement}
    \end{subfigure}%
    \begin{subfigure}{0.45\textwidth}
        \includegraphics[width=0.5\linewidth]{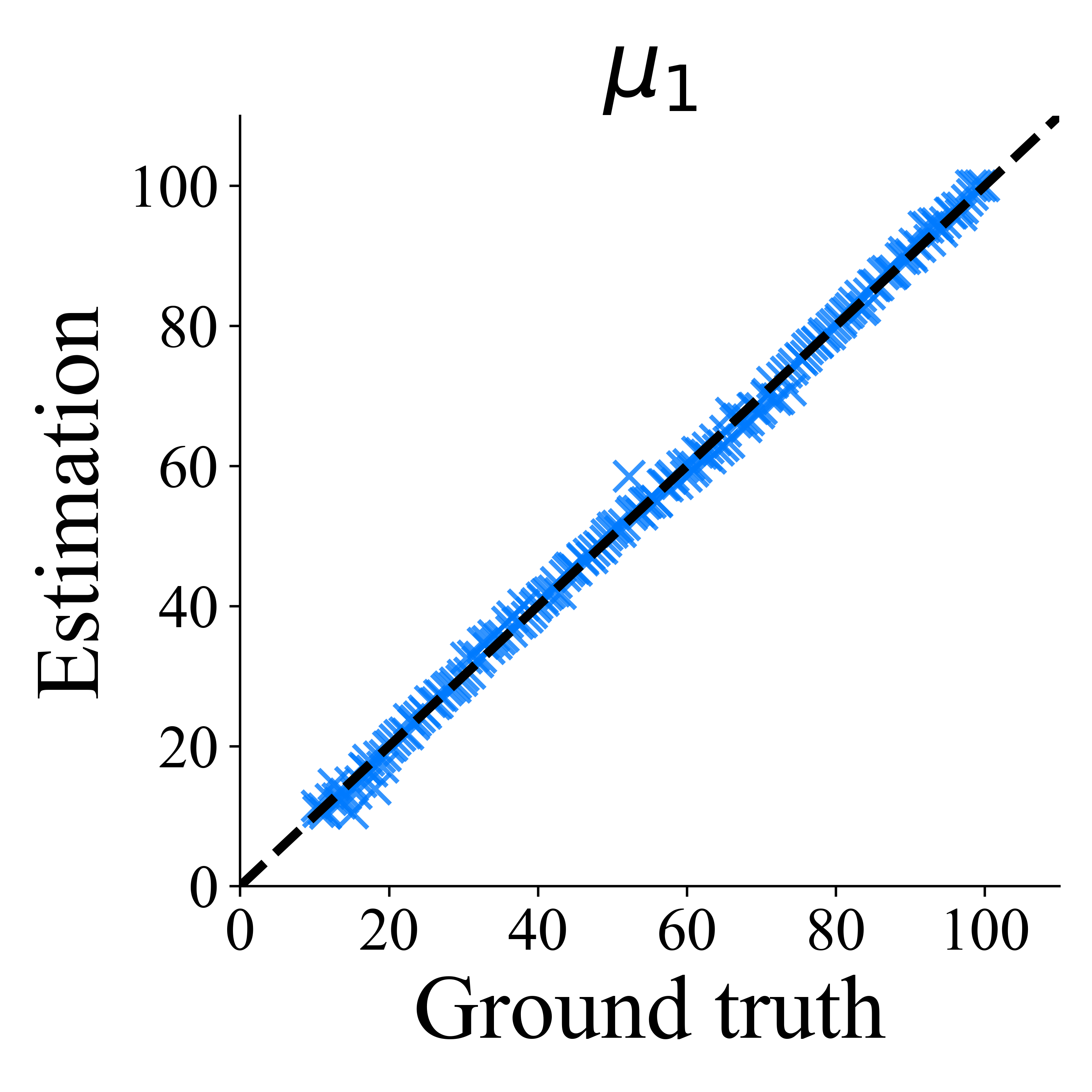}%
        \includegraphics[width=0.5\linewidth]{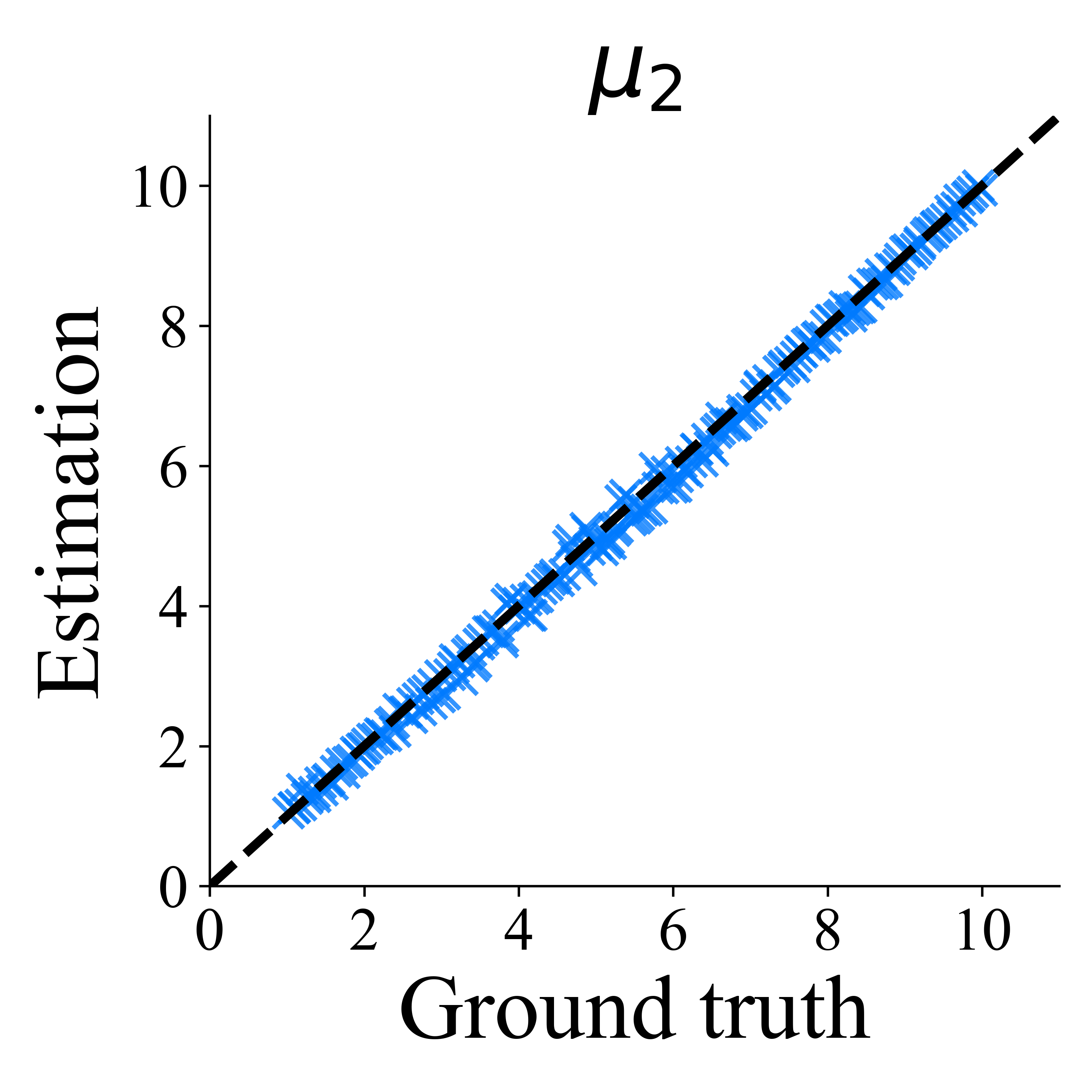}
        \caption{Case 1b, neural refinement}
    \end{subfigure}

    \begin{subfigure}{0.45\textwidth}
        \includegraphics[width=0.5\linewidth]{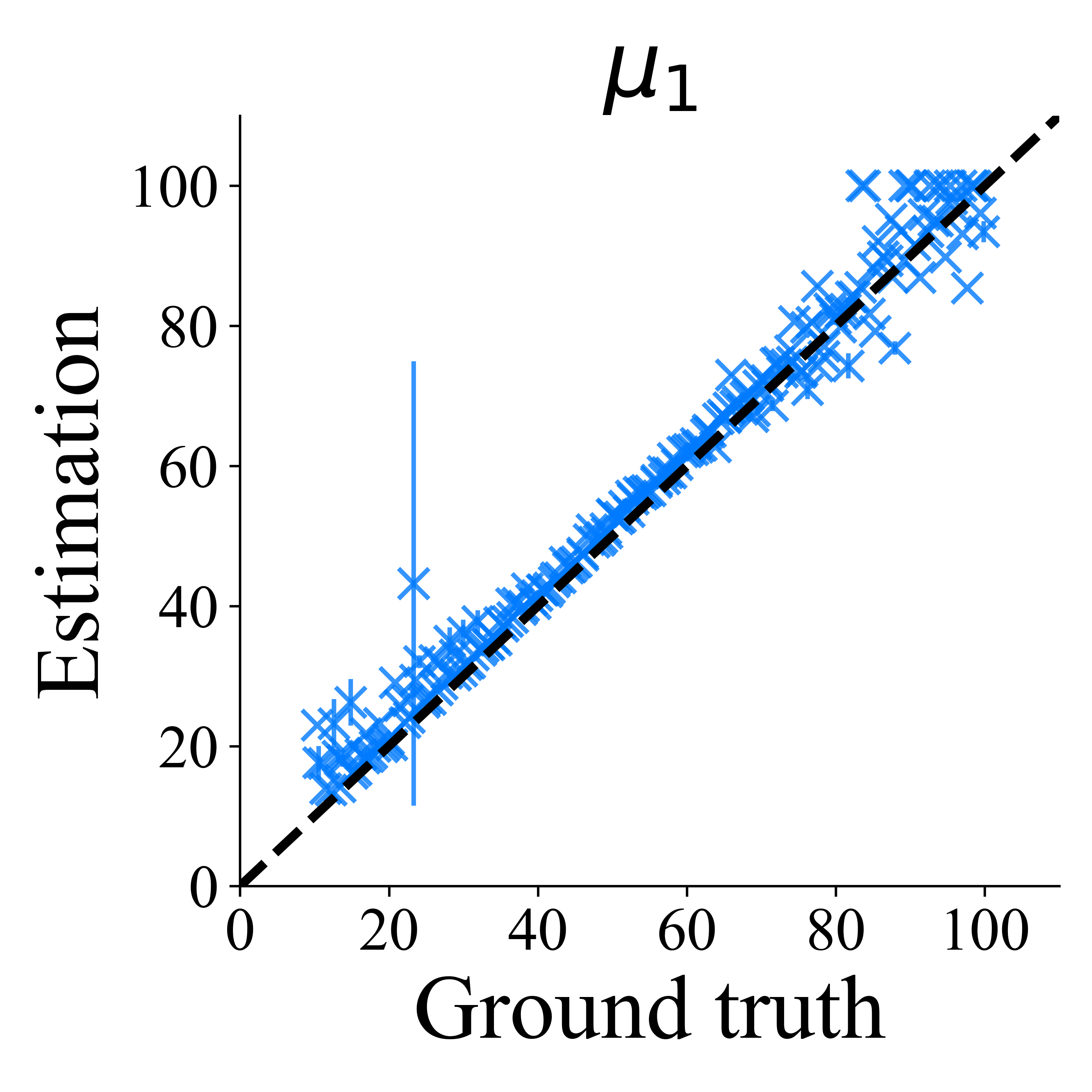}%
        \includegraphics[width=0.5\linewidth]{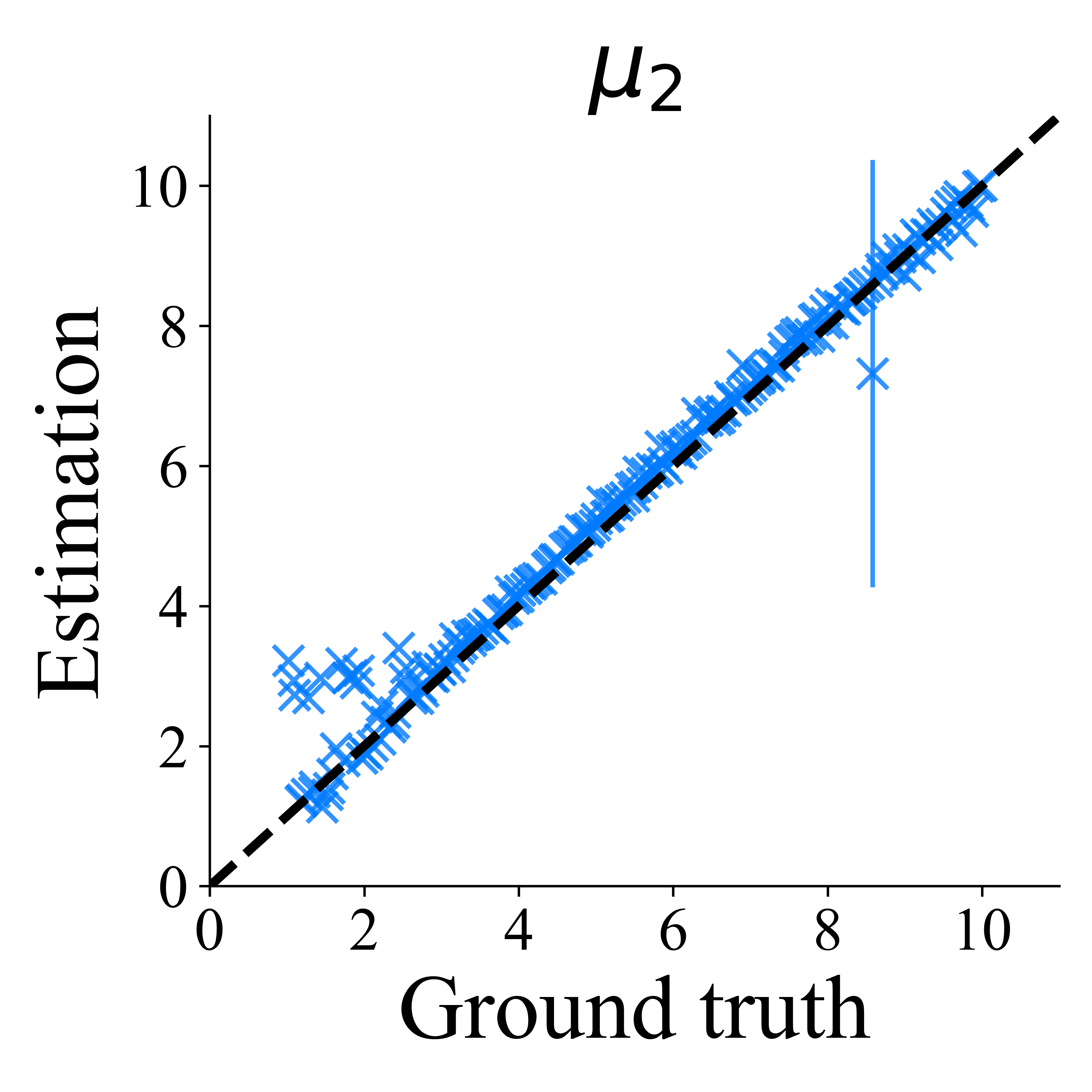}
        \caption{Case 1c, gradient-based initialization}
    \end{subfigure}%
    \begin{subfigure}{0.45\textwidth}
        \includegraphics[width=0.5\linewidth]{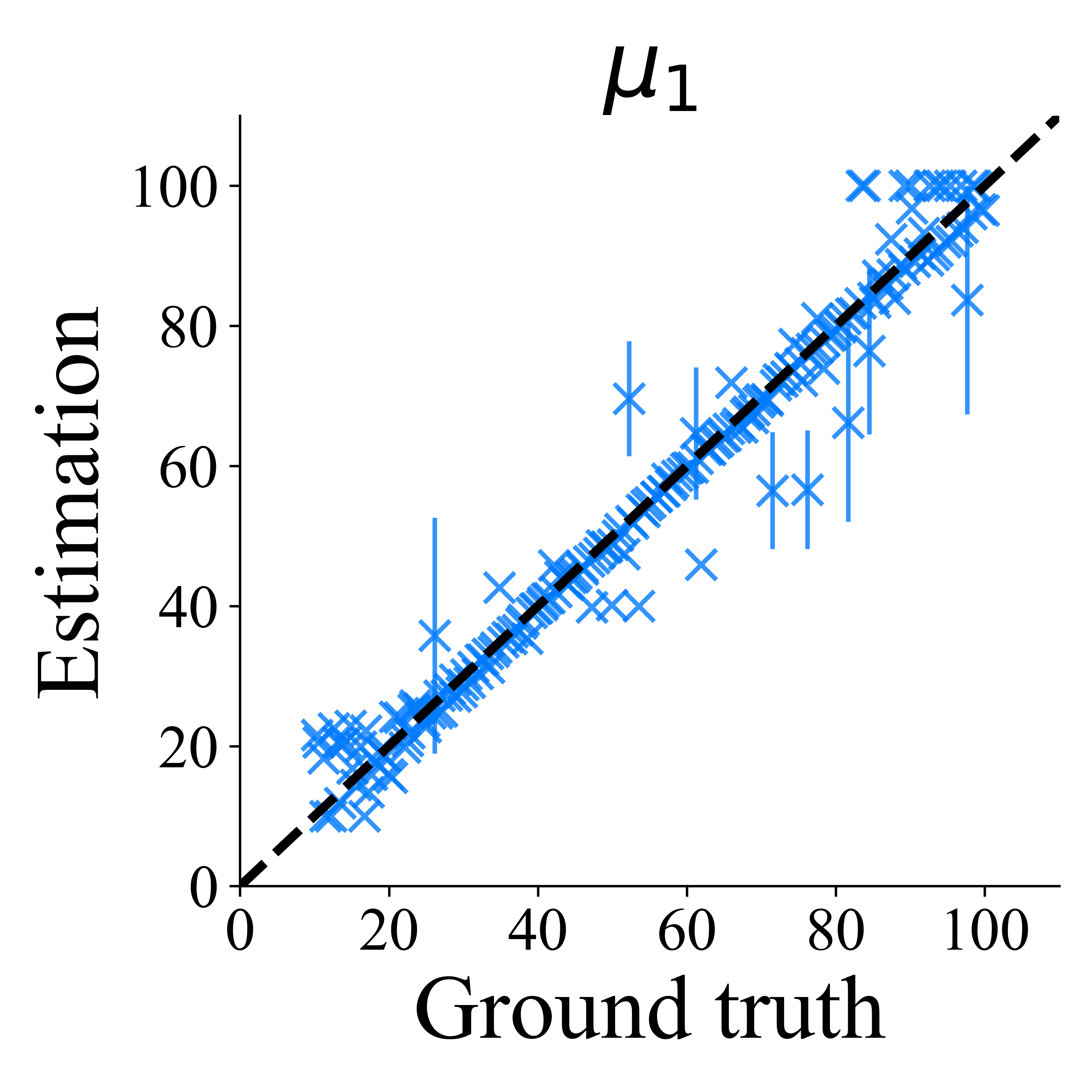}%
        \includegraphics[width=0.5\linewidth]{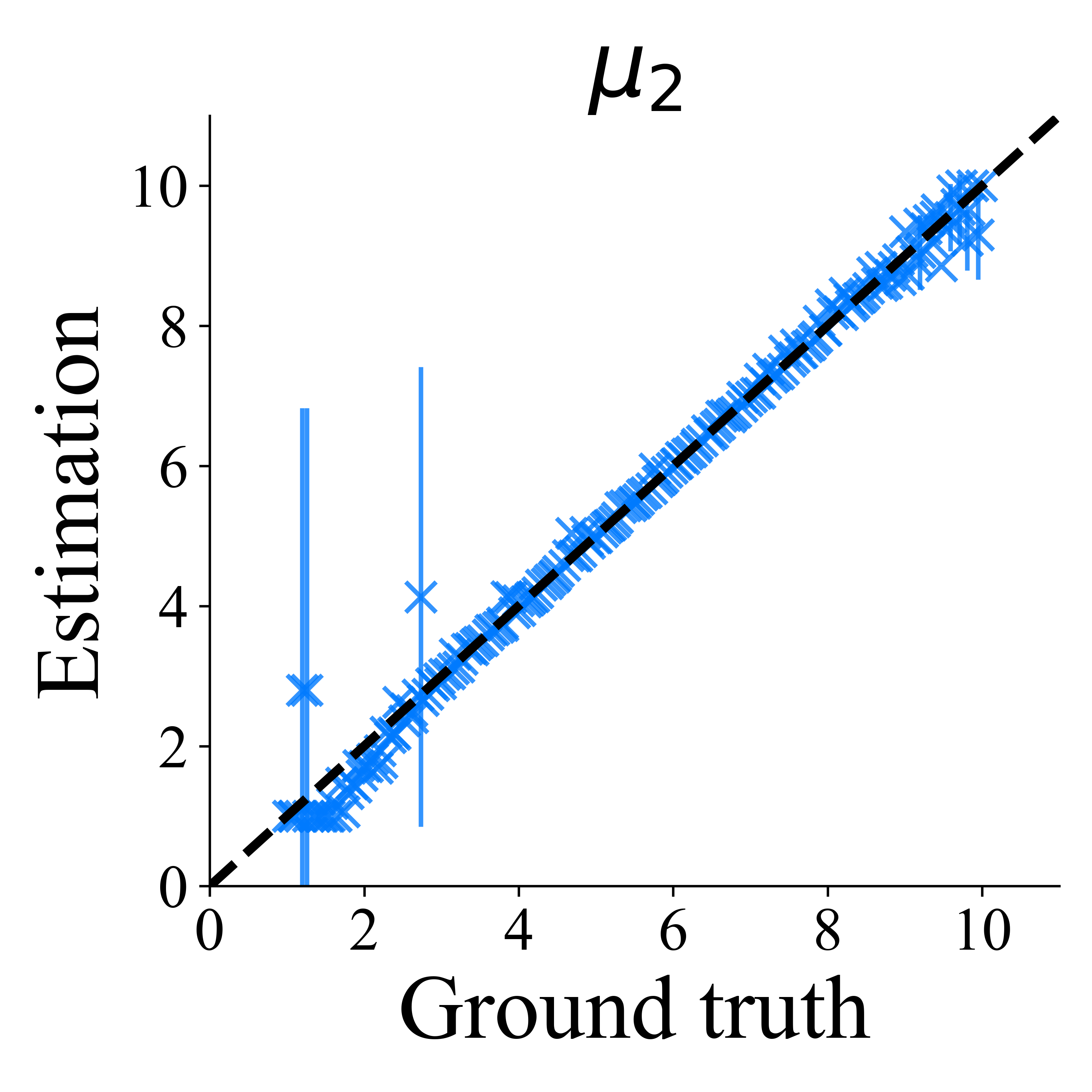}
        \caption{Case 1d, gradient-based initialization}
    \end{subfigure}

    \begin{subfigure}{0.45\textwidth}
        \includegraphics[width=0.5\linewidth]{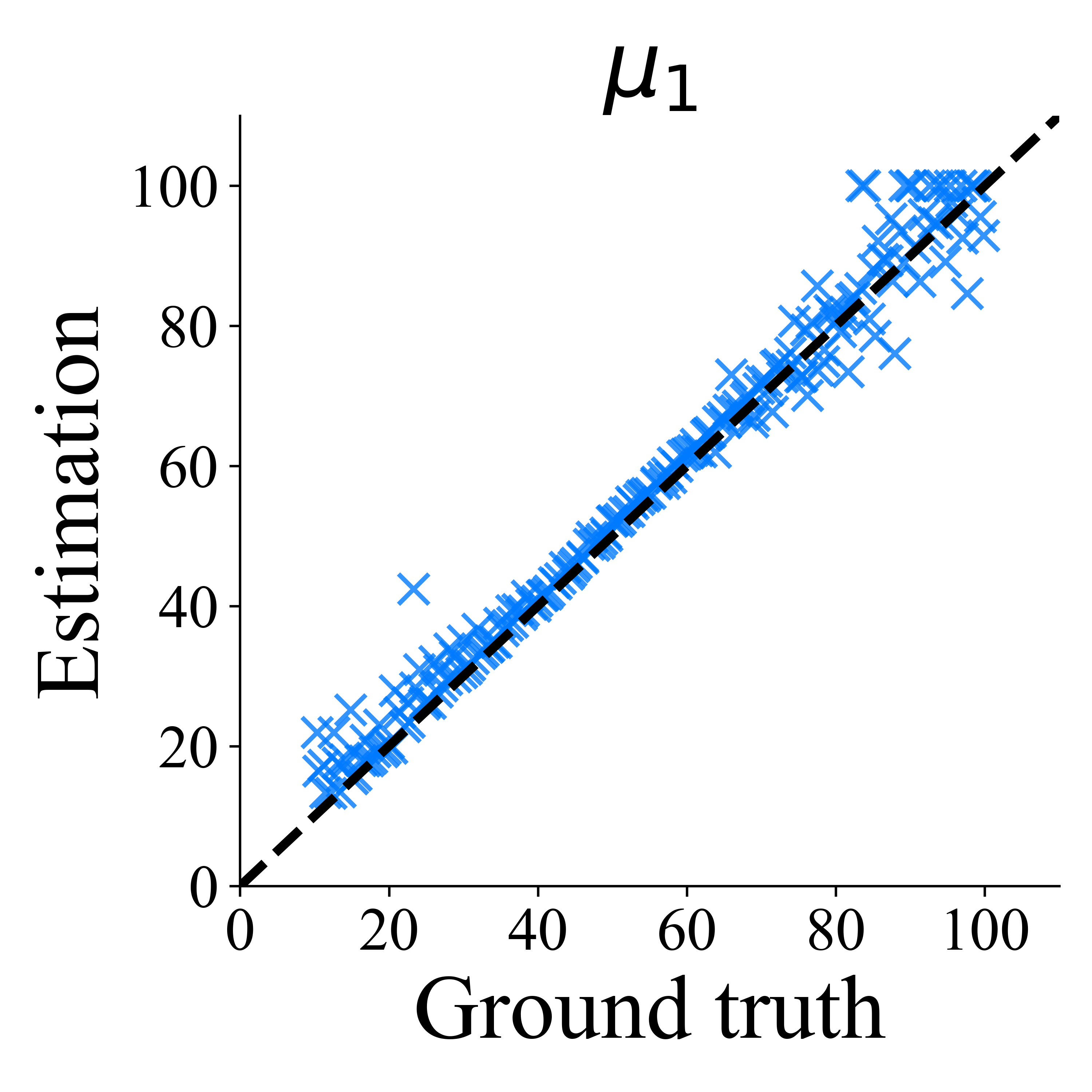}%
        \includegraphics[width=0.5\linewidth]{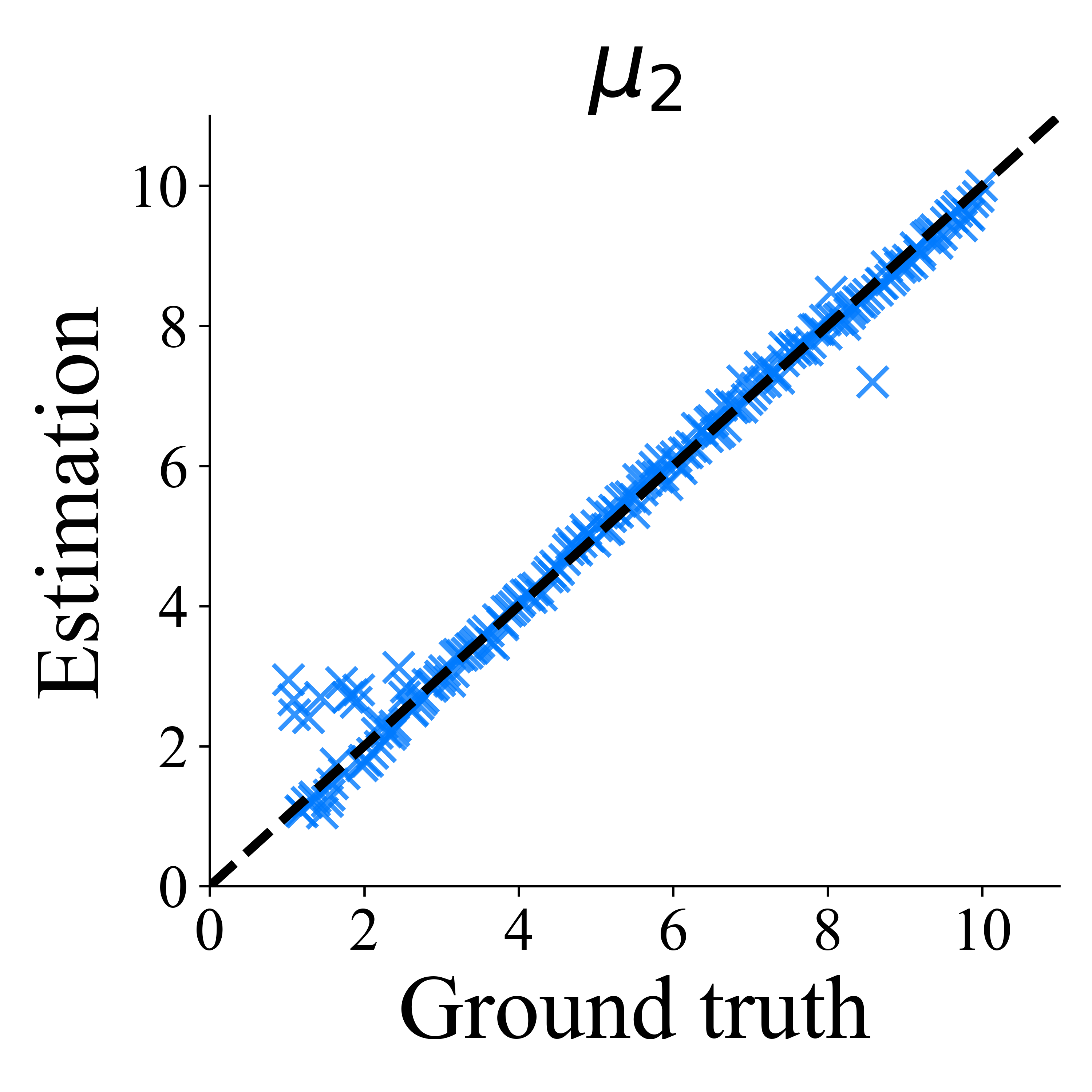}
        \caption{Case 1c, neural refinement}
    \end{subfigure}%
    \begin{subfigure}{0.45\textwidth}
        \includegraphics[width=0.5\linewidth]{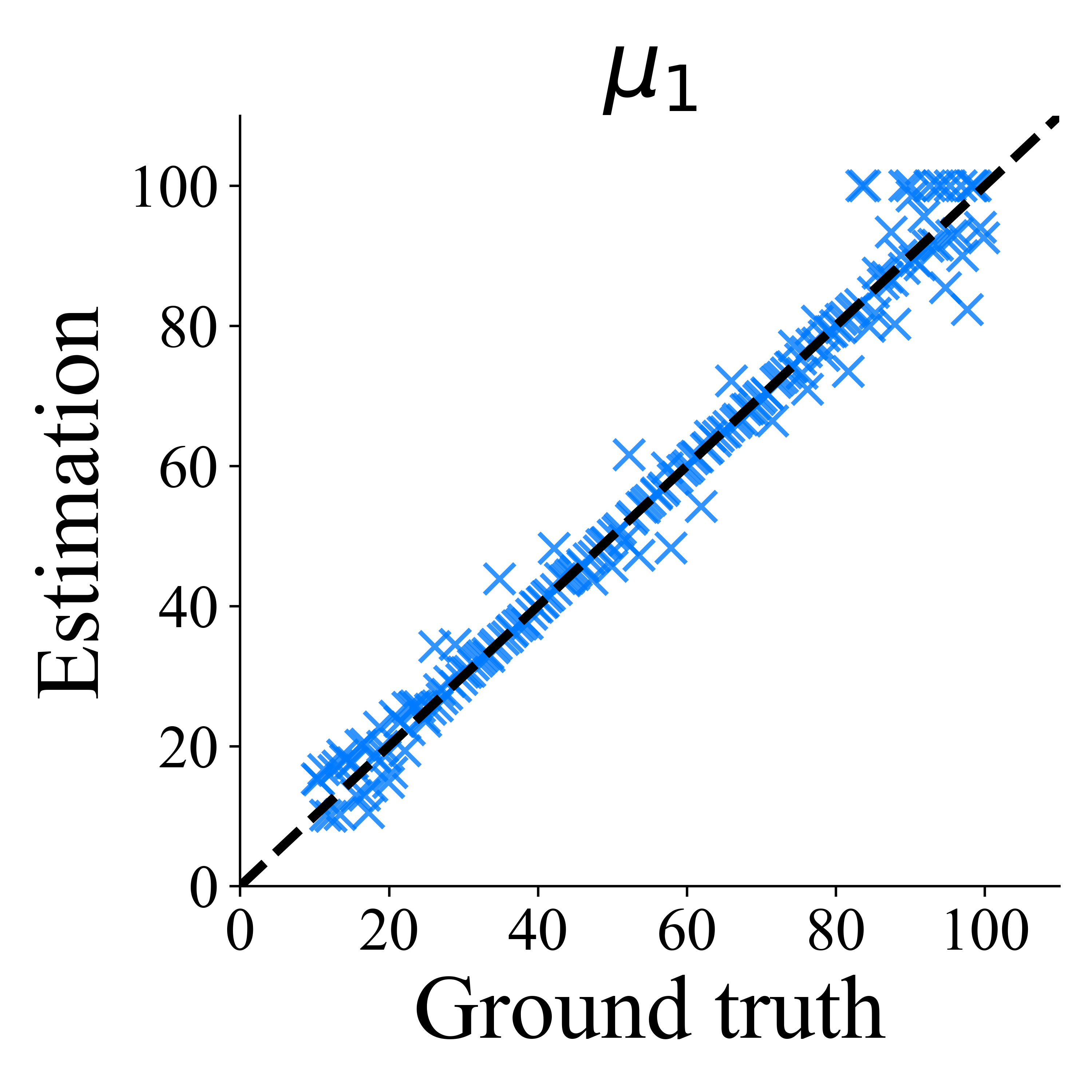}%
        \includegraphics[width=0.5\linewidth]{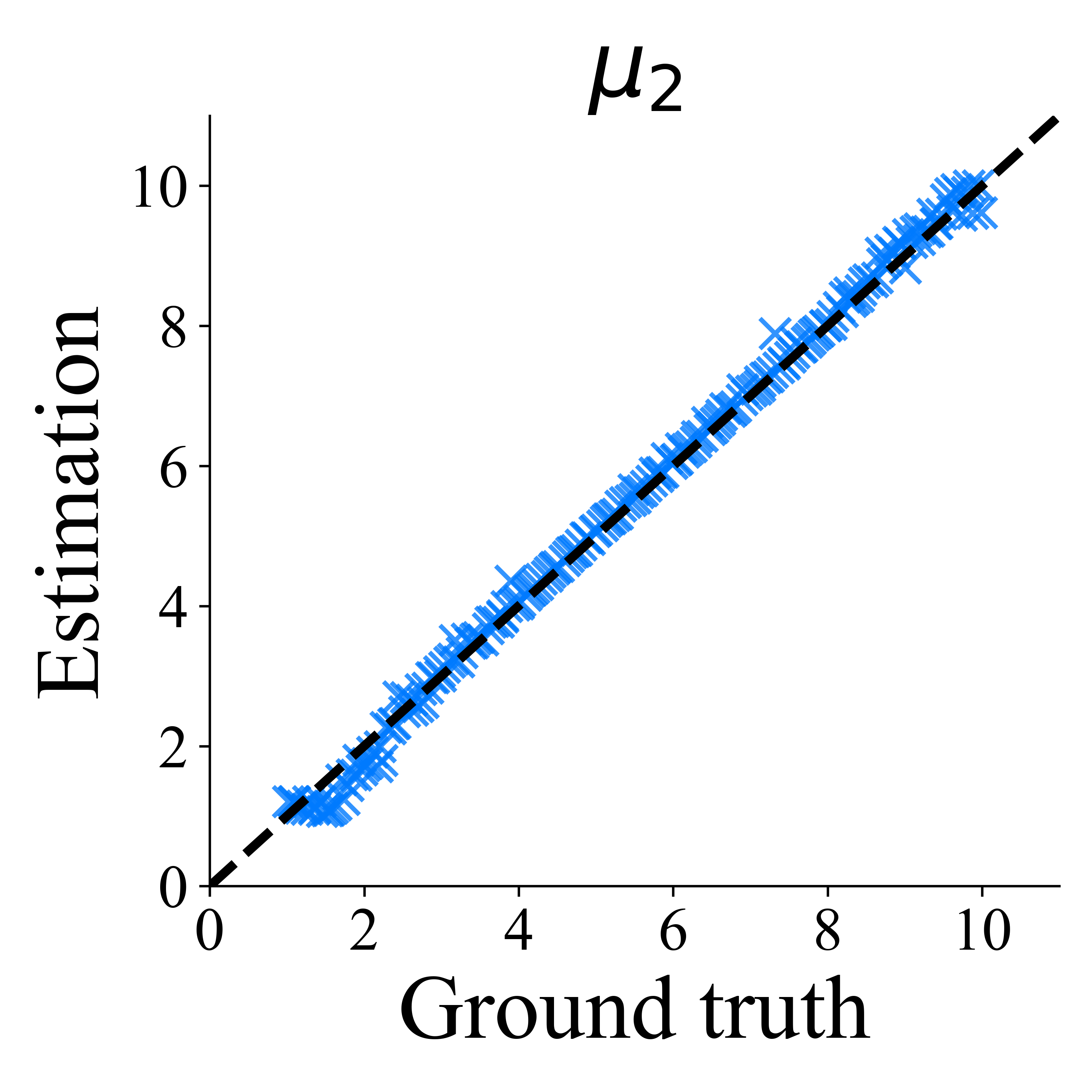}
        \caption{Case 1d, neural refinement}
    \end{subfigure}
    \caption{Parameter estimation results of \textbf{test} data samples in Case 1, gradient-based initialization is implemented over 5 times on the trained forward model, and the
    mean values and standard deviations are presented by error bars ($\mu_1$ is the stiffness, $\mu_2$ is the damping).}
    \label{fig:case1_inverse_nr}
\end{figure*}

Figure~\ref{fig:case1_inverse_nr} further presents the qualitative results of Parametric DeepONet (ND).
In Cases 1a and 1b, both the stiffness and damping are estimated with high accuracy.
Notably, Case 1b involves both inverse modeling and extrapolation, which is considerably challenging.
In more challenging extrapolation cases (Cases 1c and 1d), parameters outside the training range, such as stiffness and damping in Case 1c, and stiffness in Case 1d - are estimated less accurately while still maintained in a reasonable range.

Overall, all models except for CNNs are effective for parameter estimation in Case 1.
In interpolation (Case 1a) and mild extrapolation (Case 1b) scenarios, Parametric DeepONets and MLP exhibit competitive results.
However, all current models still have room for improvements for the strong extrapolation cases (Cases 1c and 1d).

\subsection{Case 2: Experimental wind turbine blade}
\label{sec:case2}
\begin{figure}[!h]
    \centering
    \includegraphics[width=0.9\linewidth]{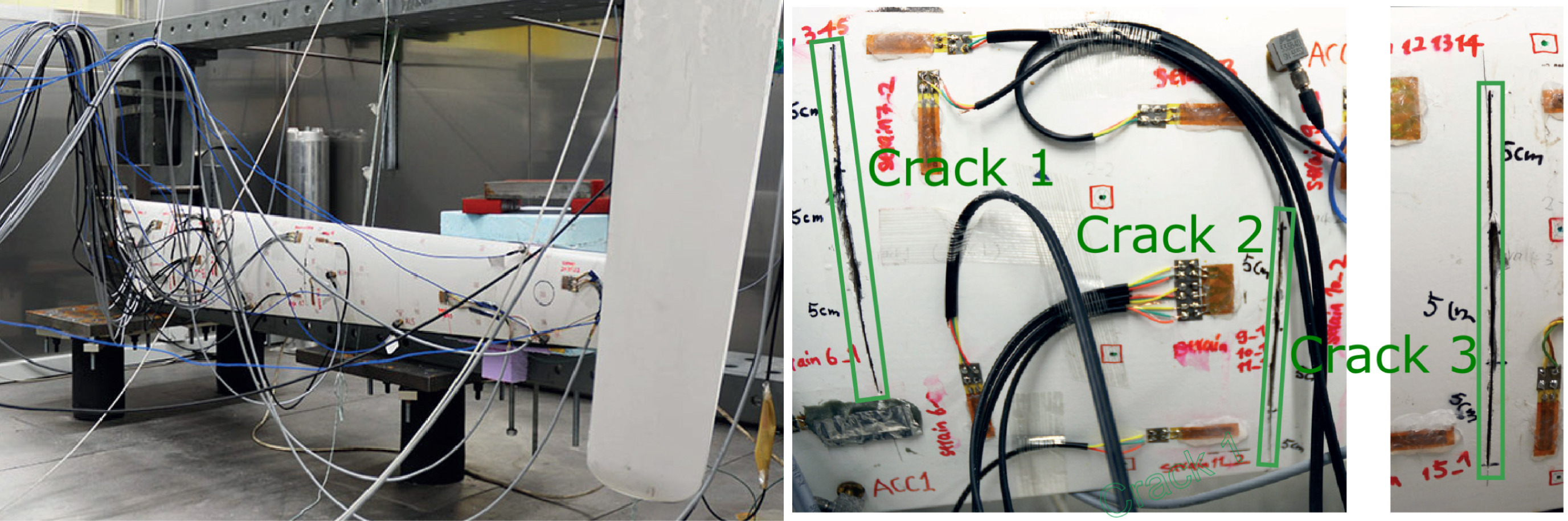}
    \caption{Left: Laboratory configuration, Right: Crack damages. (image credits: Ou, et al. from~\cite{ou2021vibration}).}
    \label{fig:crack}
\end{figure}

\begin{figure}[!h]
    \centering
\includegraphics[width=0.9\linewidth]{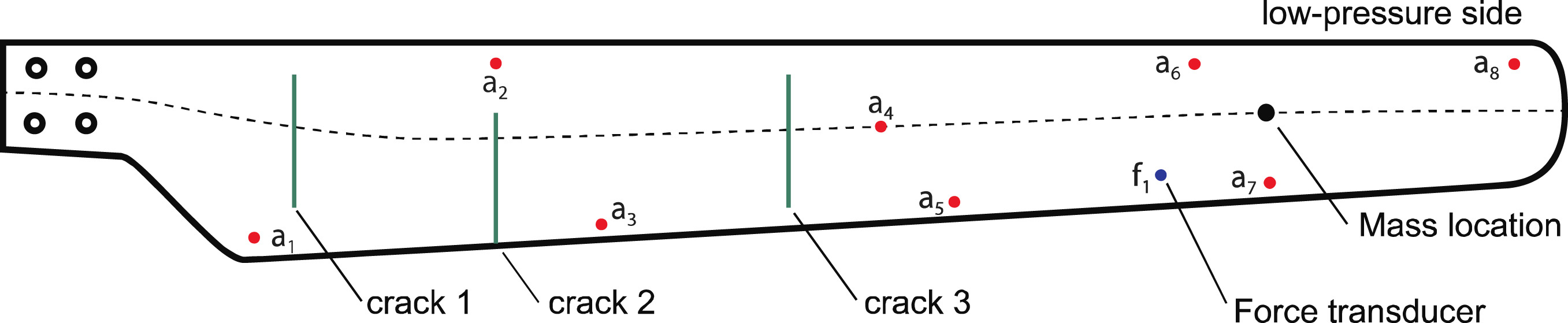}
    \caption{Sensor configuration on the wind turbine blade (image credits: Ou, et al. from~\cite{ou2021vibration}).}
    \label{fig:blade}
\end{figure}

To further validate the proposed framework in more practical scenarios, we consider an experimental wind turbine blade in~\cite{ou2021vibration}.
The wind turbine blade has a length of 1.75 m and a mass of 5.0 kg. 
As shown in Figure~\ref{fig:blade}, eight accelerometers are installed on the blade to record its accelerations at different locations,
capturing the dynamic responses under the excitation from a shaker. 
Ten structural healthy states are physically introduced into the wind turbine blade, including one healthy state and nine damaged states (cracks). 
Specifically, three different damage locations are considered at $17\%$, $30\%$, and $50\%$ of the blade length, respectively.
In each location, three different damage lengths are considered of $5, 10, 15$ cm, respectively.
The location and severity of each structural damage are illustrated in Figure~\ref{fig:crack} and~\ref{fig:blade}.

\begin{figure}[!h]
    \centering
    \begin{subfigure}{0.5\linewidth}
    \includegraphics[width=\linewidth]{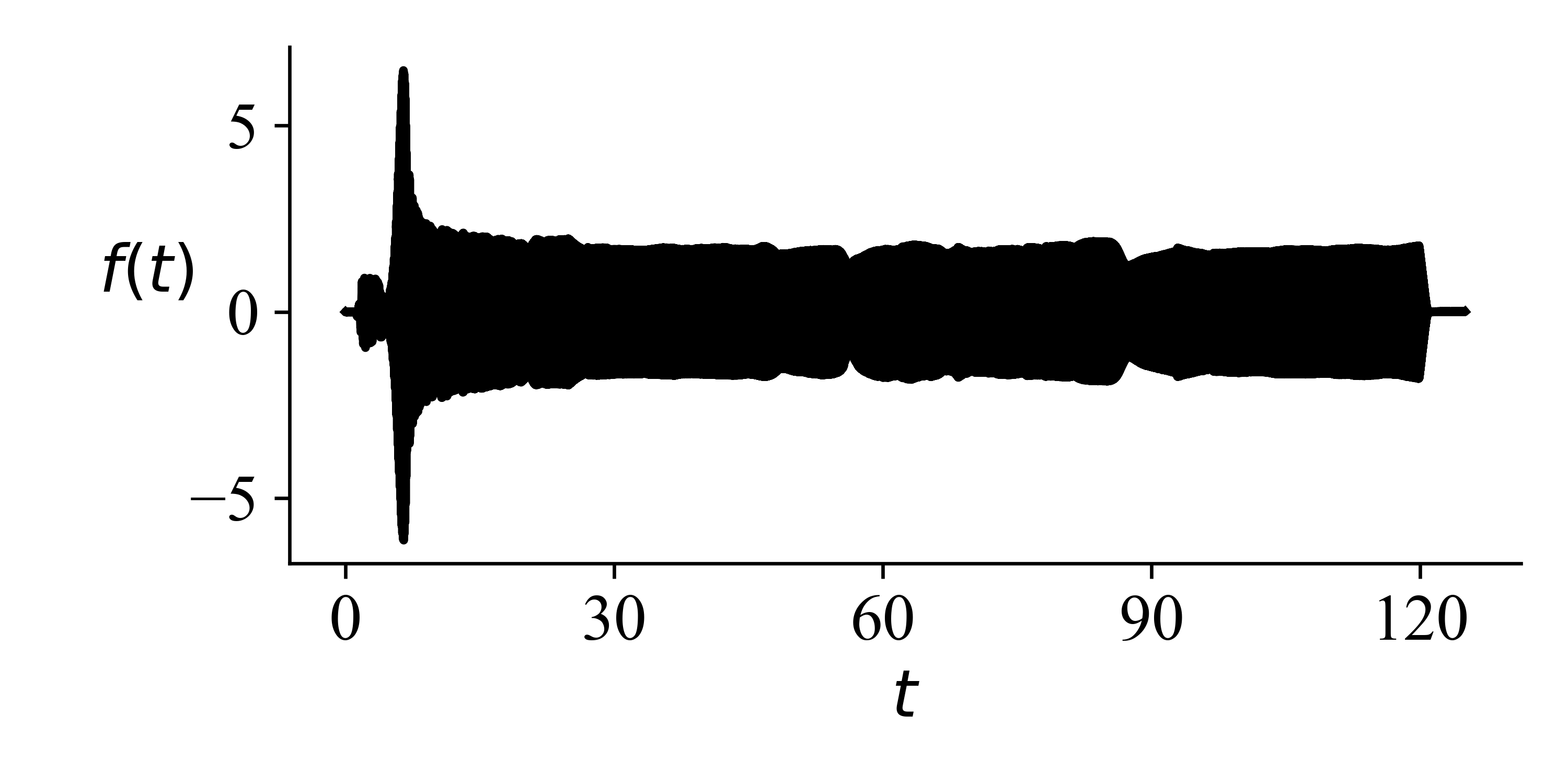}
    \caption{The excitation period of an experiment}
    \end{subfigure}%
    \begin{subfigure}{0.5\linewidth}
    \includegraphics[width=\linewidth]{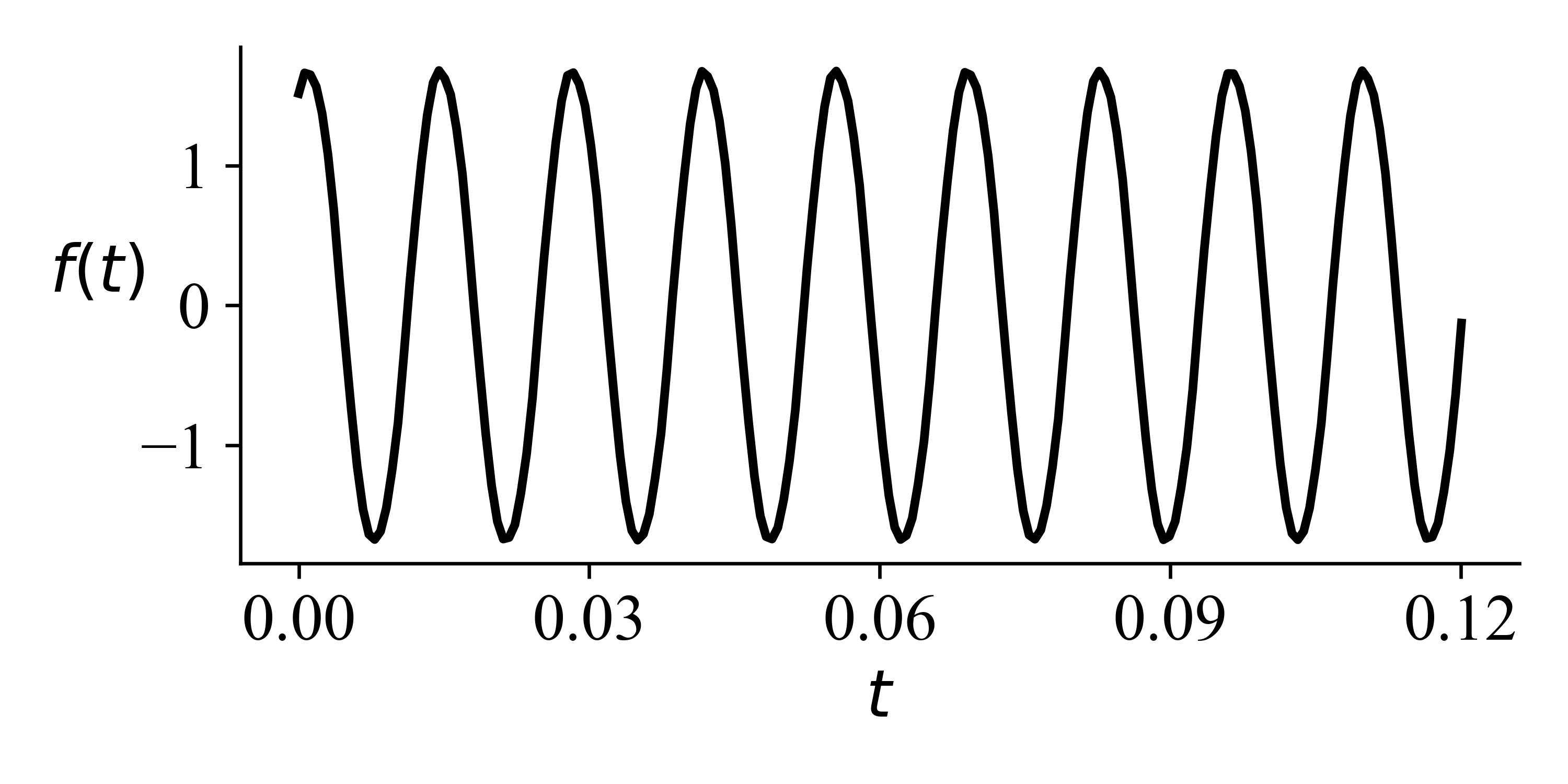}
    \caption{The excitation period of a data sample}
    \end{subfigure}
    \caption{The sine sweep excitation force in Case 2 ($\ddot{x}$ - $m^2/s$, $t$ - s).}
    \label{fig:case2-excitation}
\end{figure}

For each structural state, the wind turbine blade is excited by a sine sweep force for around $120$ seconds, with frequencies ranging from $1$ to $300$ Hz.
During excitation, both the excitation force and the response accelerations of the blade are measured with a sampling frequency of $1666$ Hz.
Figure~\ref{fig:case2-excitation} shows the total excitation period of an experiment, with each excitation having approximately $120 \times 1666 = 199,920$ time instances. 

\subsubsection{Parameterization}
\label{sec:param_case_2}
\textit{Parameterization} refers to identifying the minimal set of parameters that fully characterize a physical system~\cite{tarantola2005inverse}. 
However, it is often challenging to define a complete set of parameters for complex dynamical systems, particularly for real mechanical or civil structures.
In this regard, customizing a handful of parameters relevant to the application context becomes a pragmatic solution. 
In this case, we illustrate that the proposed method can be framed for the task of structural damage identification. 

In the context of structural health monitoring, damage identification aims to assess structural health at different levels, including damage detection, localization, and quantification~\cite{rytter1993vibrational}. 
To attain the equivalent goals, 
we define a six-dimensional parameter $\vect{\mu} = (\mu_1, \mu_2, \mu_3, \mu_4, \mu_5, \mu_6) $ to describe the structural healthy state of the blade as follows:
\begin{itemize}
    \item $\vect{\mu}$ is for damage detection, which indicates the existence of damage.
    \item $\mu_{\text{Loc}} =(\mu_1,\mu_2,\mu_3)$ is for damage localization, which describes the relative locations of cracks 1, 2, and 3.
    \item $\mu_{\text{L}} =(\mu_4,\mu_5,\mu_6)$ is for damage quantification, which describes the relative length of cracks 1, 2, and 3.
\end{itemize}
Specifically, the parameter $\vect{\mu}$ is normalized as follows:
\begin{align}
    \mu_i= \begin{cases} {L_{\text {location }}} / {L_{\text {length }}} & i=1,2,3 , \\
    {L_{\text {crack }}} / {L_{\text { width }}} & i=4,5,6 ,\end{cases}
\end{align}
where $L_{\text {length }} = 1.75  $ m and $L_{\text {width }} = 20 $ cm are the total length and mean width of the blade, and $L_{\text {location }}$ and $L_{\text {crack }}$ are the location and length of the structural damages (cracks).
Table~\ref{table:blade_damage} presents the details of the structural state and the parametrization of the wind turbine blade.
It is noted that conventional ideas to quantify the damage are to use physical metrics (such as stiffness reduction), while it is not feasible to directly derive these metrics from the presence of the cracks. As the proposed framework is a data-driven method, it offers the flexibility of customizing the parametrization.   
\begin{table*}[!h]
    \centering
    \caption{The structural states and the parameterization of the wind turbine blade (States 1,2,3,4,7,10 - training, States 5,6,8,9 - test). }
    \label{table:blade_damage}
    \begin{tabular}{*{22}{ccccccc}}
        \hline
        \multirow{3}{*}{State label} & \multicolumn{6}{c}{Parameterization} & \multicolumn{3}{c}{Damage location and serverity}\\ 
        \cline{2-10} 
        & \multicolumn{3}{c}{$\mu_{\text{Loc}}$}  & \multicolumn{3}{c}{$\mu_{\text{L}}$} & \multirow{2}{*}{Crack 1}  & \multirow{2}{*}{Crack 2}  & \multirow{2}{*}{Crack 3}  \\
        \cline{2-4} \cline{5-7}
        &  $\mu_1$ & $\mu_2$ & $\mu_3$ & $\mu_4$ & $\mu_5$ & $\mu_6$ \\
        \hline
        1  & 0.17 & 0.3 & 0.5 & 0 & 0 & 0 & \multicolumn{3}{c}{healthy}  \\
        2  & 0.17 & 0.3 & 0.5 & 0.25 & 0 & 0  &   5cm &- & - \\
        3  & 0.17 & 0.3 & 0.5 & 0.25 & 0.25 & 0.25 & 5cm &   5cm &- \\
        4  & 0.17 & 0.3 & 0.5& 0.25& 0.25& 0.25 & 5cm &   5cm &   5cm \\
        5  & 0.17 & 0.3 & 0.5& 0.5 & 0.25 & 0.25 &   10cm &   5cm &   5cm  \\
        6  & 0.17 & 0.3 & 0.5& 0.5& 0.5& 0.25   &  10cm &  10cm &   5cm\\
        7  & 0.17 & 0.3 & 0.5& 0.5& 0.5& 0.5  &   10cm &   10cm &  10cm\\
        8  & 0.17 & 0.3 & 0.5& 0.75& 0.5& 0.5  &   15cm &  10cm &   10cm\\
        9  & 0.17 & 0.3 & 0.5& 0.75& 0.75& 0.5 &  15cm &   15cm &   10cm\\
        10 & 0.17 & 0.3 & 0.5& 0.75& 0.75& 0.75 &   15cm &  15cm &   15cm\\
        \hline
    \end{tabular}
\end{table*}

\begin{figure*}[!h]
    \centering
    \begin{subfigure}{0.4\textwidth}
        \centering
        \includegraphics[width=\linewidth]{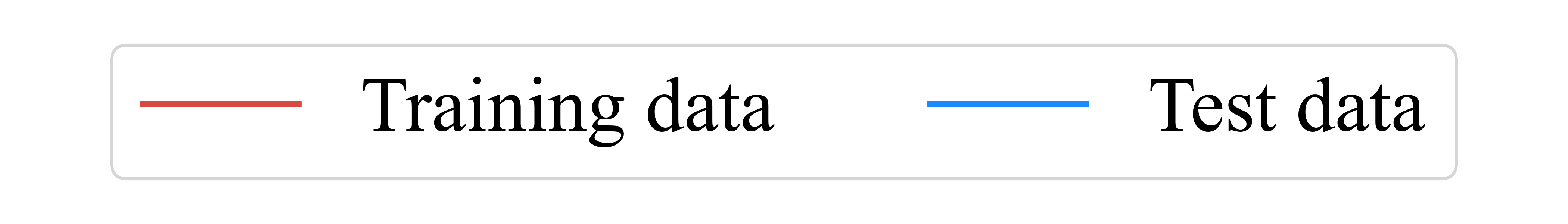}
    \end{subfigure}
    
    \begin{subfigure}{0.33\textwidth}
        \includegraphics[width=\linewidth]{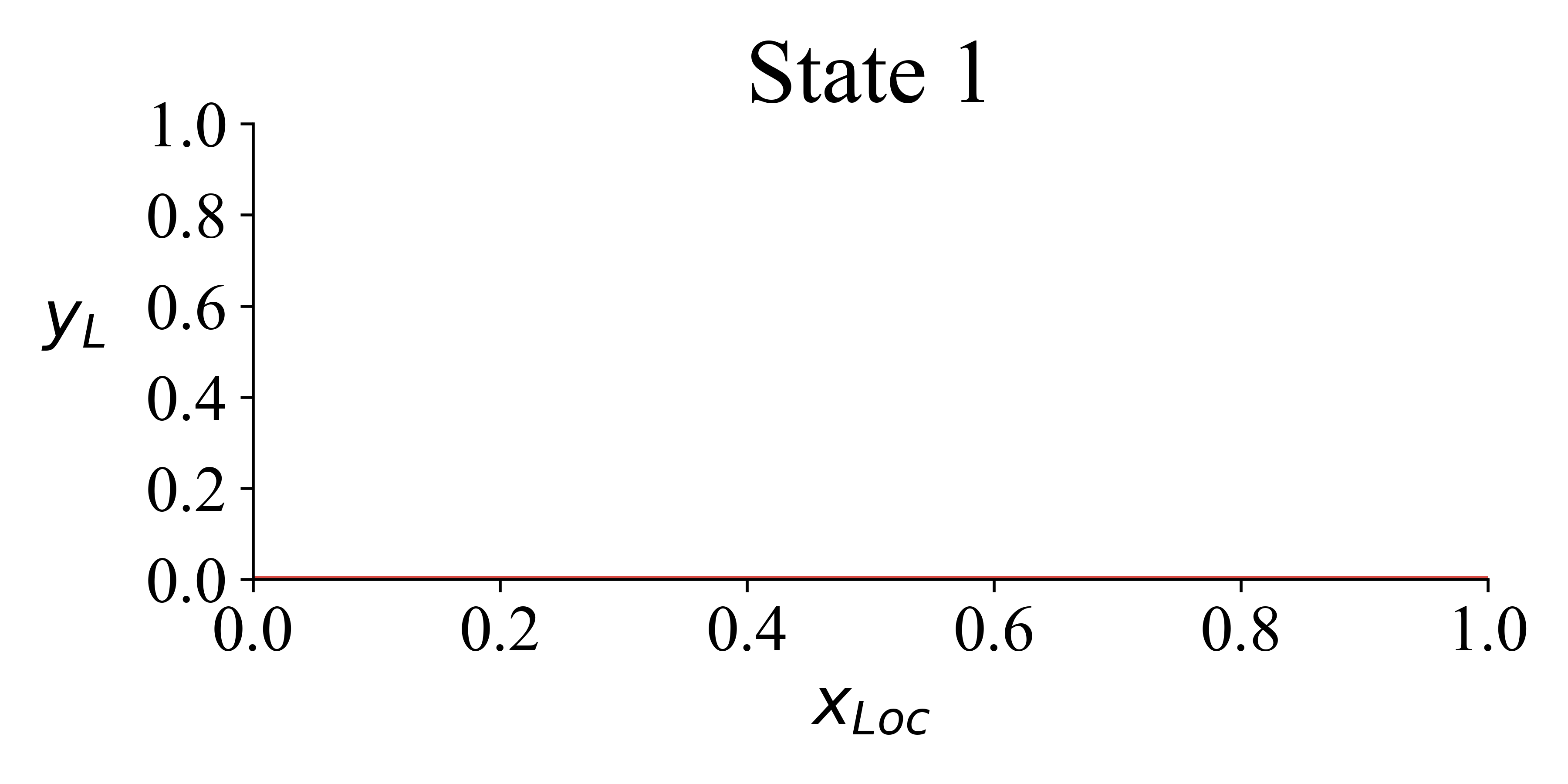}
    \end{subfigure}%
    
    \begin{subfigure}{0.33\textwidth}
        \includegraphics[width=\linewidth]{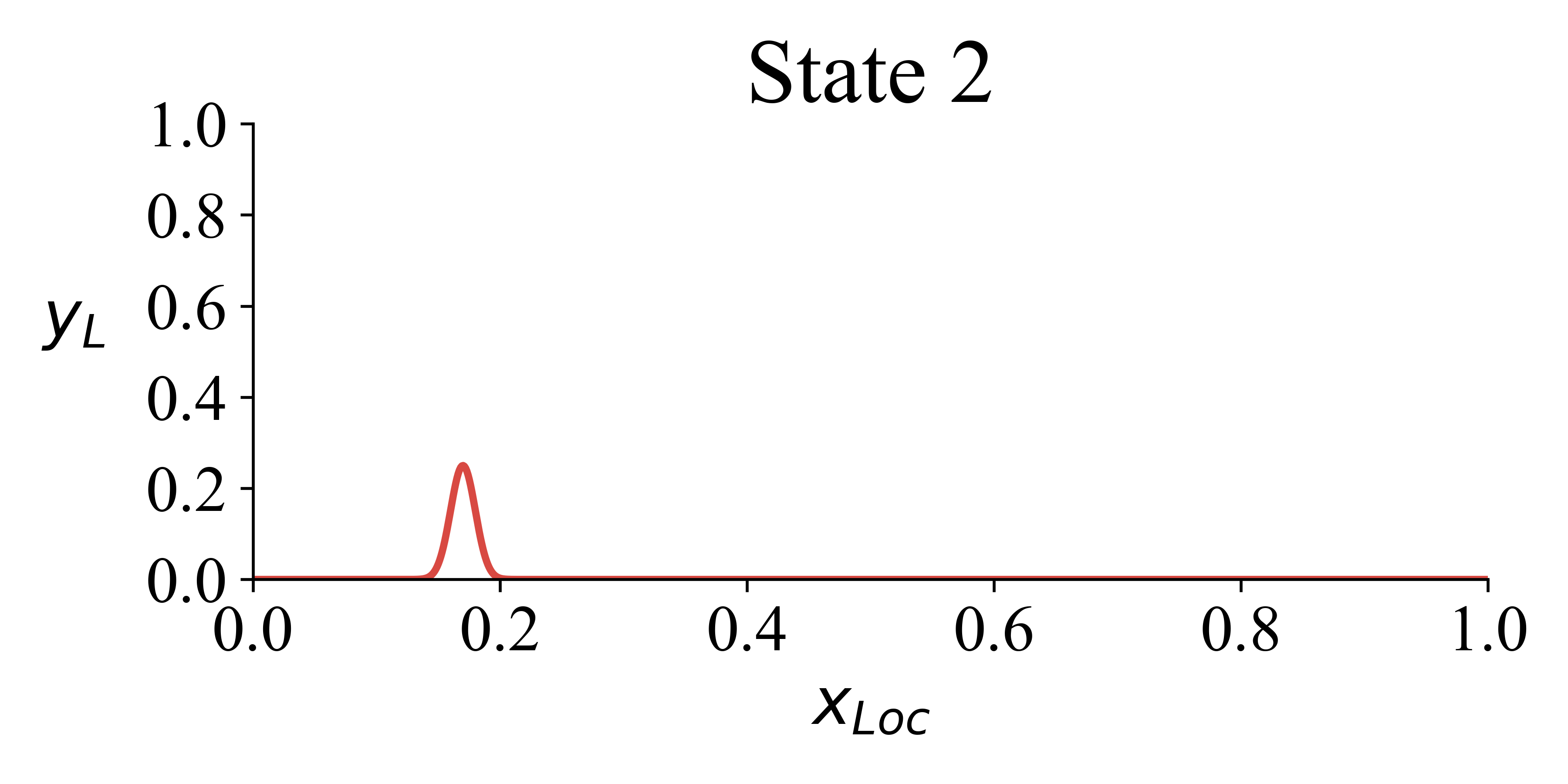}
    \end{subfigure}%
    \begin{subfigure}{0.33\textwidth}
        \includegraphics[width=\linewidth]{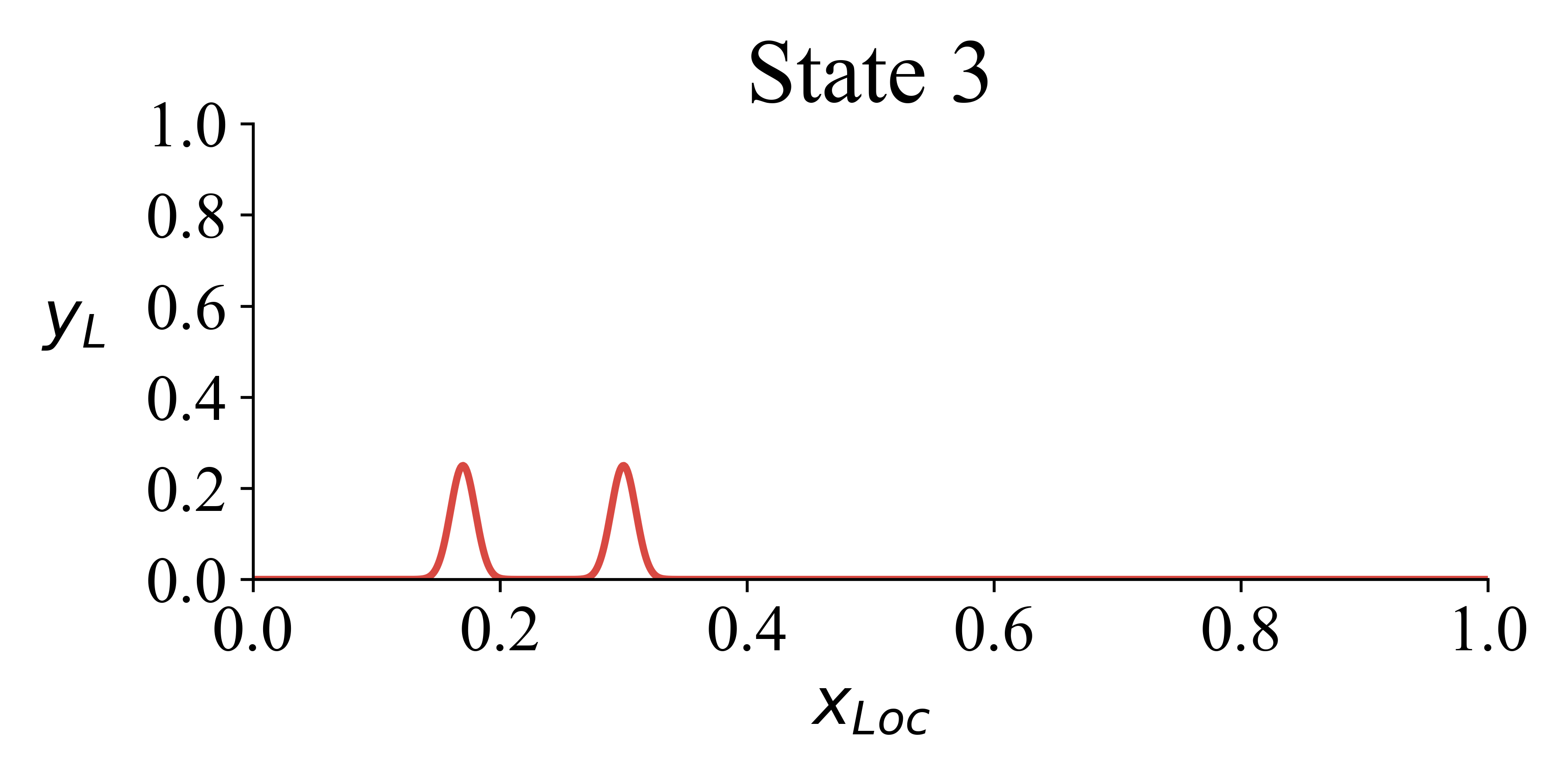}
    \end{subfigure}%
    \begin{subfigure}{0.33\textwidth}
        \includegraphics[width=\linewidth]{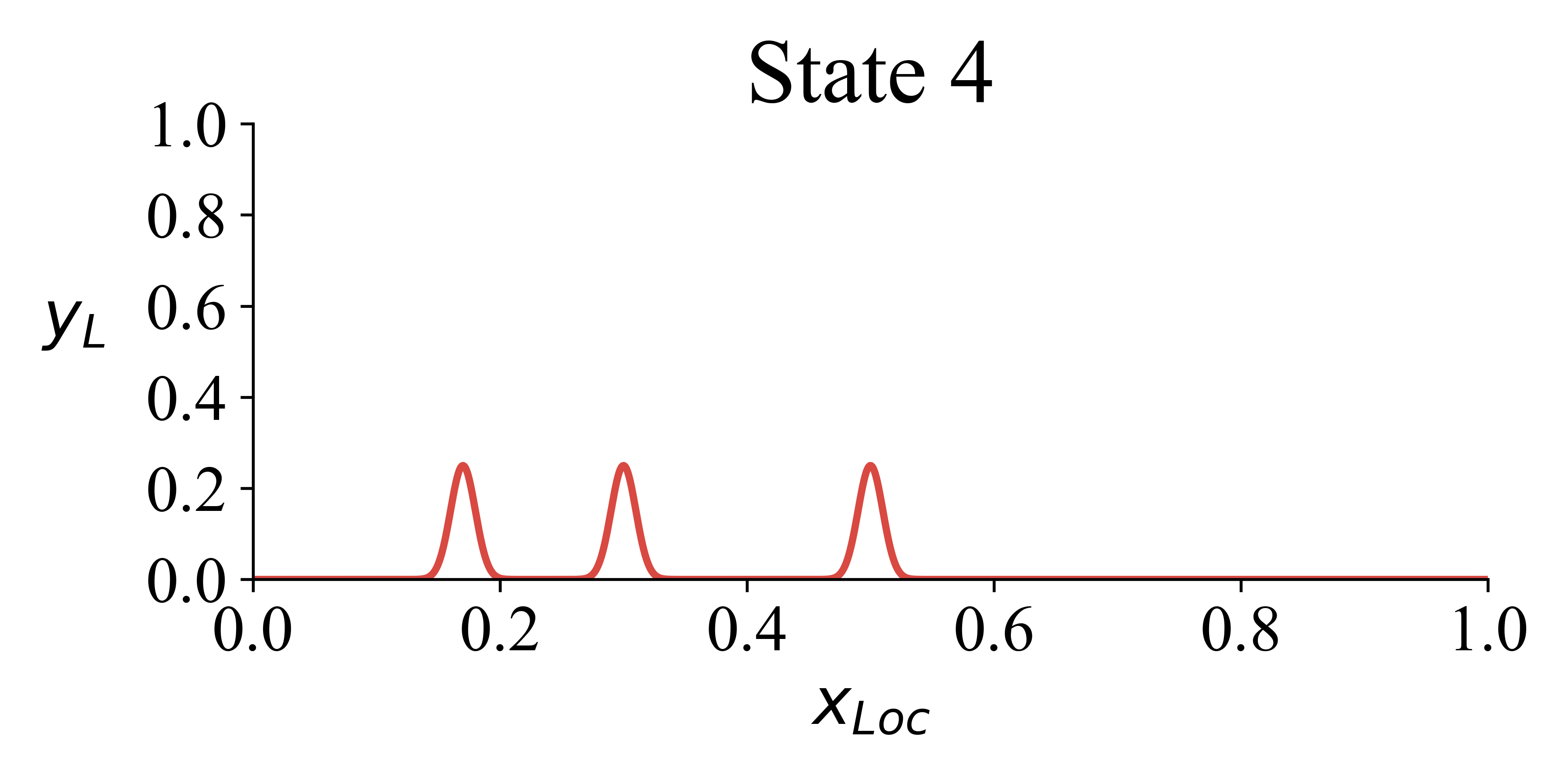}
    \end{subfigure}
    
    \begin{subfigure}{0.33\textwidth}
        \includegraphics[width=\linewidth]{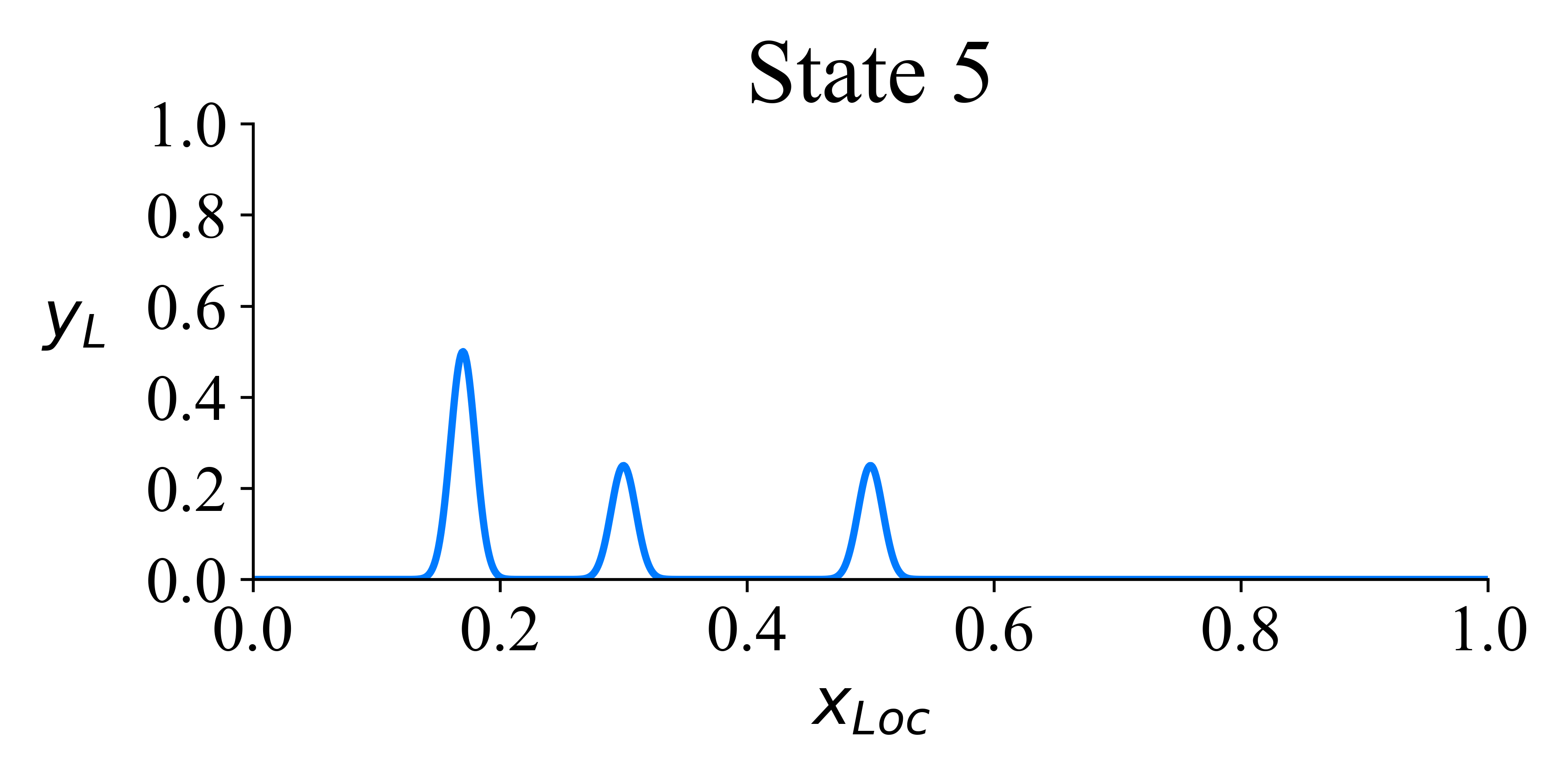}
    \end{subfigure}%
    \begin{subfigure}{0.33\textwidth}
        \includegraphics[width=\linewidth]{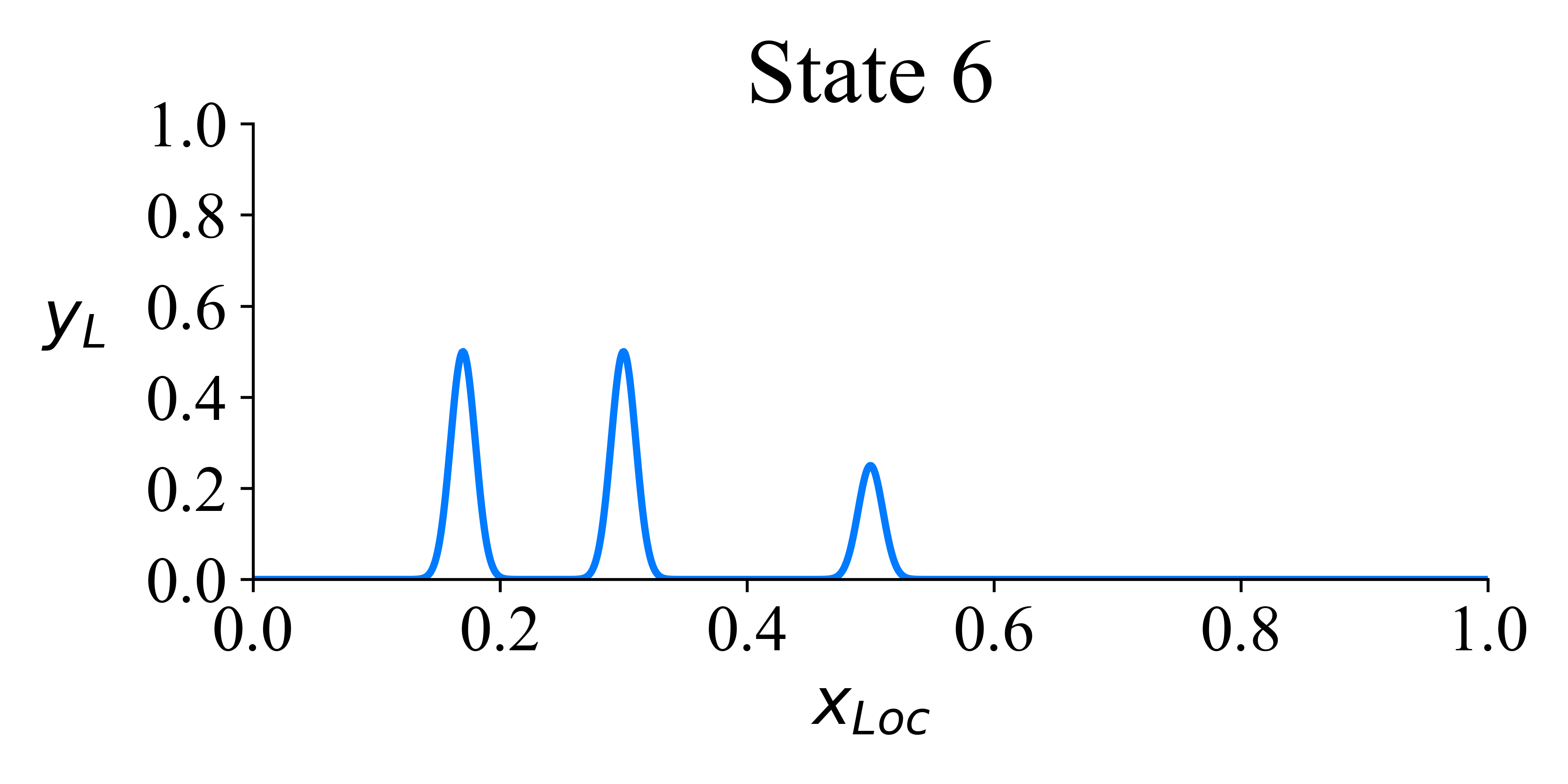}
    \end{subfigure}%
    \begin{subfigure}{0.33\textwidth}
        \includegraphics[width=\linewidth]{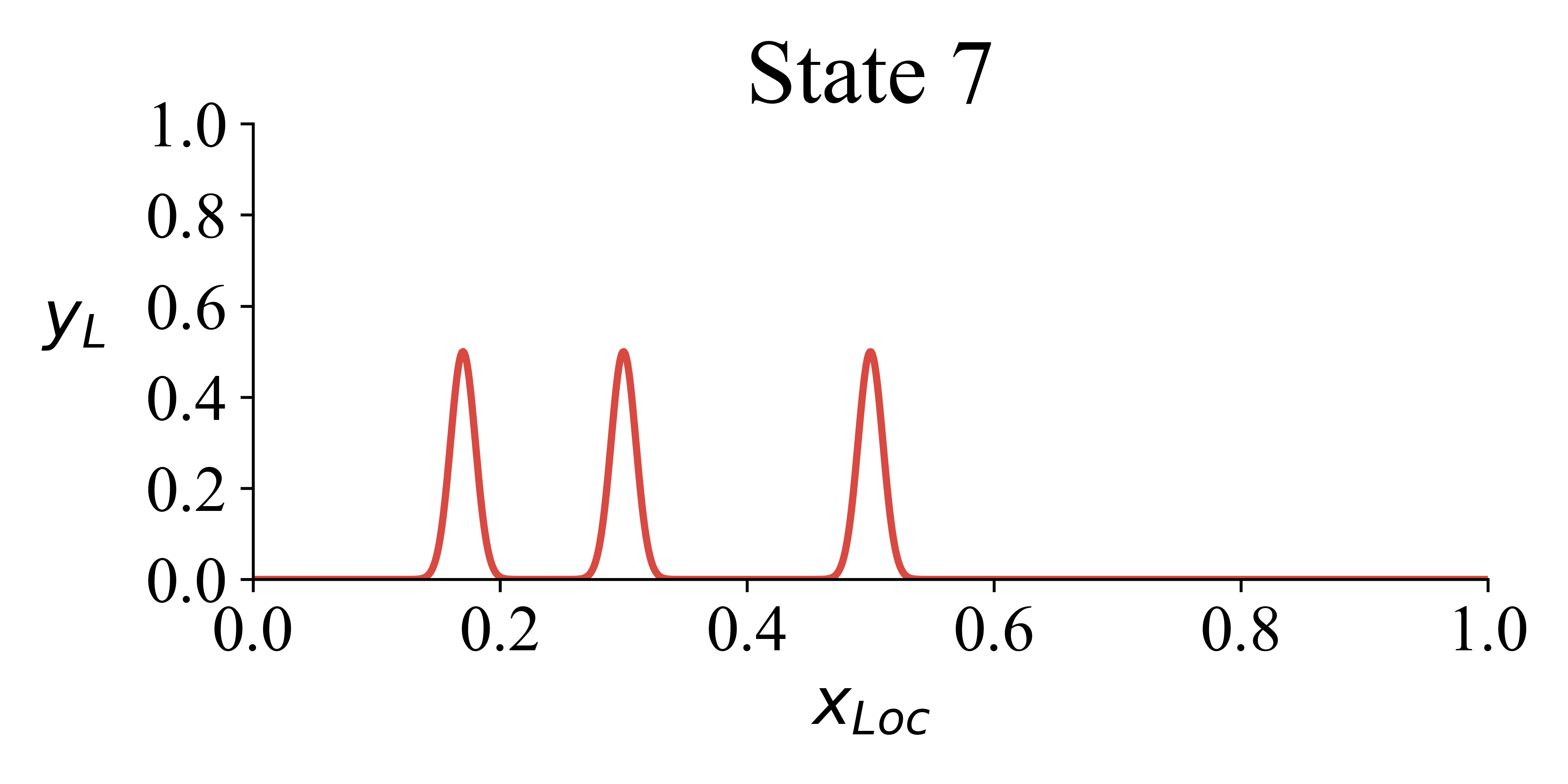}
    \end{subfigure}
    
    \begin{subfigure}{0.33\textwidth}
        \includegraphics[width=\linewidth]{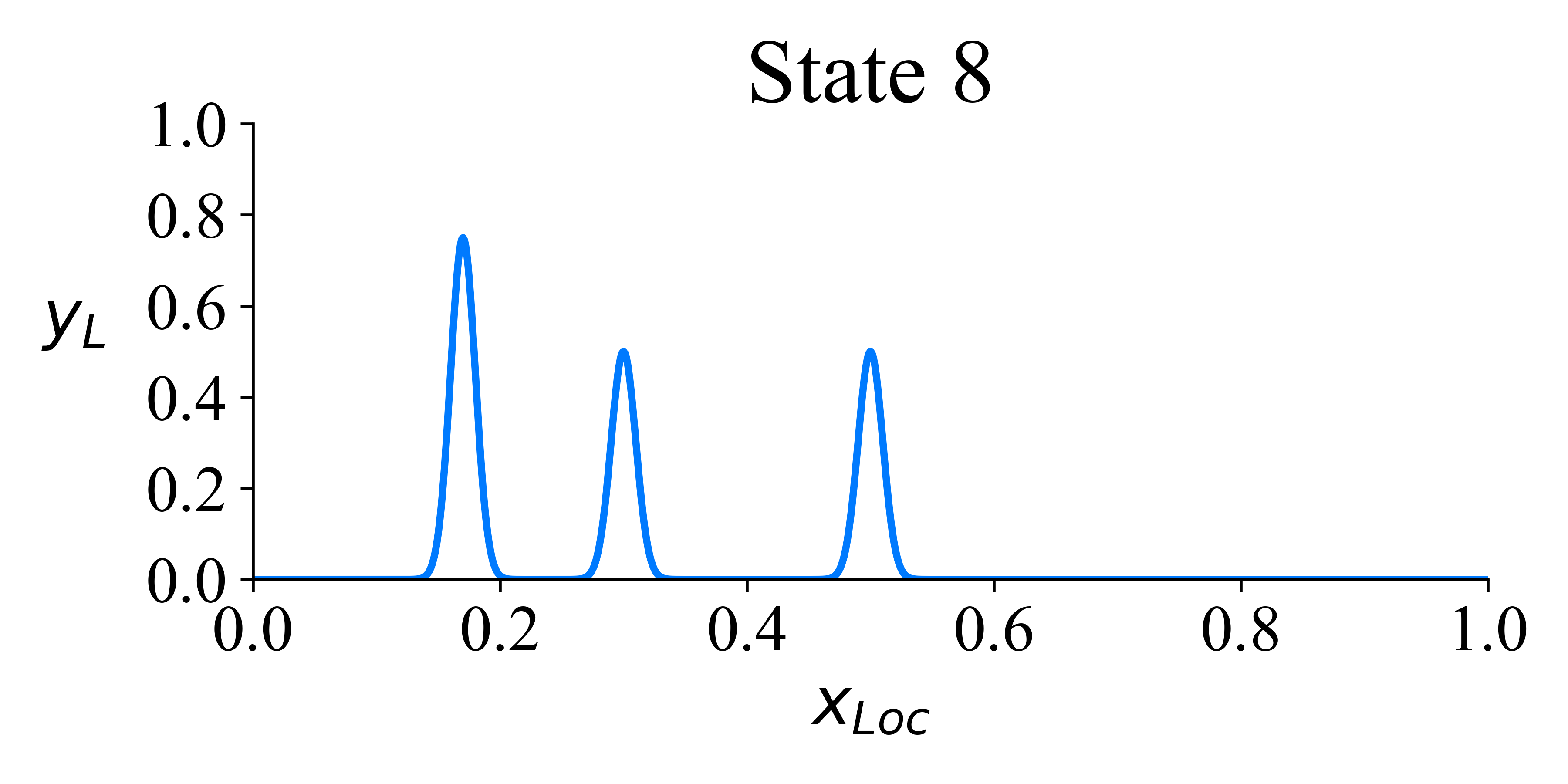}
    \end{subfigure}%
    \begin{subfigure}{0.33\textwidth}
        \includegraphics[width=\linewidth]{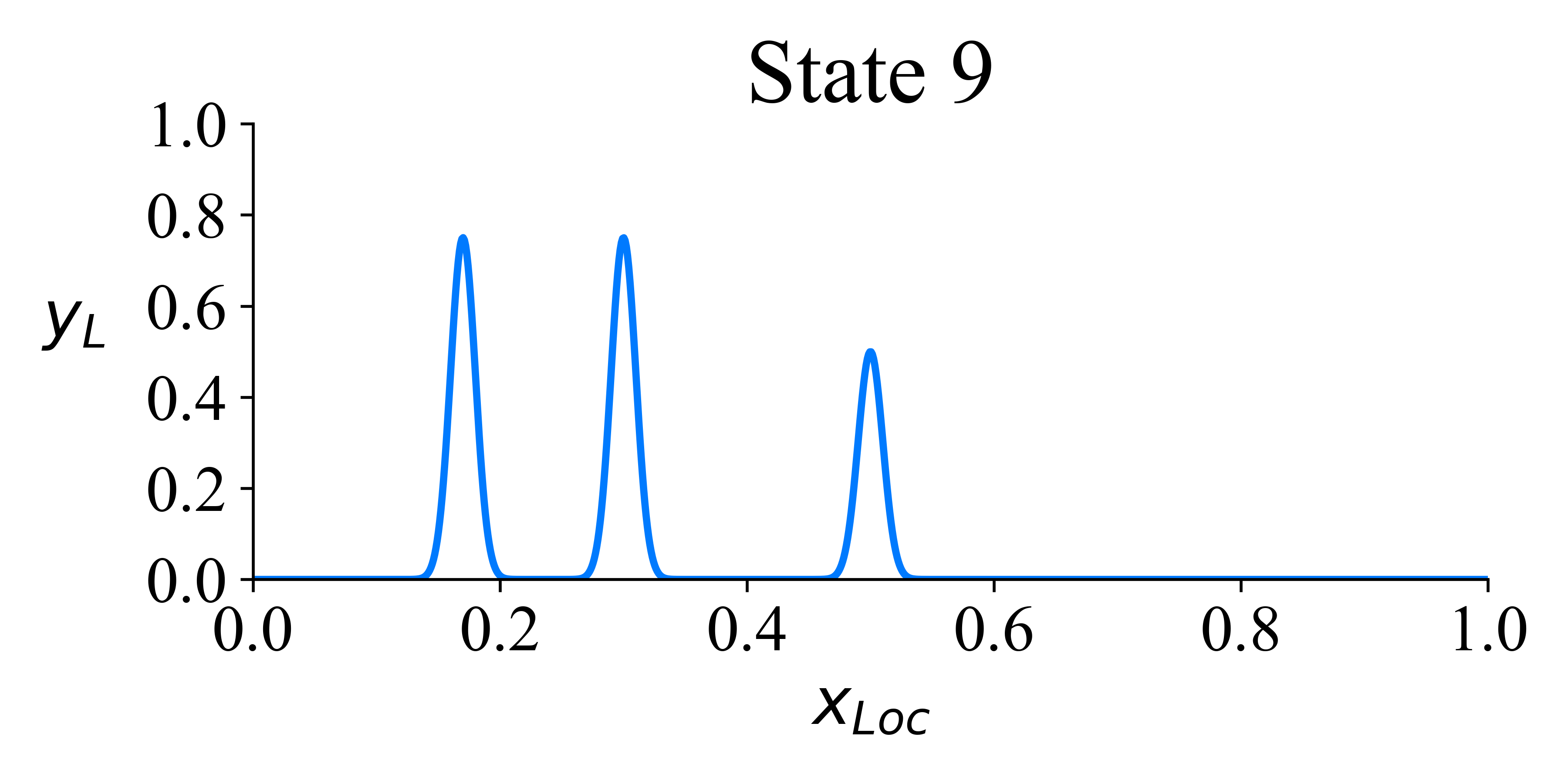}
    \end{subfigure}%
    \begin{subfigure}{0.33\textwidth}
        \includegraphics[width=\linewidth]{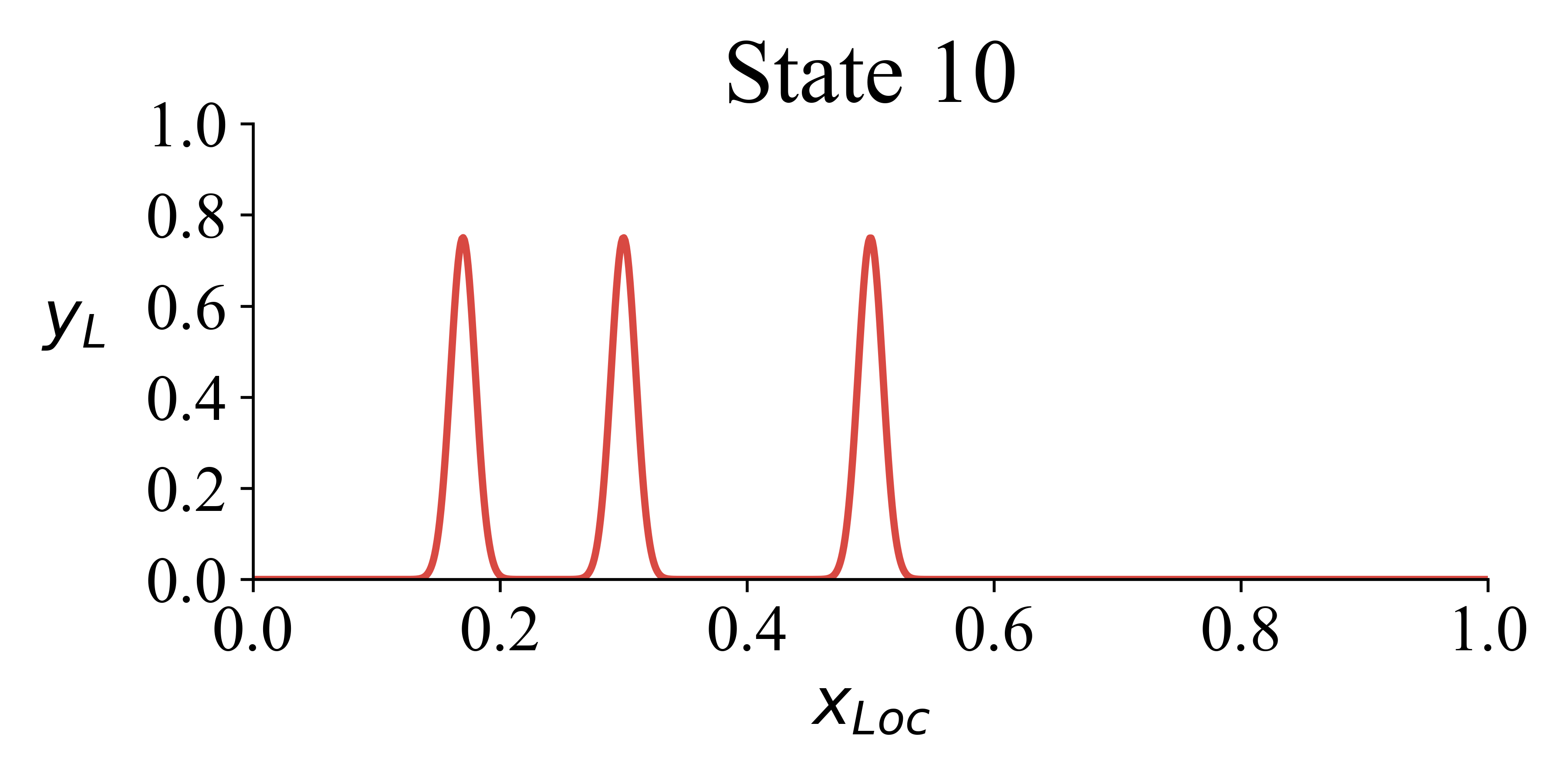}
    \end{subfigure}
    \caption{Qualitative visualization of system parameters of the blade, for different structural healthy states ($x_{\text{Loc}}$ and $y_{\text{L}}$ describe the relative location and length of damage).}
    \label{fig:case2_damage}
\end{figure*}

To visualize the structural state of the blade, we construct a damage profile function by summation of three 1D-Gaussian basis functions:
\begin{align*}
    y_{\text{L}}(x_{\text{Loc}}, \mu_1, \mu_2, \mu_3, \mu_4, \mu_5, \mu_6) = & \frac{\mu_4}{\sigma \sqrt{2 \pi}} \exp  \left(\frac{-(x_{\text{Loc}}-\mu_1)^2}{2 \sigma^2}\right) +
    \frac{\mu_5}{\sigma \sqrt{2 \pi}} \exp \left(\frac{-(x_{\text{Loc}}-\mu_2)^2}{2 \sigma^2}\right)\\ 
    + &\frac{\mu_6}{\sigma \sqrt{2 \pi}} \exp \left(\frac{-(x_{\text{Loc}}-\mu_3)^2}{2 \sigma^2}\right),
\end{align*}
where $x_{\text{Loc}}$ is the spatial location (the axis where damage is observed), $y_{\text{L}}$ is the relative damage length at each location along the blade, $\sigma = 0.01$ is the standard deviation of the Gaussian function and $\exp(\cdot)$ is the natural exponential function.
Figure~\ref{fig:case2_damage} shows the qualitative visualization of each structural state, where $x_{\text{Loc}}$ is the horizontal axis and $y_{\text{L}}$ is the vertical axis.
It is noted that relative damage location parameters $\mu_1, \mu_2, \mu_3$ are constant in all structural states, allowing us to validate whether the framework can accurately estimate both constant and varying parameters in the system parameters space.

For data pre-processing, a low-pass filter with a cut-off frequency of 380 Hz is applied to the raw acceleration measurements.
For dataset preparation, data samples are generated by windowing the force and acceleration data between 50,000 to 70,000 time instances in an experiment, with a fixed length of 200 (as shown in Figure~\ref{fig:case2-excitation}).
Within this period, each structural state is excited under similar external forces. 
We utilize the acceleration measurements of $a_1$ to $a_4$, whose locations are presented in Figure~\ref{fig:blade}. 
Each data sample is a triple of $(\mathbf{f}, \vect{\mu}, \ddot{\mathbf{x}})$ with $\mathbf{f} \in \mathbb{R}^{200\times1} $ and $ \vect{\mu} \in \mathbb{R}^{6 }$ and $ \ddot{\mathbf{x}} \in \mathbb{R}^{200\times4}$. 
The training dataset includes six structural states with state labels $1,2,3,4,7$ and $10$,
while the test dataset includes four structural states with state labels $5,6,8$ and $9$.
Acceleration data of the training and test datasets is normalized to [-1, 1].
Each structural state includes 100 data samples for either training or testing.
Thus, the training and test datasets include 600 and 400 data samples, respectively.

\subsubsection{Forward modeling results}
\label{sec:case2_forward} 
\begin{figure*}[!htb]
    \centering
    \includegraphics[width=0.4\linewidth]{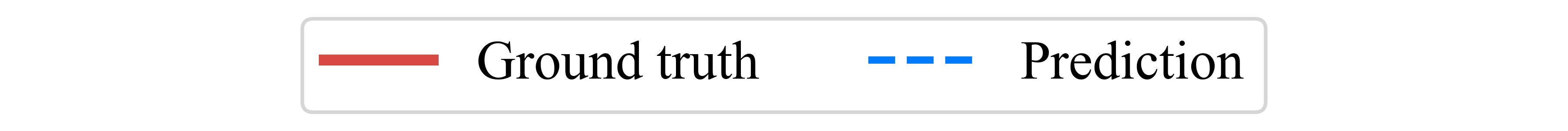}
    \begin{subfigure}{\textwidth}
        \includegraphics[width=\linewidth]{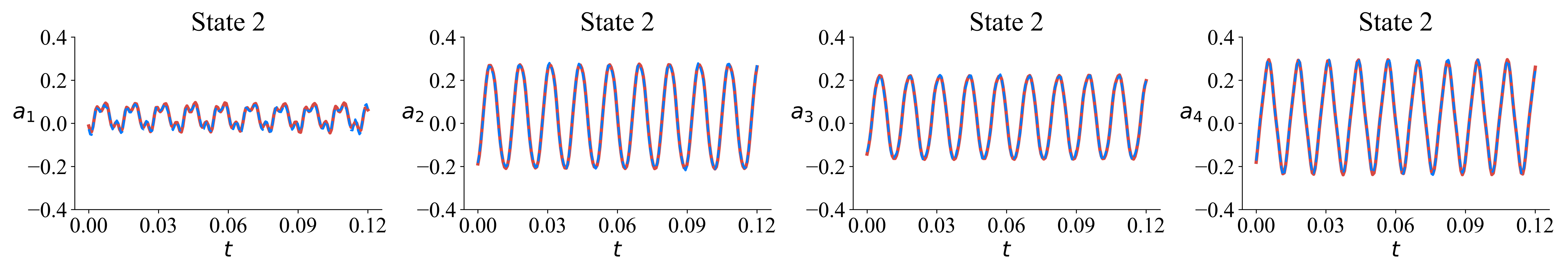}
        \caption{Results of a selected \textbf{training} data sample}
        \label{fig:case2_inter}
    \end{subfigure}
    \vfill
    \begin{subfigure}{\textwidth}
        \includegraphics[width=\linewidth]{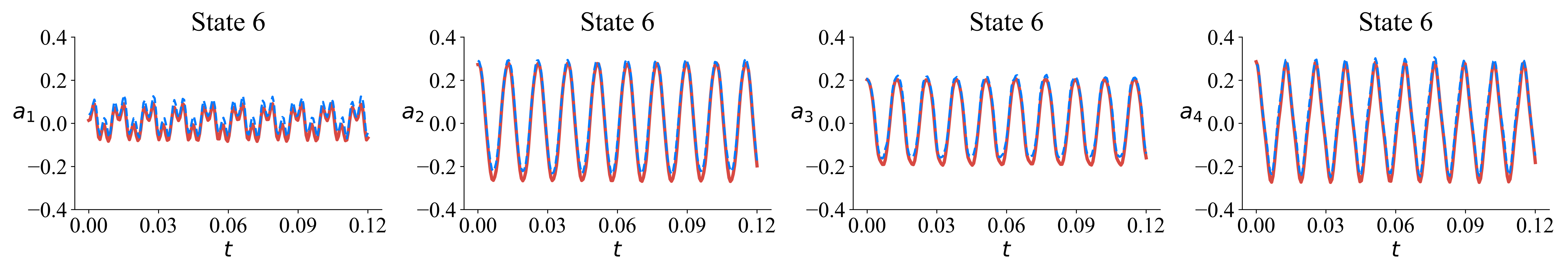}
        \includegraphics[width=\linewidth]{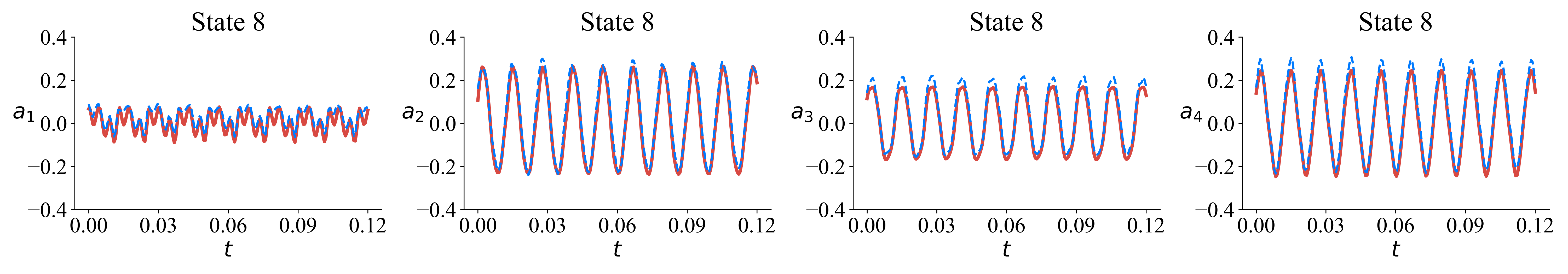}
        \includegraphics[width=\linewidth]{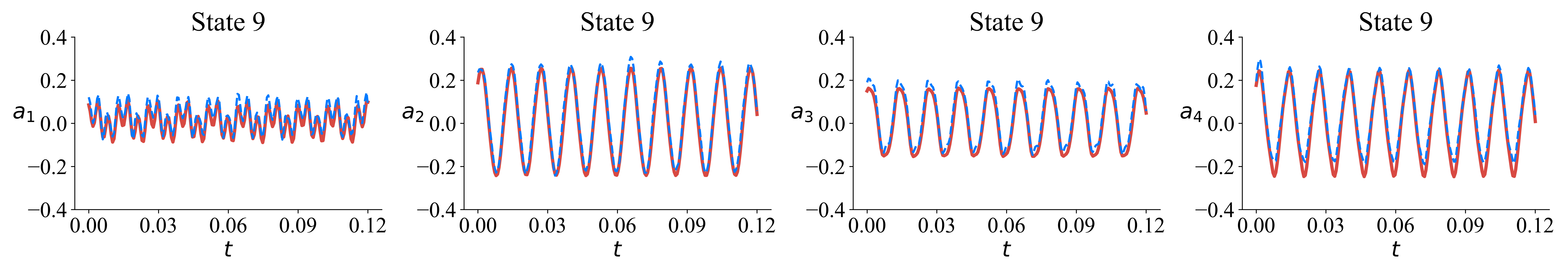}
        \caption{Results of selected \textbf{test} data samples}
        \label{fig:case2_extra}
    \end{subfigure}
    \caption{Response prediction results of data samples in Case 2 ($a_1 - a_4$ denote normalized acceleration).}
    \label{fig:case2_forward}
\end{figure*}
Table~\ref{tab:case2_forward} presents the NRMSE for response prediction of different models in Case 2. 
Parametric DeepONet (ND) achieves the best performance on both training and test data.
It achieves the NRMSE of $0.48 \times 10^{-1}$ on the training data and $1.69 \times 10^{-1}$ on the test data. 
CNN follows closely with NRMSE of $0.64 \times 10^{-1}$ for training and $2.32 \times 10^{-1}$ for test data.
Parametric DeepONet (LD) and MLP are also effective in the training and test stages.
In contrast, vanilla DeepONet is less effective in capturing the forward dynamics in Case 2.
\begin{table*}[!htb] 
    \centering
    \caption{NRMSE of forward response prediction in Case 2, for different datasets and models. All values are scaled by $10^{-1}$.}
    \begin{tabular}{lcc>{\centering\arraybackslash}m{2cm}>{\centering\arraybackslash}m{2cm}>{\centering\arraybackslash}m{2cm}} 
        \hline
       Dataset & \makecell{Parametric   \\ DeepONet (LD)} & \makecell{Parametric\\ DeepONet (ND)} &  DeepONet  & MLP & CNN \\
        \hline
        Training & $ 0.70  $ & $ 0.48   $ & $ 5.39 $ & $ 0.63  $ & 0.64 \\
        Test     & $ 2.69  $ & $ 1.69   $ & $ 5.60  $ & $ 2.89  $ & 2.32\\
        \hline
    \end{tabular}
    \label{tab:case2_forward}
\end{table*}

Figure~\ref{fig:case2_forward} presents qualitative results of response prediction by Parametric DeepONet (ND).
In particular,
Figure~\ref{fig:case2_inter} shows the response prediction of a selected train data sample, with very high accuracy.
Figure~\ref{fig:case2_extra} shows the response predictions of selected test data samples, where Parametric DeepONet (ND) successfully captures the major dynamics, though some minor errors remain in local details.

In summary, Parametric DeepONet (ND) demonstrates a significant advantage in response prediction in Case 2. Parametric DeepONet (LD), CNN, and MLP also show robustness and effectiveness with real experimental data, with comparable performance in response prediction.
Notably, CNN can also capture the major dynamics of the response without necessarily encoding the customized system parameters, according to the gradient-based initialization results in Section~\ref{sec:case2_inverse}.

\subsubsection{Inverse modeling results}
\label{sec:case2_inverse}
\begin{table*}[!htb] 
    \centering
    \caption{NRMSE of inverse parameter estimation in Case 2, for different datasets and models. All values are scaled by $10^{-1}$. }
    \begin{tabular}{lcccccccccc}
        \hline
        \multirow{2}{*}{Data} & \multicolumn{2}{c}{\makecell{Parametric DeepONet  \\ (LD)}} & \multicolumn{2}{c}{\makecell{Parametric DeepONet  \\ (ND)}} & \multicolumn{2}{c}{DeepONet} & \multicolumn{2}{c}{MLP} & \multicolumn{2}{c}{CNN}  \\
        \cline{2-11}
        &  $\mu_{\text{Loc}}$ & $\mu_{\text{L}}$  &  $\mu_{\text{Loc}}$ & $\mu_{\text{L}}$  &  $\mu_{\text{Loc}}$ & $\mu_{\text{L}}$ &  $\mu_{\text{Loc}}$ & $\mu_{\text{L}}$  &  $\mu_{\text{Loc}}$ & $\mu_{\text{L}}$ \\
        \hline
        \multicolumn{5}{l}{\textit{Gradient-based intialization}} \\
        Training & $1.59  $ & $2.59  $ &1.49 & 7.27 & $13.8  $ & $6.80  $ & $8.86   $ & $7.35  $ & $14,96   $ & $15.26   $ \\
        Test     & $5.17  $ & $5.16  $ & 1.64& 6.21& $14.6  $ & $7.49  $ & $11.75  $   & $8.71  $ & $15.15   $ & $12.05   $\\
        \hline
        \multicolumn{5}{l}{\textit{Neural refinement}}  \\
        Training & $0.17 $ & $0.51 $ & 0.32& 1.34 & $ 0.28 $ & $1.87 $ & $0.20  $ & $0.40  $ & $0.25  $ &  $3.66  $ \\
        Test     & $1.25 $ & $4.19 $ & 0.42 & 5.13 & $ 0.41 $ & $6.90  $ & $0.25  $ & $3.66  $ & $0.33   $&  $4.87  $ \\
        \hline
        \label{tab:case2_inverse}
    \end{tabular}
\end{table*}

\begin{figure*}[!h]
    \centering
    \includegraphics[width=\linewidth]{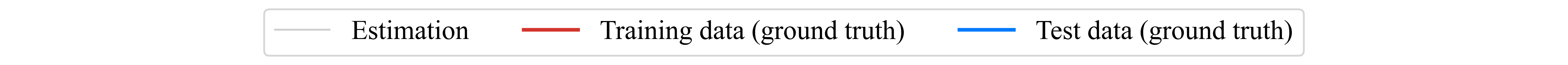}
    \begin{subfigure}{\linewidth}
        \centering
        \includegraphics[width=0.33\linewidth]{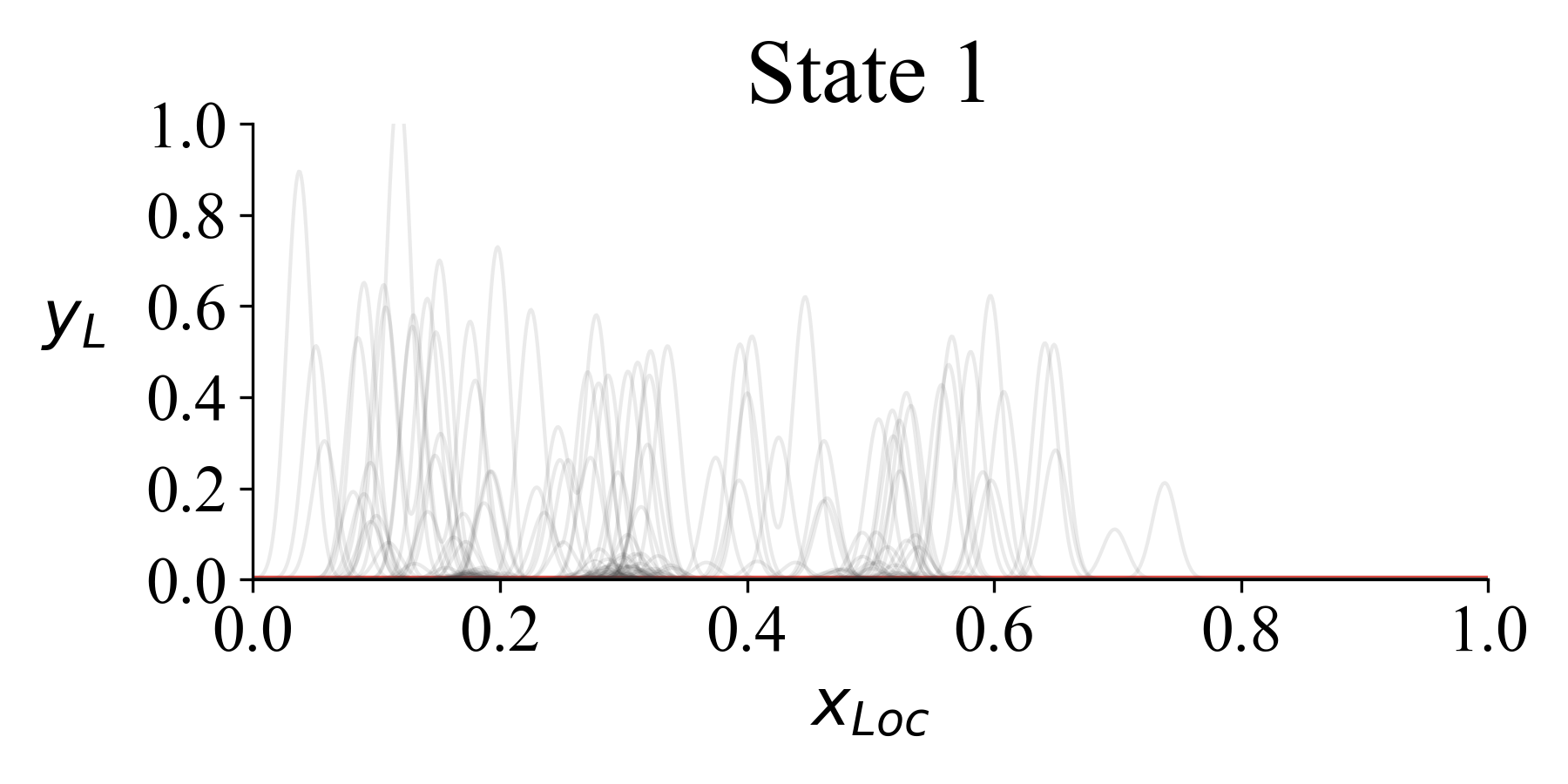}
    \end{subfigure}
        \includegraphics[width=0.33\linewidth]{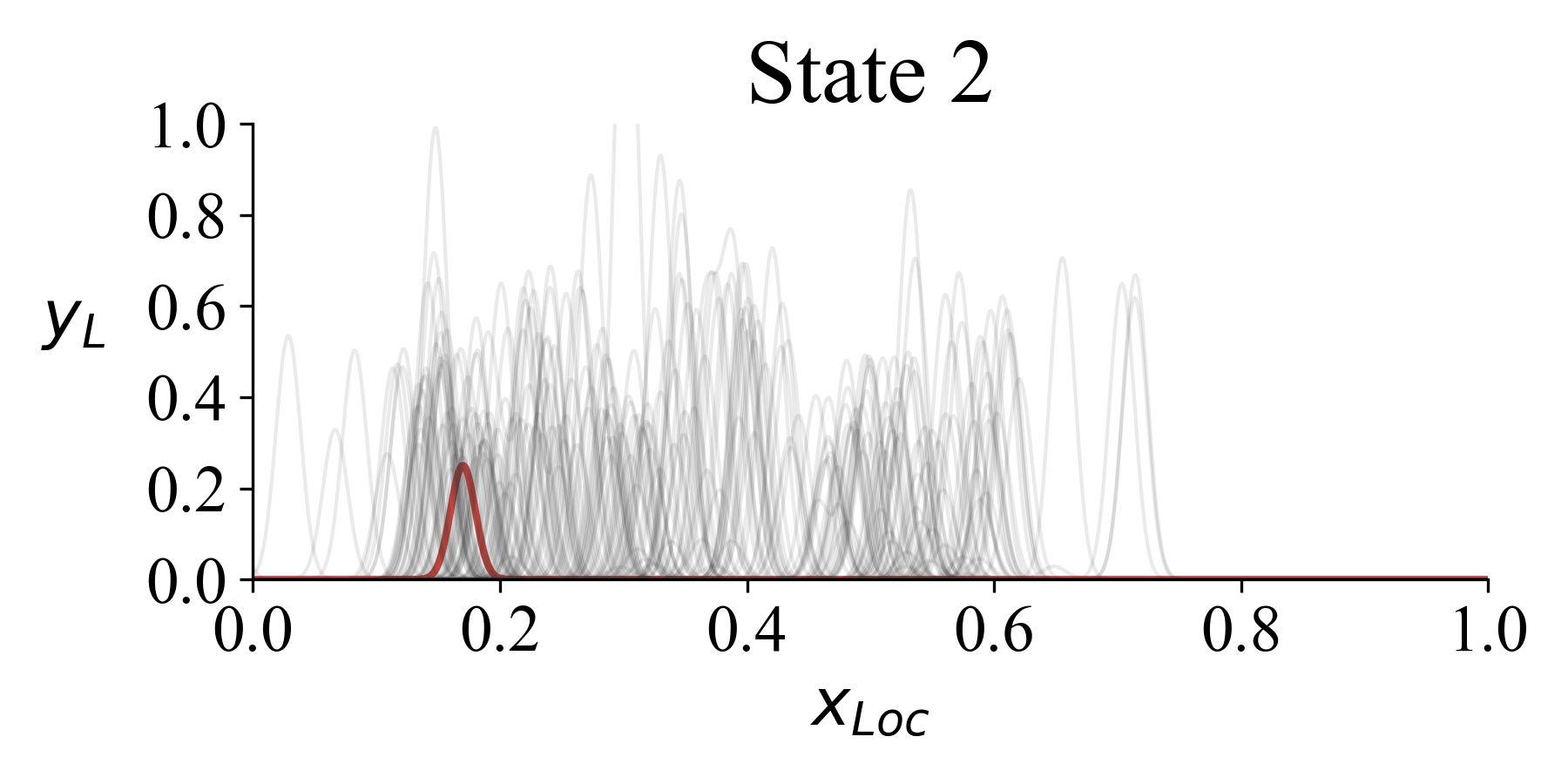}%
        \includegraphics[width=0.33\linewidth]{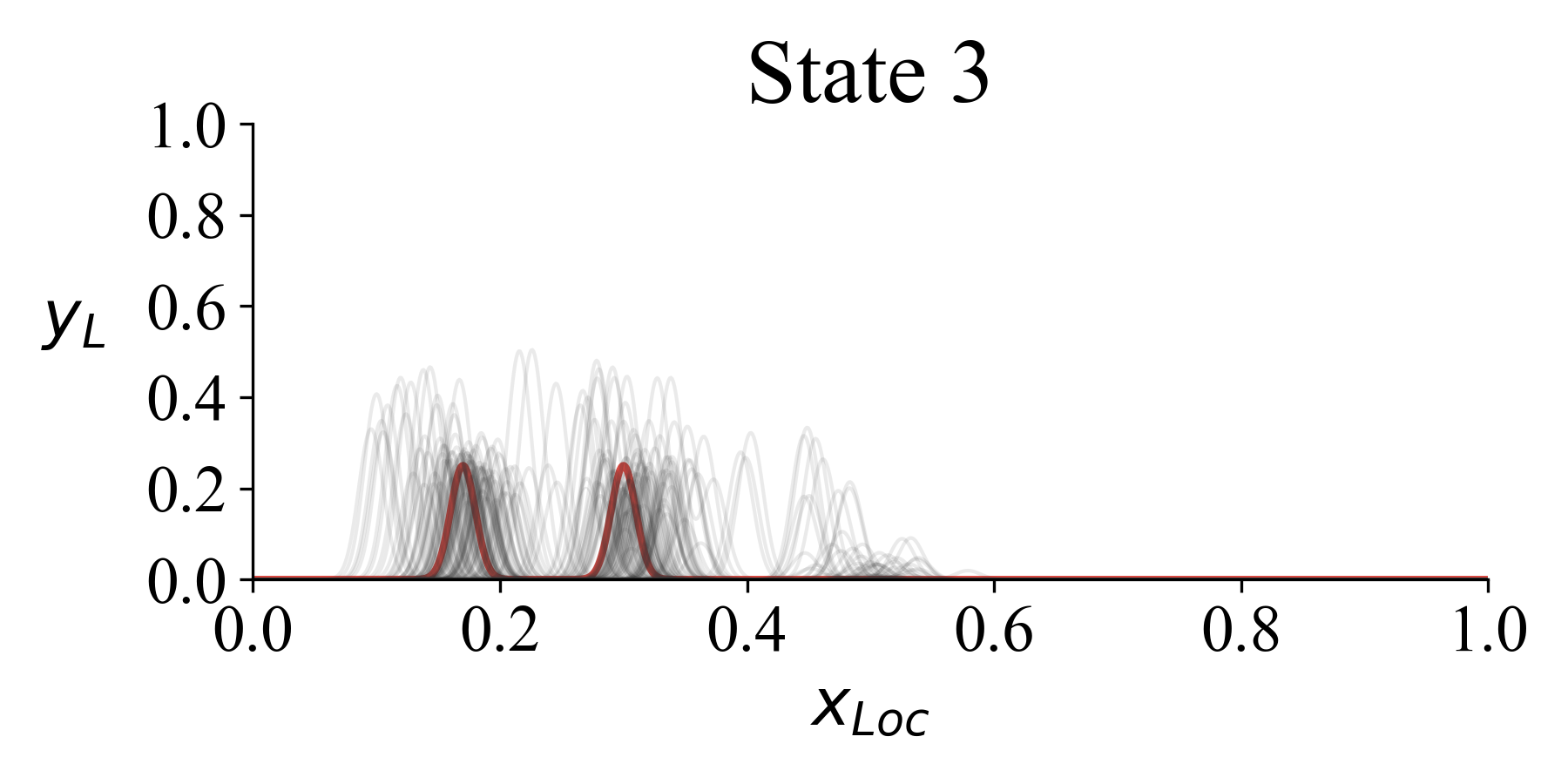}%
        \includegraphics[width=0.33\linewidth]{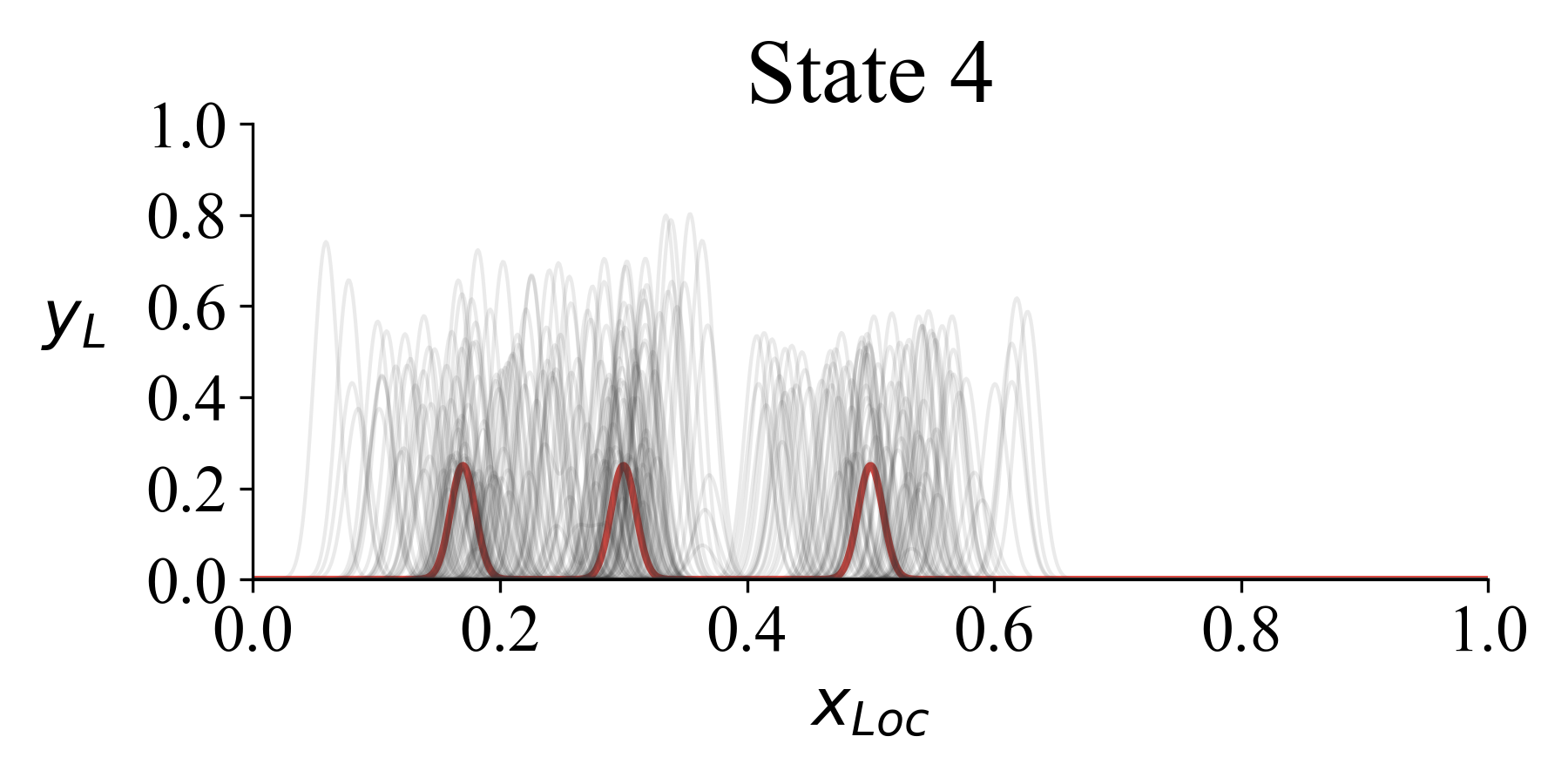}
        
        \includegraphics[width=0.33\linewidth]{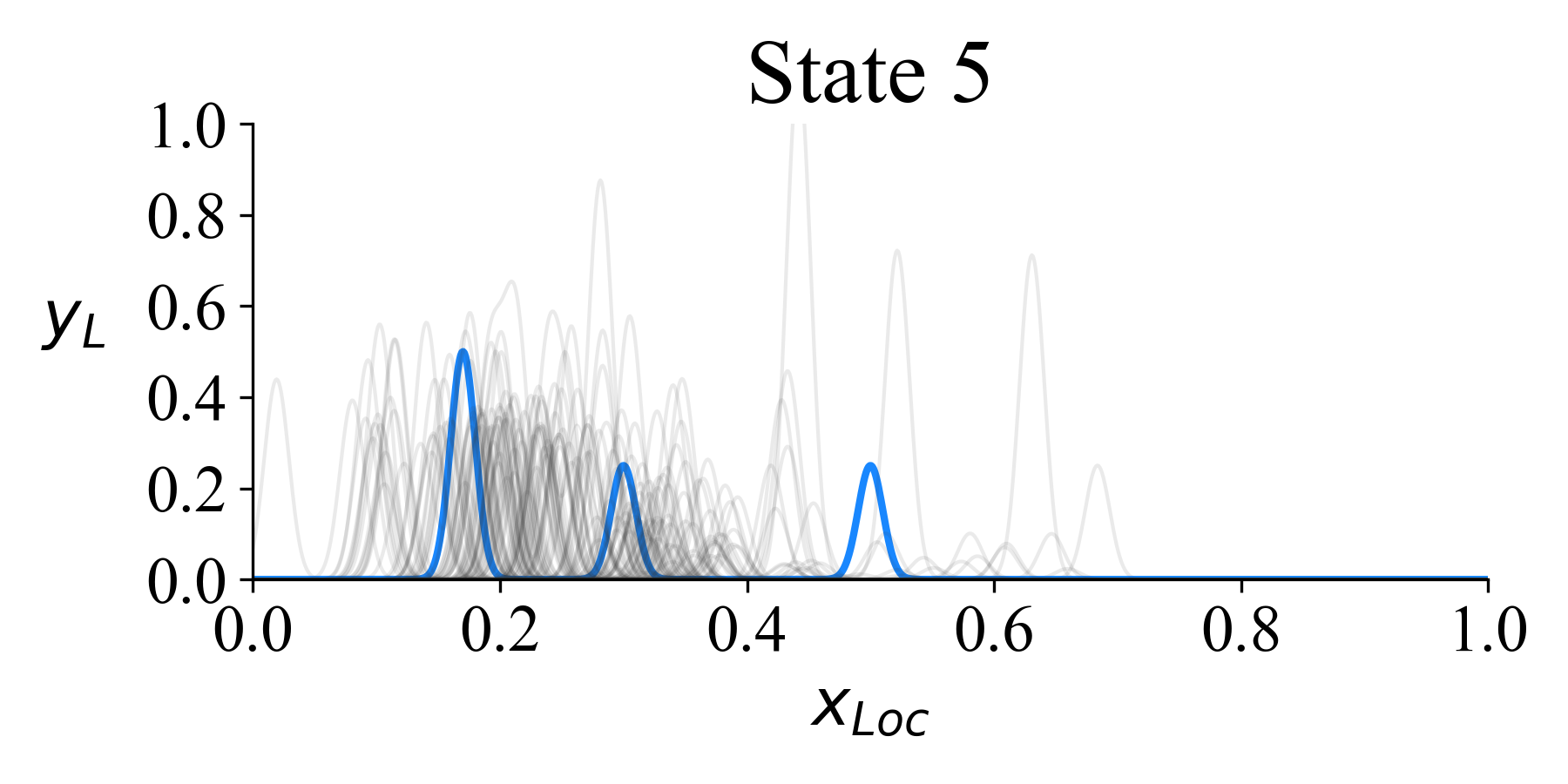}%
        \includegraphics[width=0.33\linewidth]{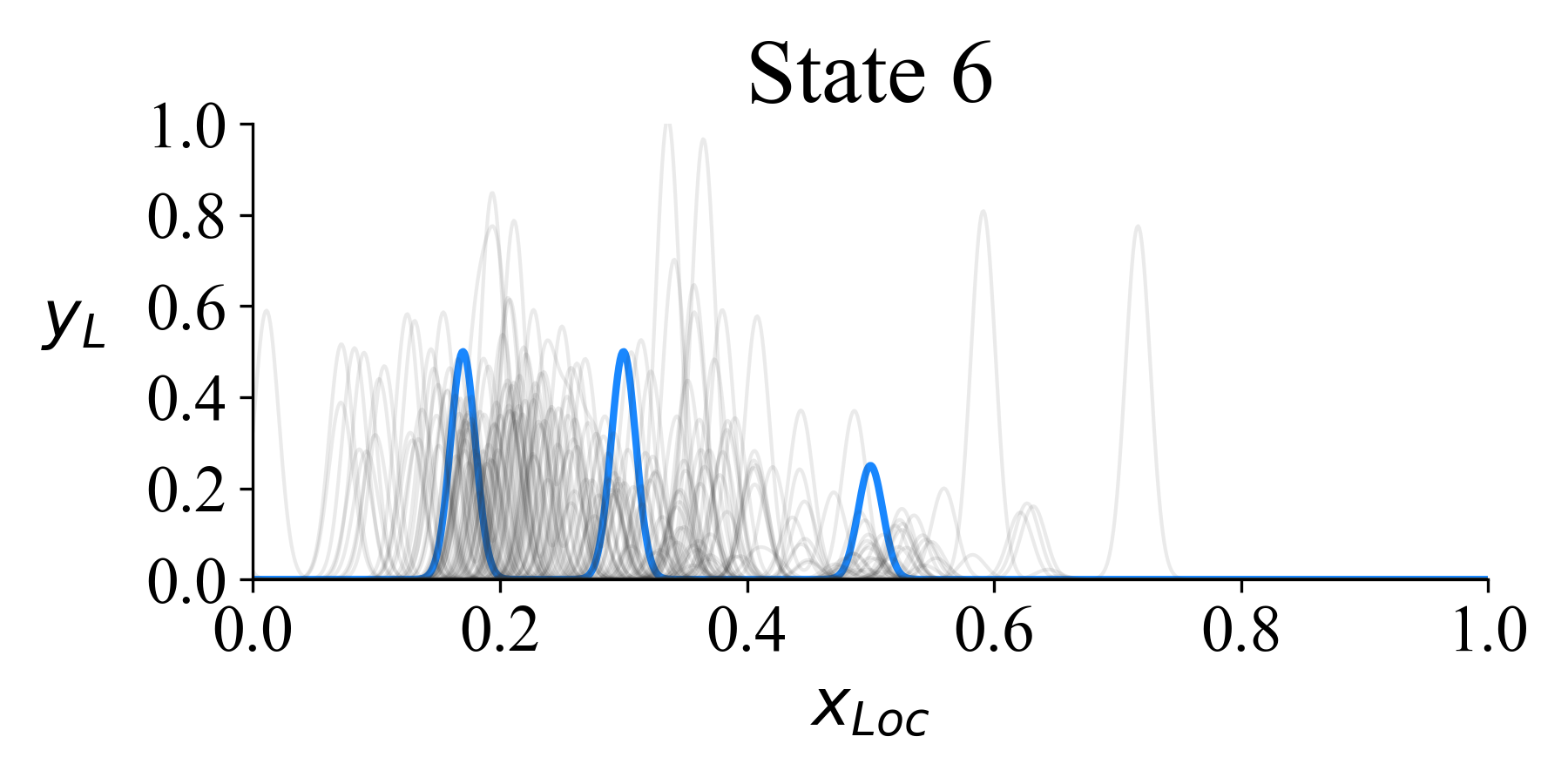}%
        \includegraphics[width=0.33\linewidth]{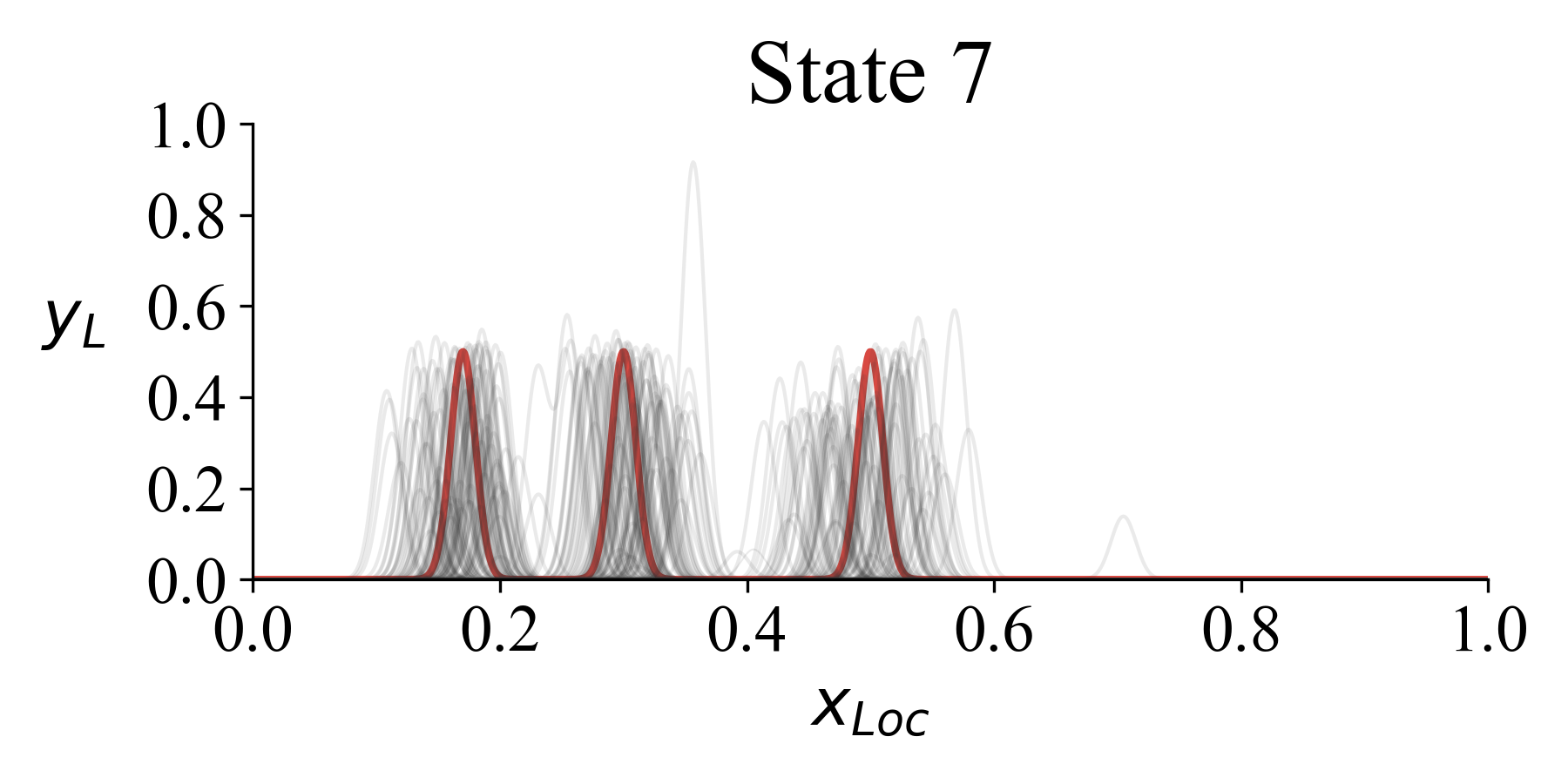}
        
        \includegraphics[width=0.33\linewidth]{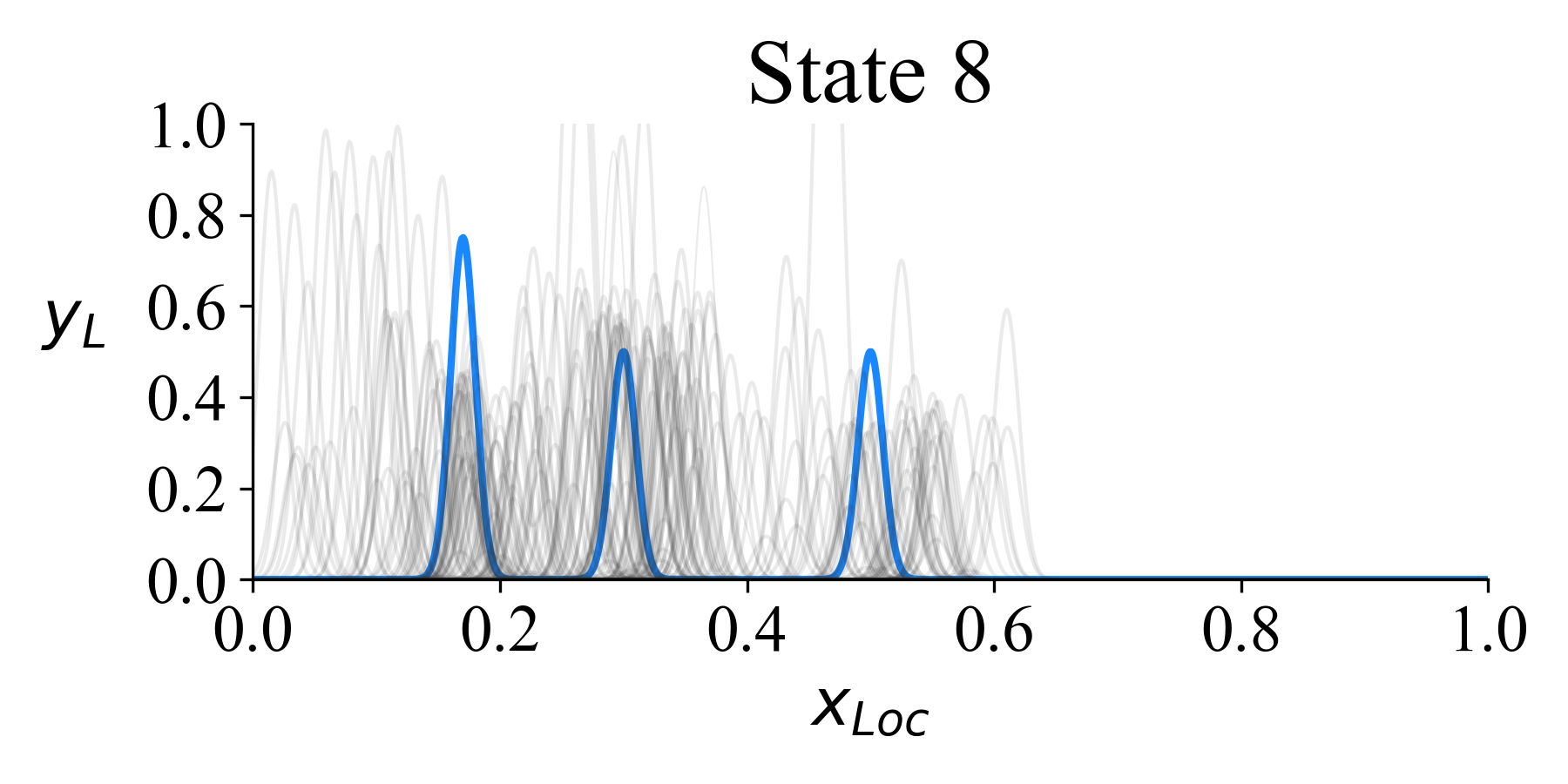}%
        \includegraphics[width=0.33\linewidth]{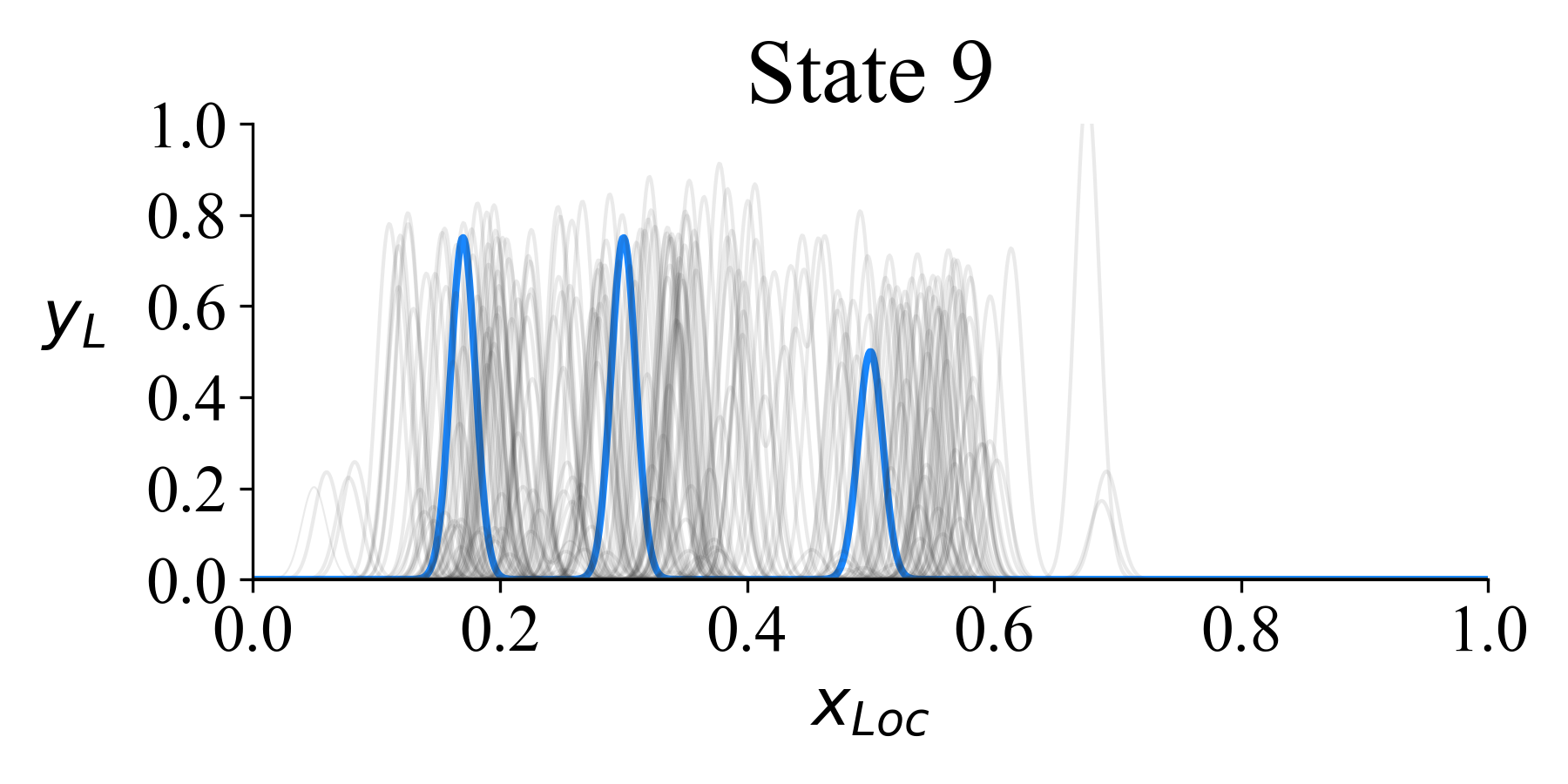}%
        \includegraphics[width=0.33\linewidth]{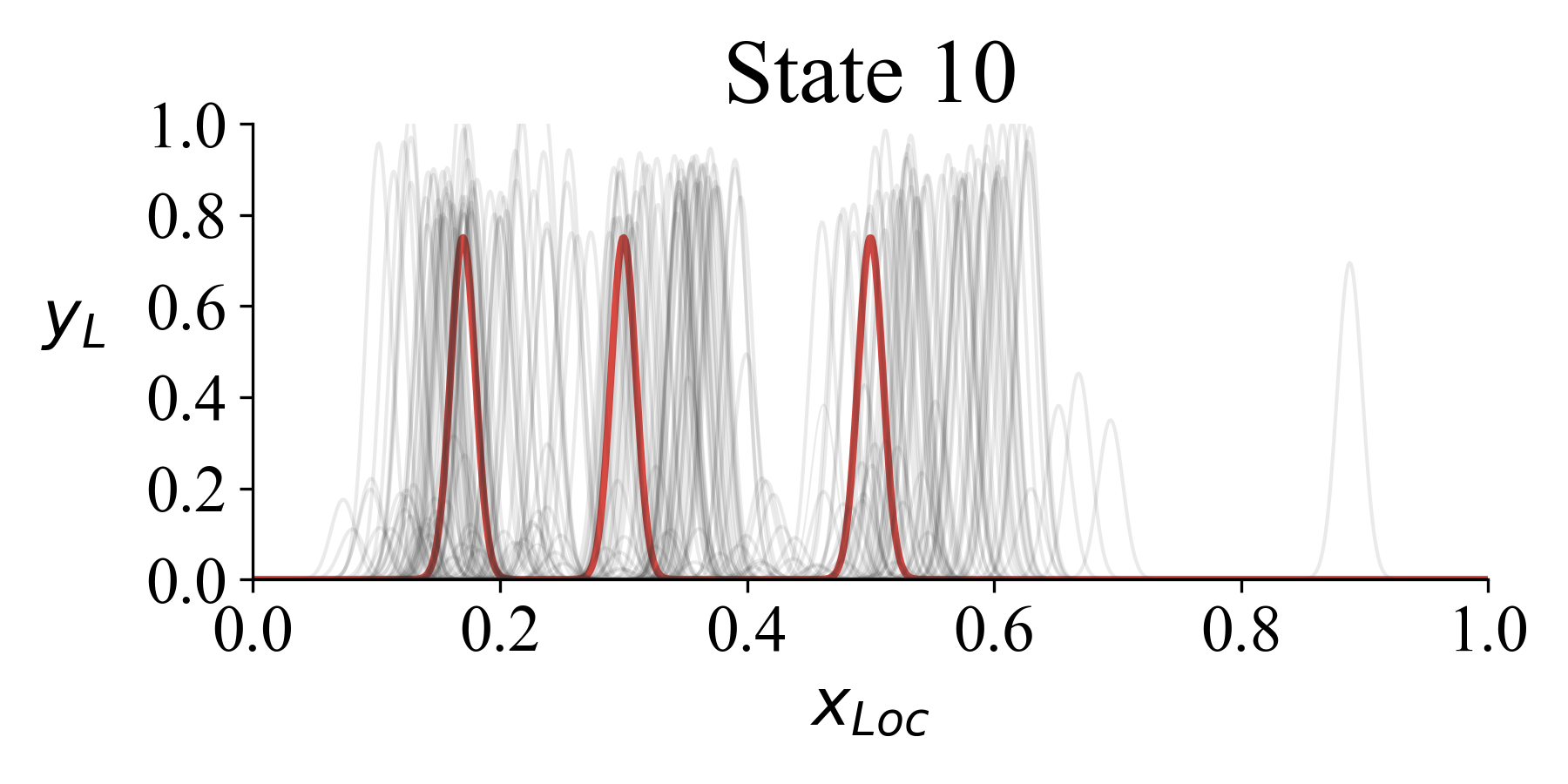}
    \caption{Parameter estimation results of structural states 1-10 in Case 2, after gradient-based initialization.}
     \label{fig:case2_inverse_gi}
\end{figure*}

\begin{figure*}[!h]
    \includegraphics[width=\linewidth]{case2/case2_inverse_legend.png}
    \begin{subfigure}{\linewidth}
    \centering
        \includegraphics[width=0.33\linewidth]{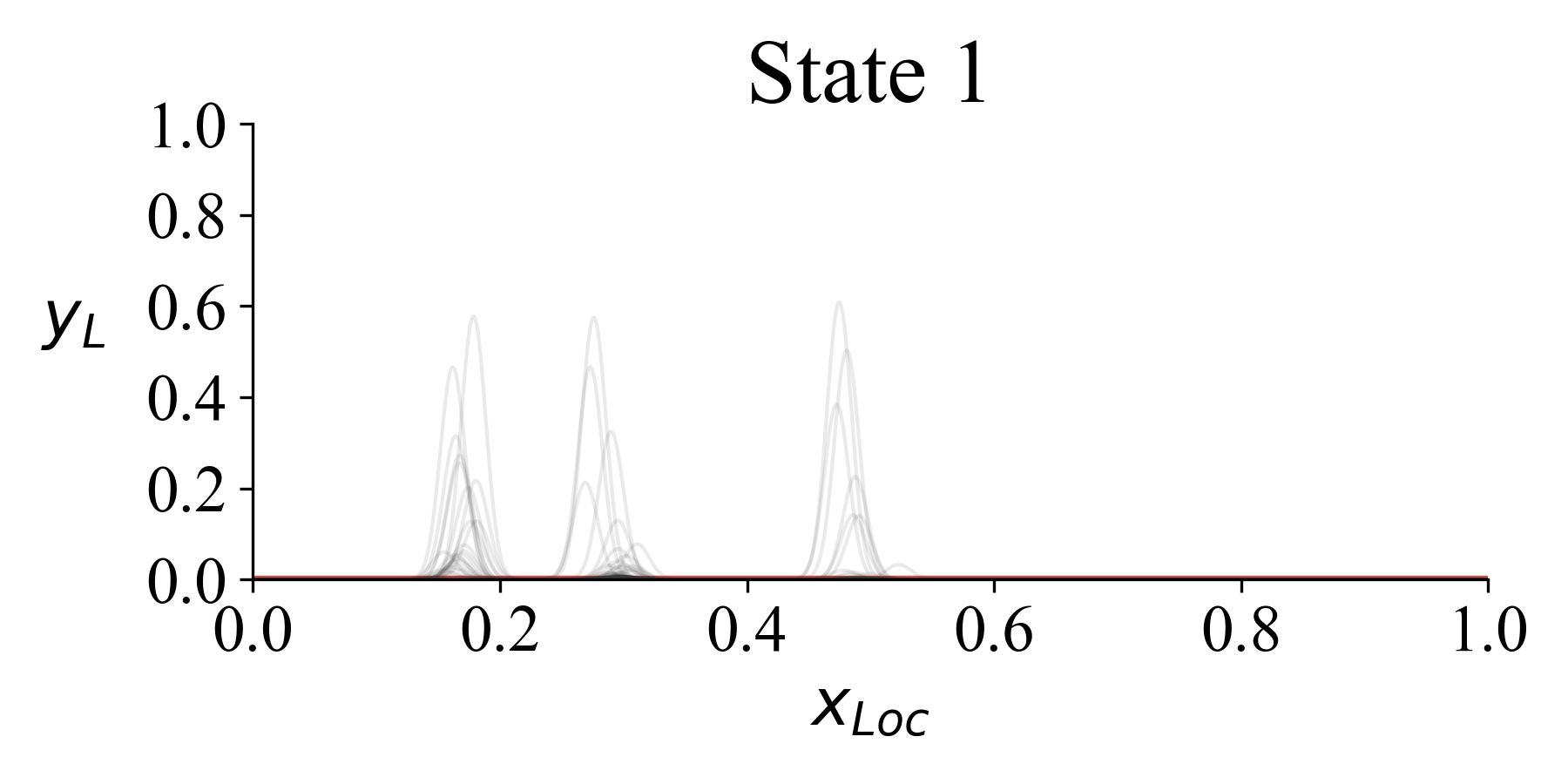}
    \end{subfigure}
        \includegraphics[width=0.33\linewidth]{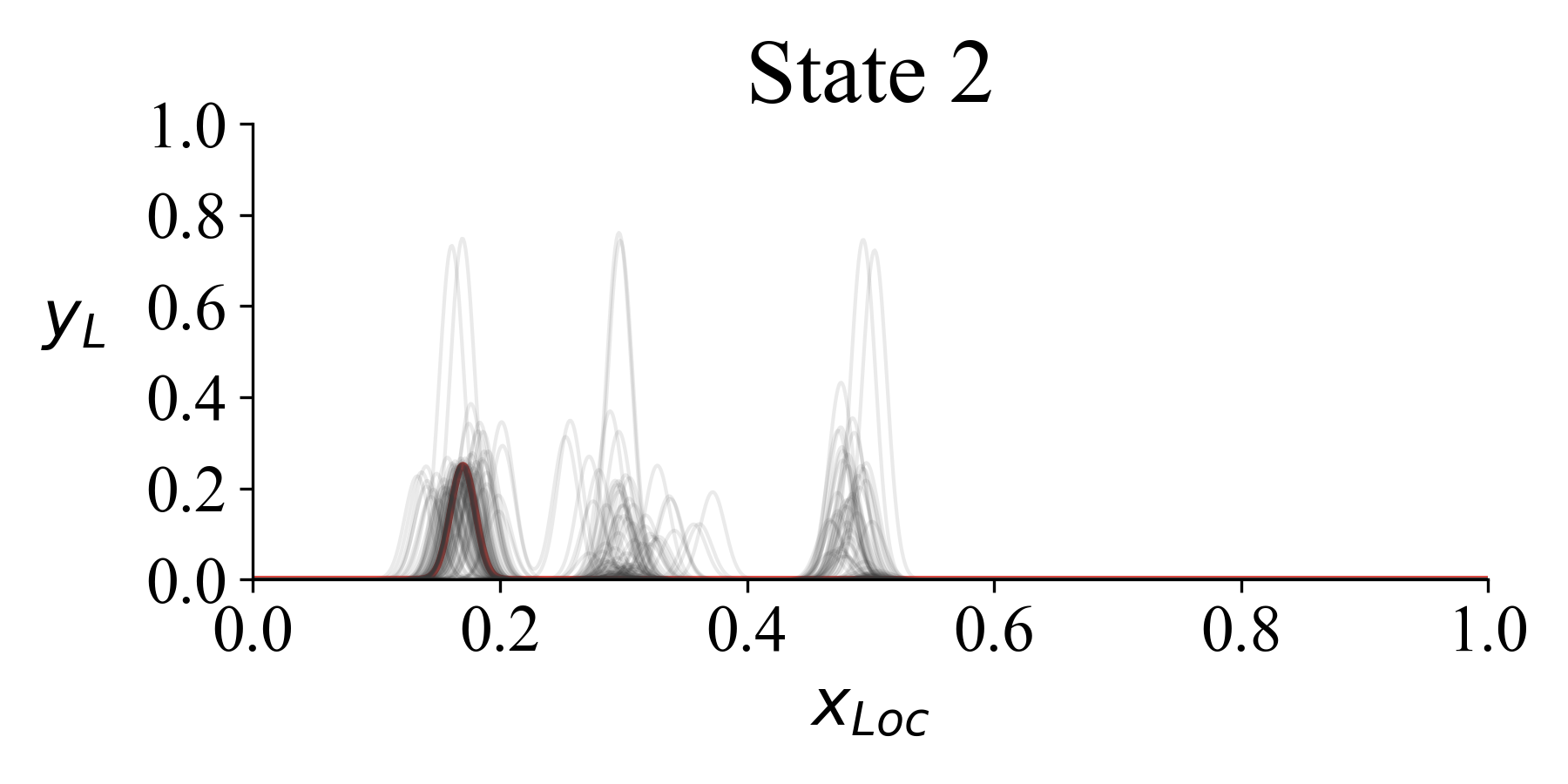}%
        \includegraphics[width=0.33\linewidth]{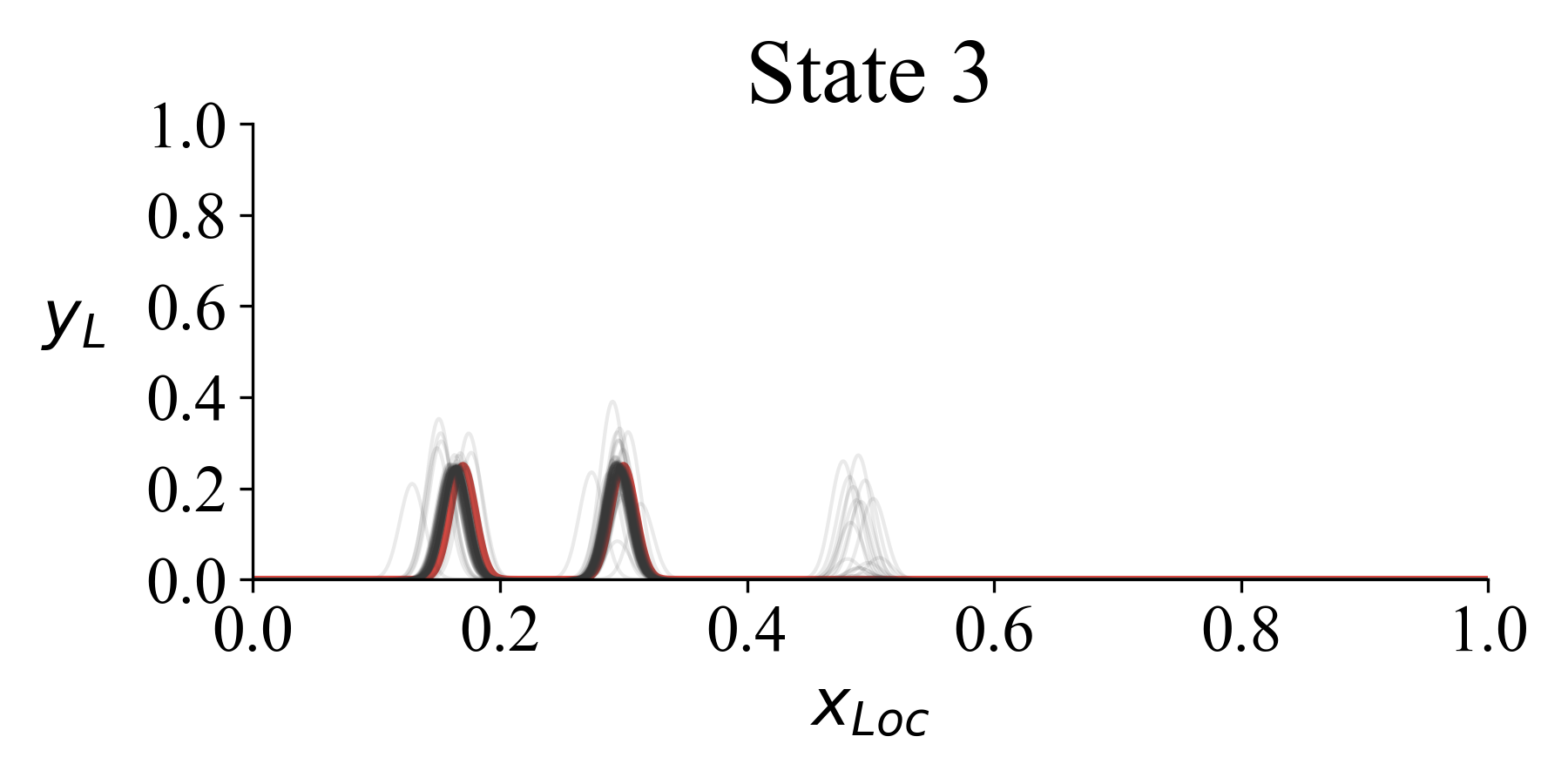}%
        \includegraphics[width=0.33\linewidth]{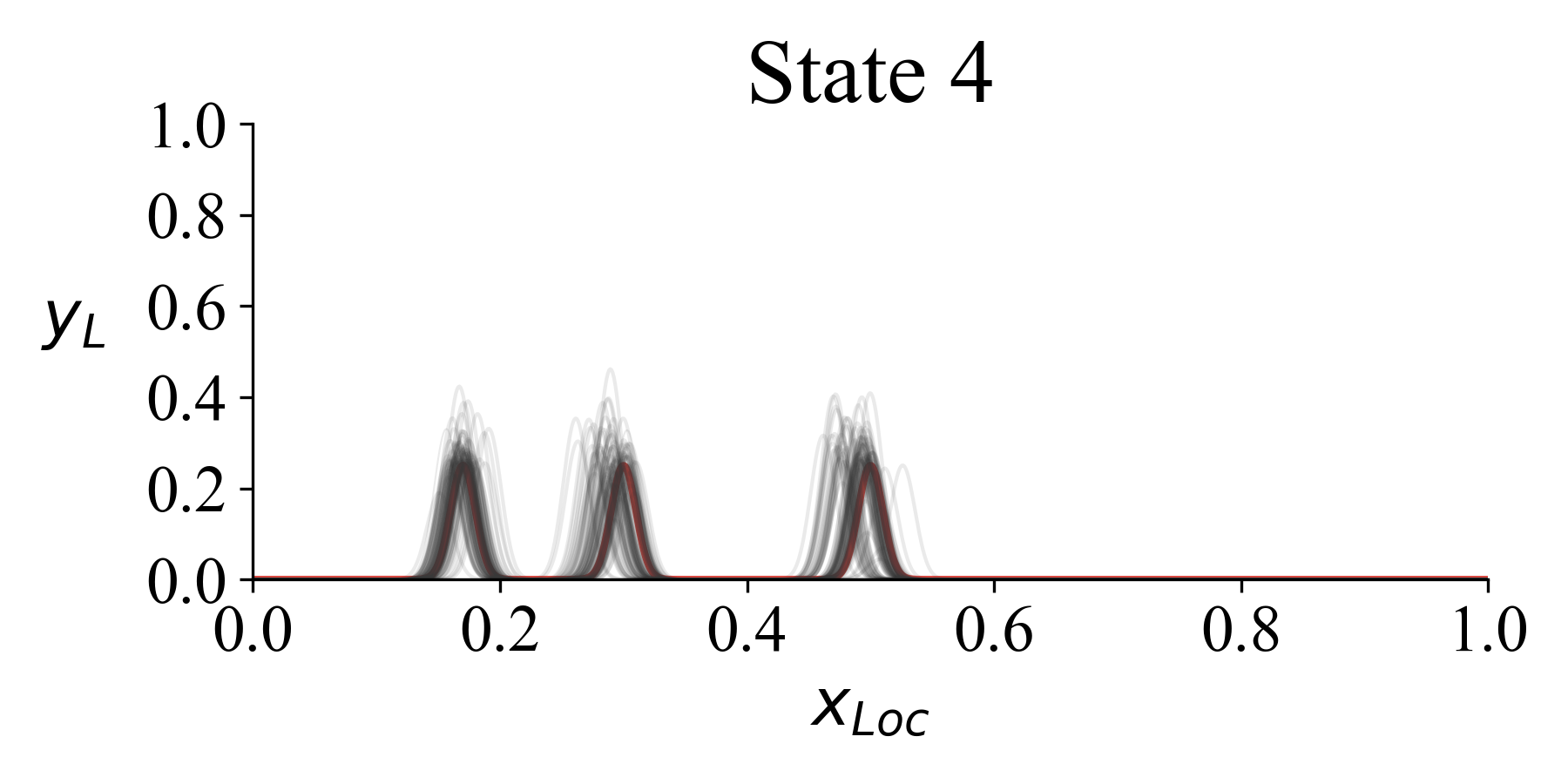}
        
        \includegraphics[width=0.33\linewidth]{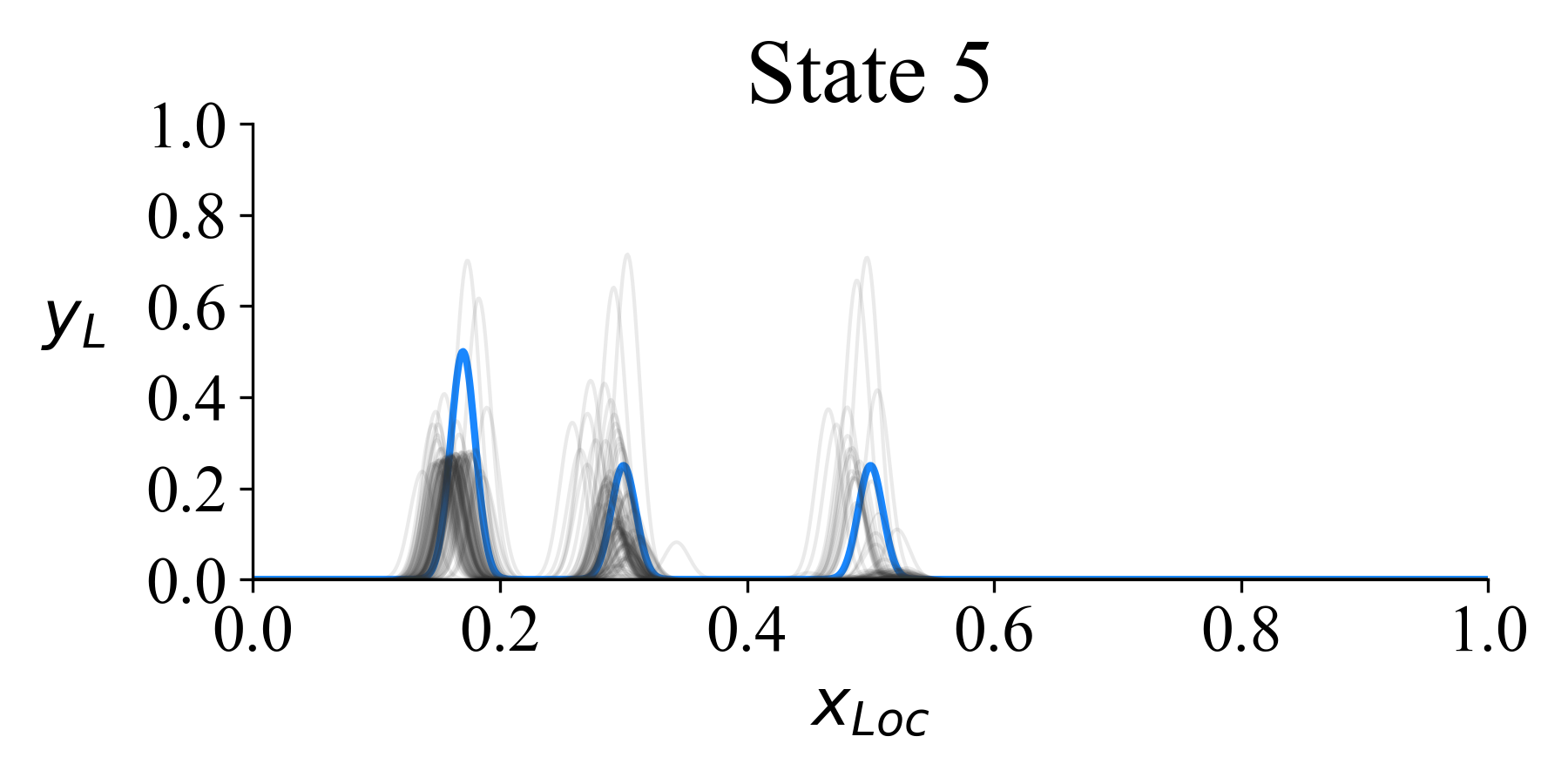}%
        \includegraphics[width=0.33\linewidth]{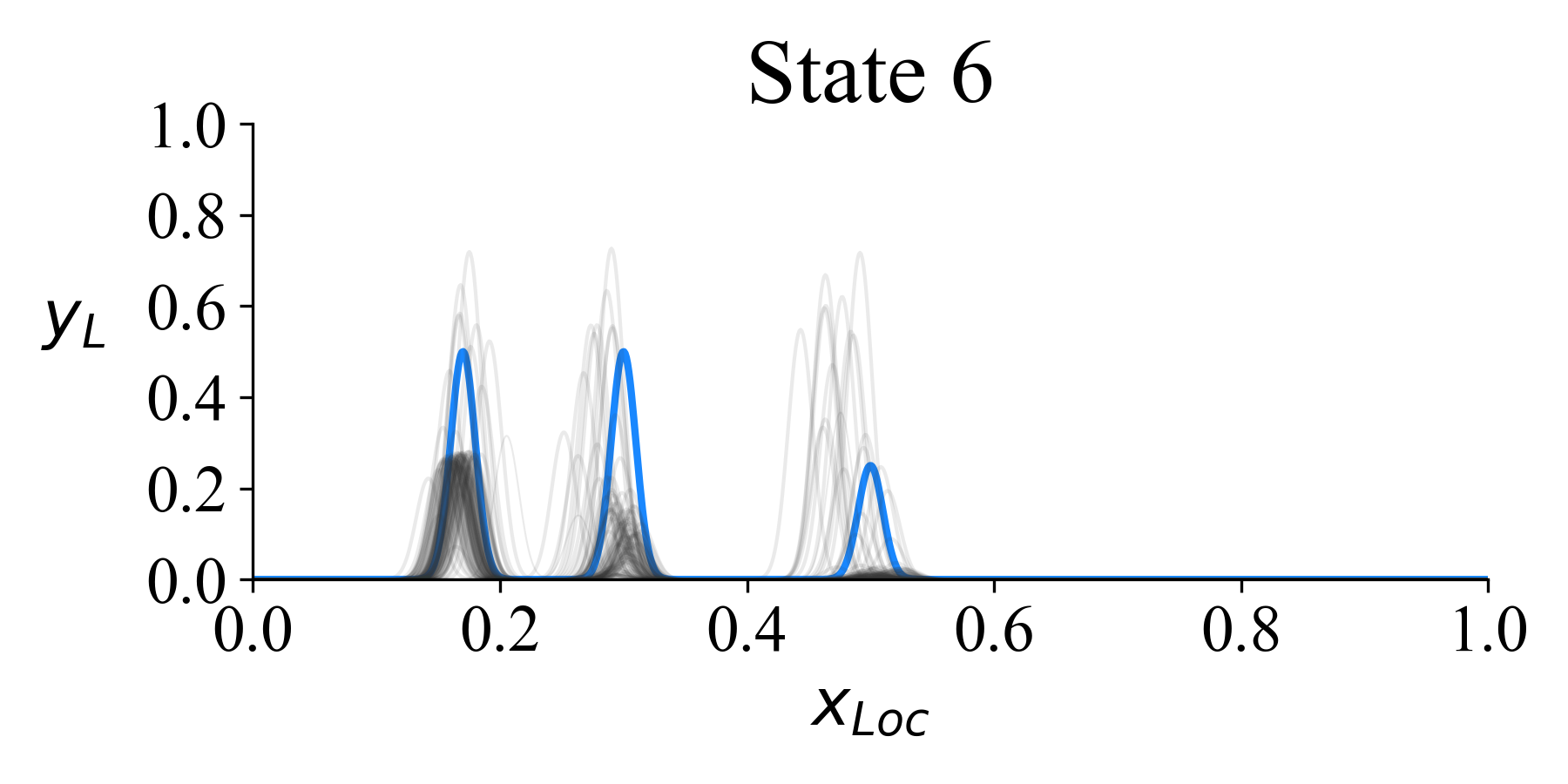}%
        \includegraphics[width=0.33\linewidth]{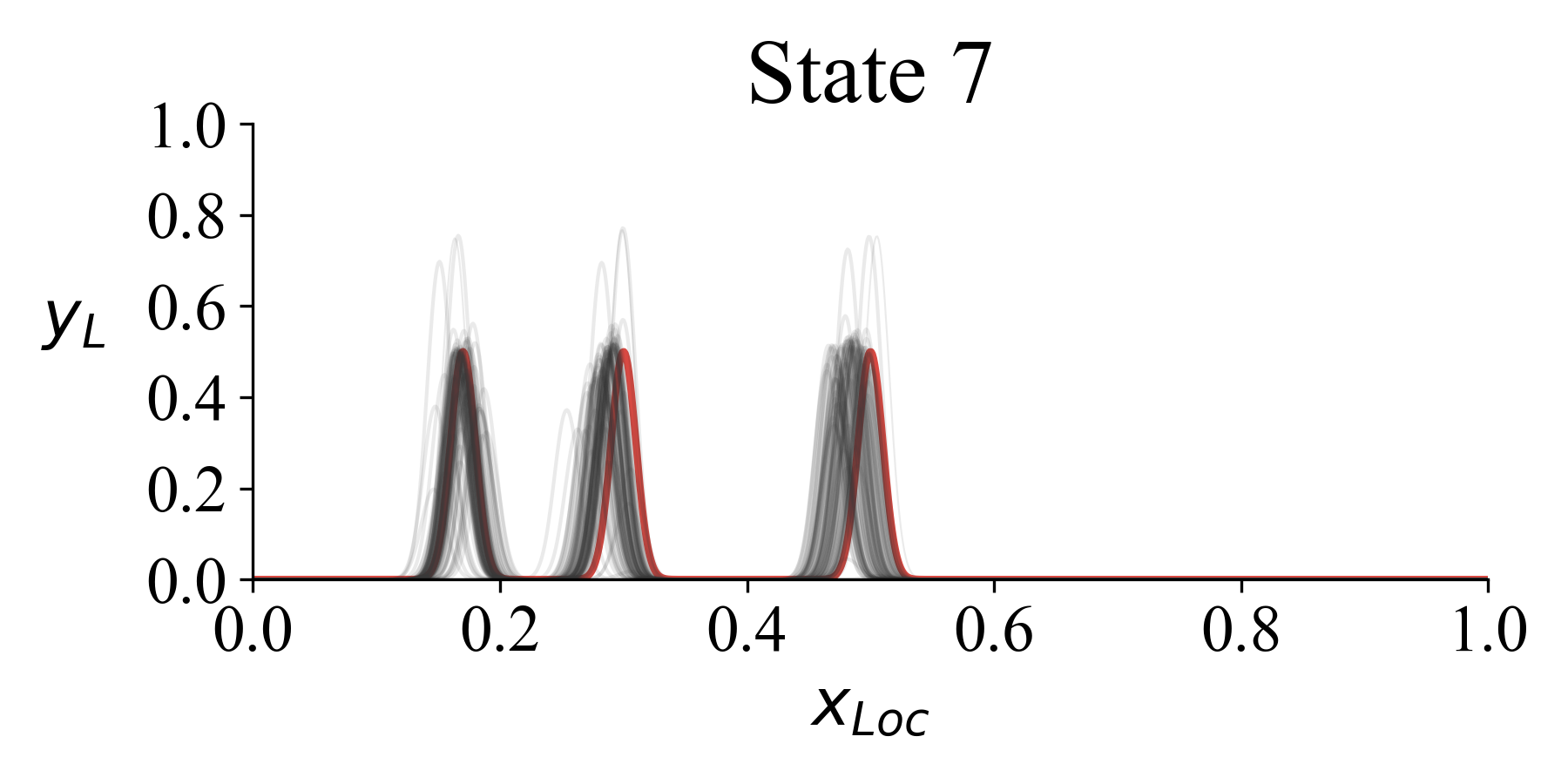}%
        
        \includegraphics[width=0.33\linewidth]{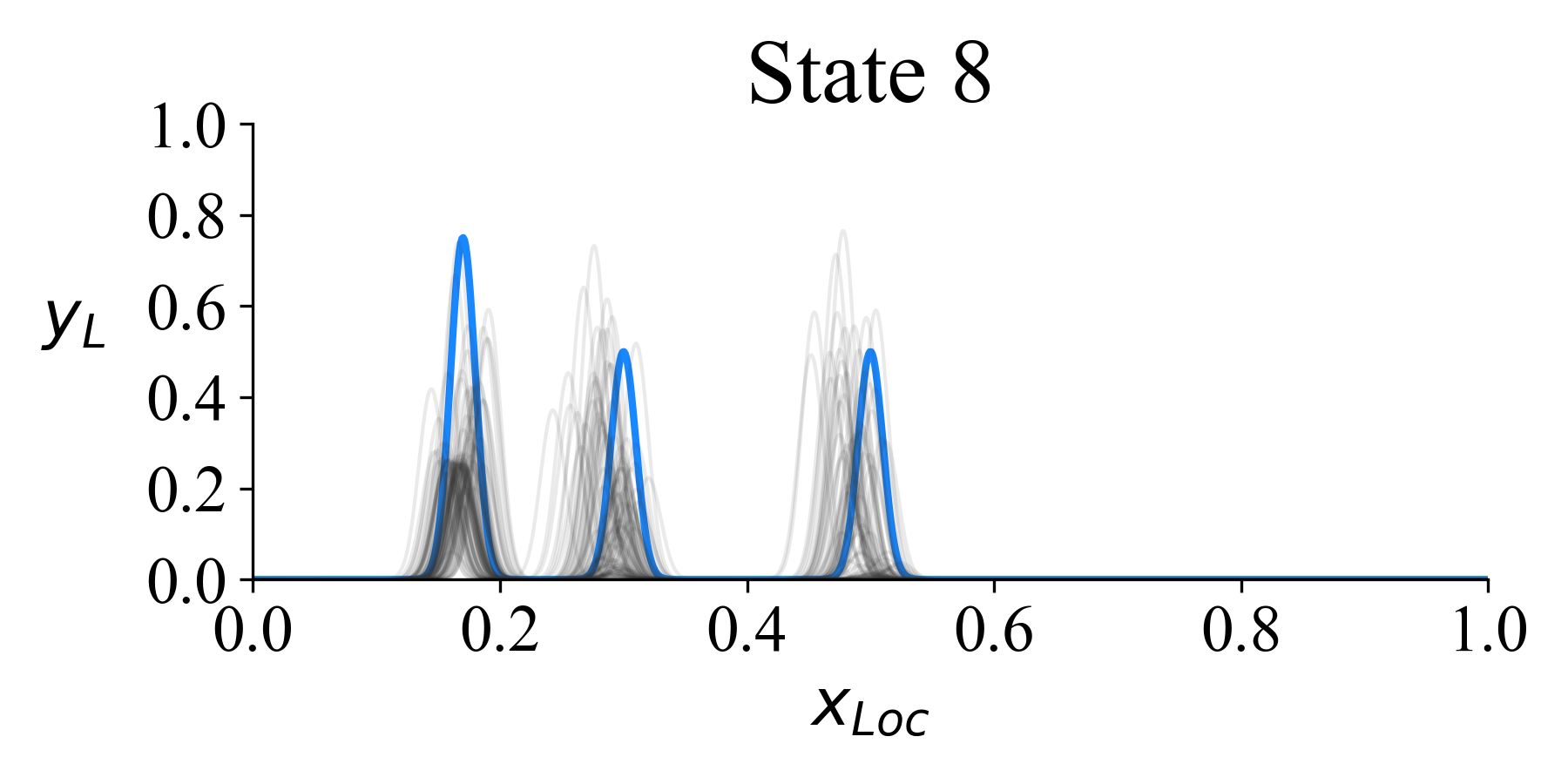}%
        \includegraphics[width=0.33\linewidth]{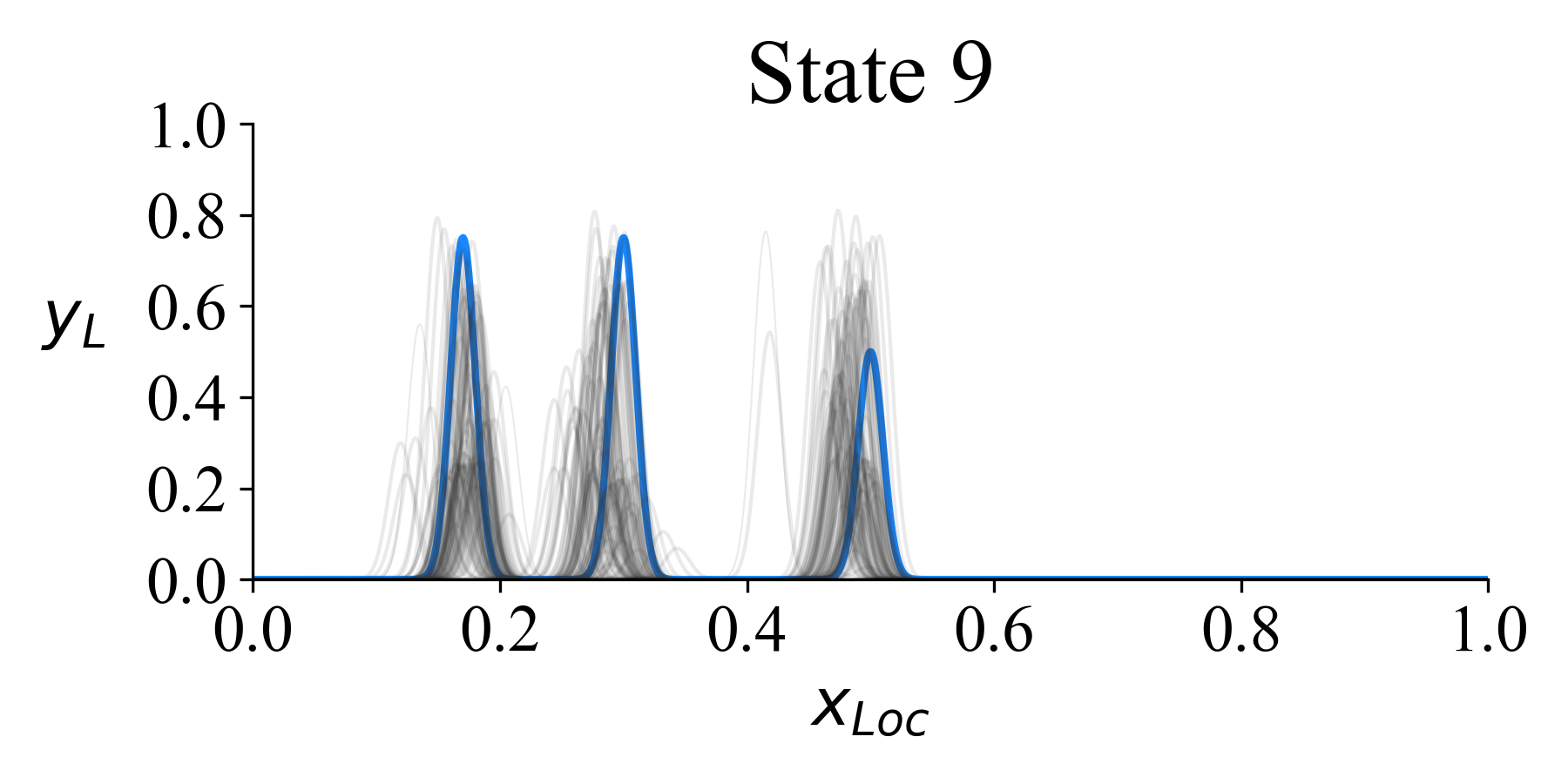}%
        \includegraphics[width=0.33\linewidth]{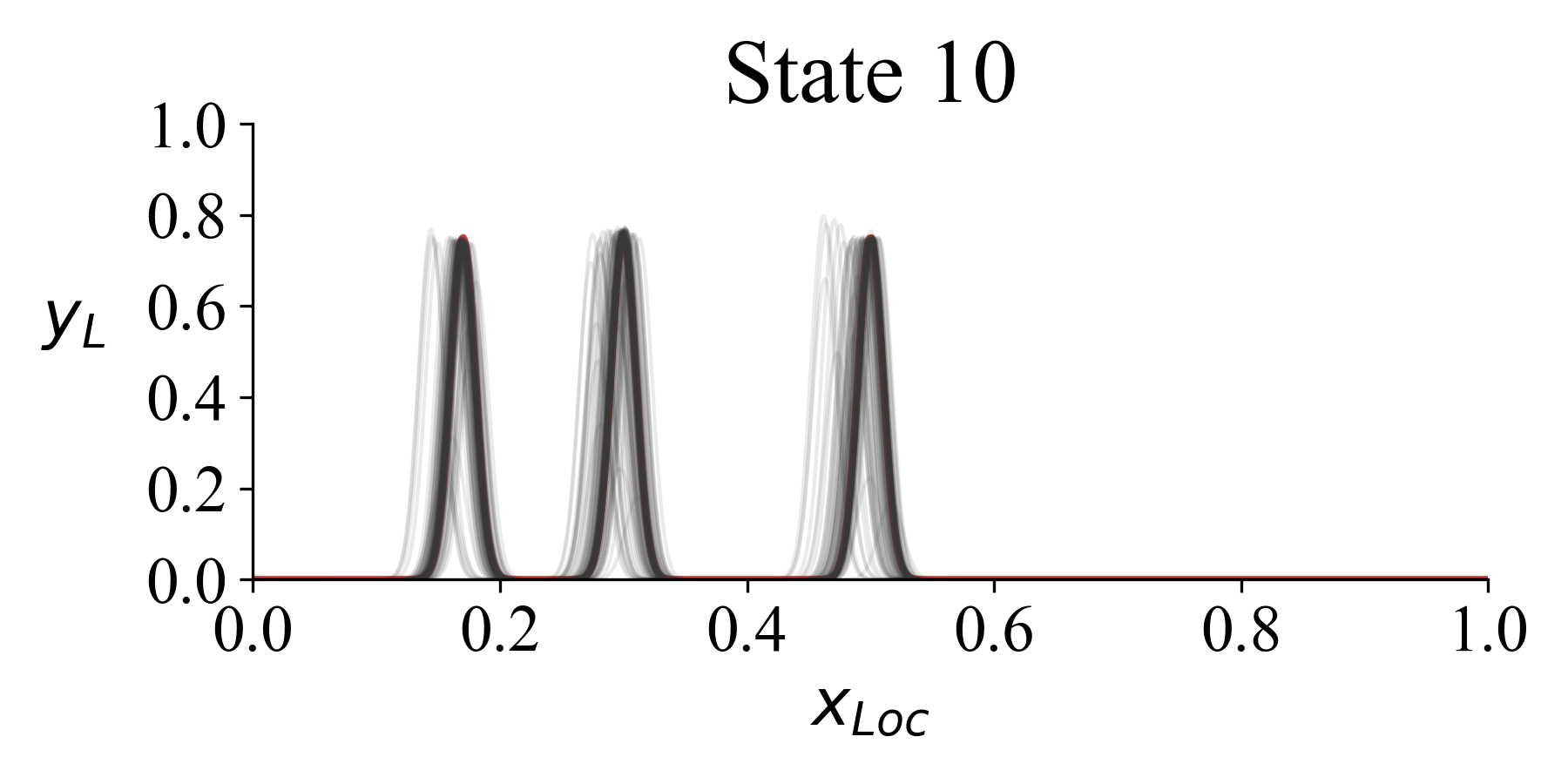}
    \caption{Parameter estimation results of structural states 1-10 in Case 2, after neural refinement.}
    \label{fig:case2_inverse_nr}
\end{figure*}

Table~\ref{tab:case2_inverse} presents the quantitative results of inverse parameter estimation for different models in Case 2.
For the system parameters $\mu_{\text{Loc}}$, which describe the relative location of damages, Parametric DeepONet (ND) achieves the best NEMSE in gradient-based initialization, with the NRMSE of $1.49 \times 10^{-1}$ on the training data and $1.64 \times 10^{-1}$ on the test data.
After employing the neural refinement, the performance further improves, with the NRMSE of $0.32 \times 10^{-1}$ for training and $0.42 \times 10^{-1}$ for test data.
For the system parameters $\mu_{\text{L}}$, which describe the relative length of the damages, Parametric DeepONet (LD) achieves the best performance in gradient-based initialization, with the NRMSE of $2.59 \times 10^{-1}$ training data and $5.16 \times 10^{-1}$ on test data.
With the neural refinement, the performance improves as well, achieving the NRMSE of $0.51 \times 10^{-1}$ for training and $4.19 \times 10^{-1}$ for test data.
Since the system parameters are only with several possible values in training data, all models achieve satisfactory performances after the supervised neural refinement.

Figure~\ref{fig:case2_inverse_gi} presents the qualitative results of Parametric DeepONet (ND) in parameter estimation.
For gradient-based initialization, the relative location $x_{\text{Loc}}$ and length $y_{\text{L}}$ of the damage deliver inconsistent results for the 100 data samples of each structural state. This is due to the random initialization of the two parameters, and the estimation is sensitive to the initializations. Figure~\ref{fig:case2_inverse_nr} presents the qualitative results of Parametric DeepONet (ND) after neural refinement, which significantly enhances the estimation, with more consistent estimation for all the data samples. One can observe that the estimation is much less dispersive than the former. 
It is noted that States 5,6,8 and 9 are not included in the training data, and serve as extrapolation tests.  Similar estimation results can be concluded, albeit the estimation should be perceived in a statistical way to make more sense. 

Parametric DeepONet has effectively encoded the system parameter into the well-trained forward models, enabling better initialization using the gradient-based scheme.
In contrast, although CNN and MLP have achieved good performance in forward modeling, their gradient-based initialization results indicate the system parameters are not effectively encoded by the neural network.
In summary, Parametric DeepONet clearly outperforms other baseline models in the gradient-based initialization of inverse parameter estimation.
With the supervised training, all models are effective in parameter estimation after the neural refinement.
Even in challenging extrapolation test scenarios, Parametric DeepONet proves effective in estimating system parameters. Notably, these system parameters are non-physical and customized, being user-defined for the task of structural damage identification.
The results indicate the capability of the proposed framework for inverse modeling of the real structural system.

\section{Conclusion}
In this work, we present a deep learning-based framework for parameter estimation, which begins by learning a neural operator enabled surrogate model for structural response prediction. The proposed framework offers a unified approach for both forward and inverse problems associated with structural dynamics via neural networks.
Within this framework, the Parametric DeepONet is proposed to combine parametric input and operator learning.
In addition, Parametric DeepONet is able to learn resolution-invariant forward structural dynamics, achieving reasonable super-resolution performance.
It achieves competitive results in forward modeling and demonstrates consistent effectiveness in inverse problems, boosted by the introduction of neural refinement. 
The framework is validated through both a simple numerical oscillator and a real experimental wind turbine structure.

In our current attempt, we consider the low-dimensional parameter space (up to 6 dimensions in the validation cases), and future work will explore more general settings with high-dimensional parameter spaces.
In such cases, dimension reduction techniques are most likely to be added to the current architecture.
Additionally, excitation and response measurements of real-world structural systems are generally subject to uncertainties, such as measurement errors and environmental disturbances.
Since the proposed framework is a data-driven approach, in such cases, high-quality data and appropriate parameterization of the system are essential for the training of an effective forward model.

Moreover, this framework is architecture-agnostic, with its sub-modules can be easily adaptive to other advanced neural networks like Transformers and graph neural networks, etc.
We hope the framework's performance and flexibility will encourage a new paradigm of addressing inverse problems in structural dynamics by first learning differentiable forward surrogate models via deep learning.

\section*{Acknowledgement}
The authors wish to express their gratitude for the financial support
received from the Guangzhou-HKUST(GZ) Joint Funding Grant (No.2023A03J0105), and the Guangdong Provincial Key Lab of Integrated Communication, Sensing and Computation for Ubiquitous Internet of Things (No.2023B1212010007). 

\appendix

\section{Further analysis on Parametric DeepONet}
\subsection{Positional encoding}
\label{sec:pe}
\begin{figure*}[!h]
    \centering
    \begin{subfigure}{.33\linewidth}
        \includegraphics[width=\linewidth]{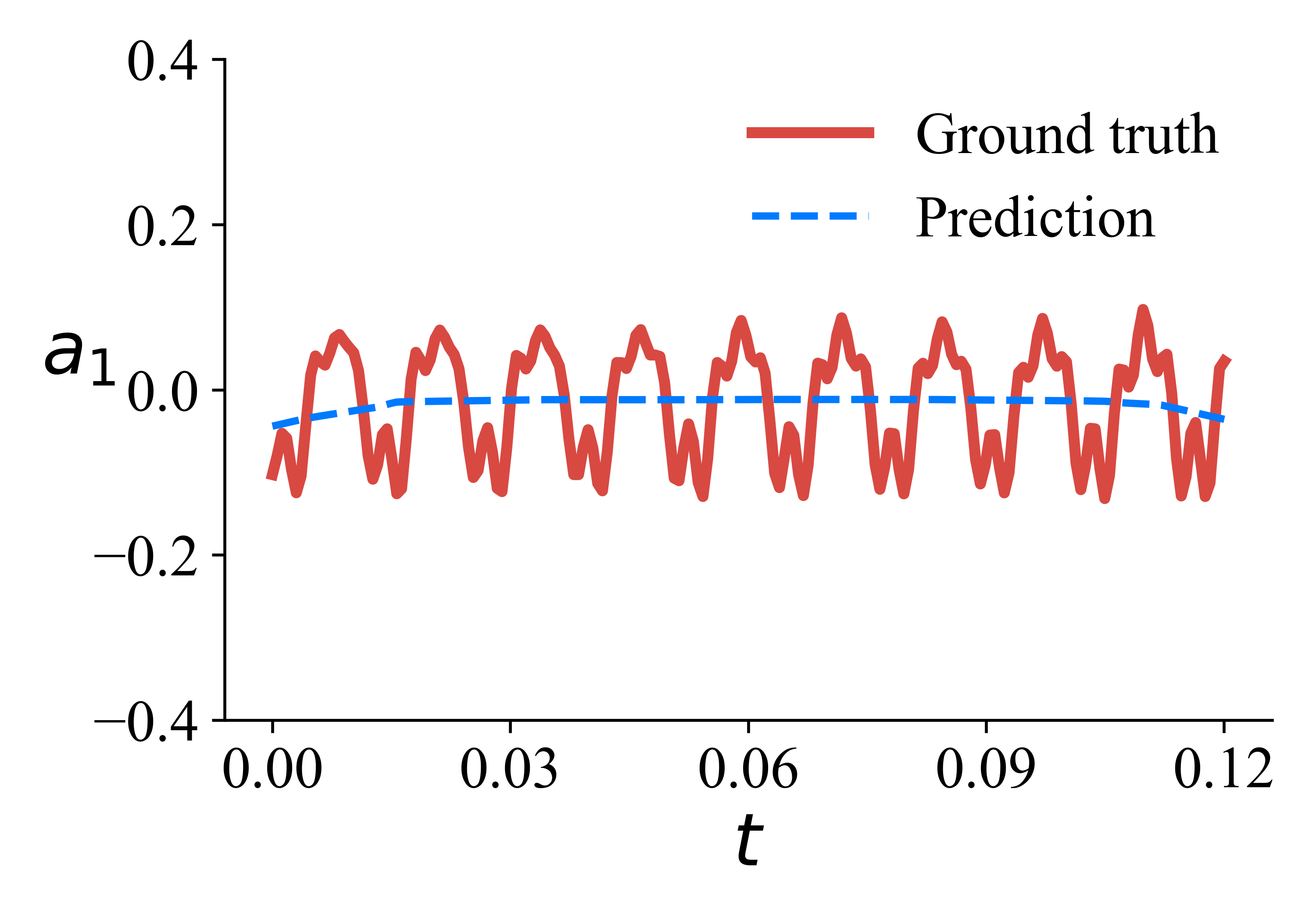}
        \caption{}
        \label{fig:pe_1}
    \end{subfigure}%
    \begin{subfigure}{.33\linewidth}
        \includegraphics[width=\linewidth]{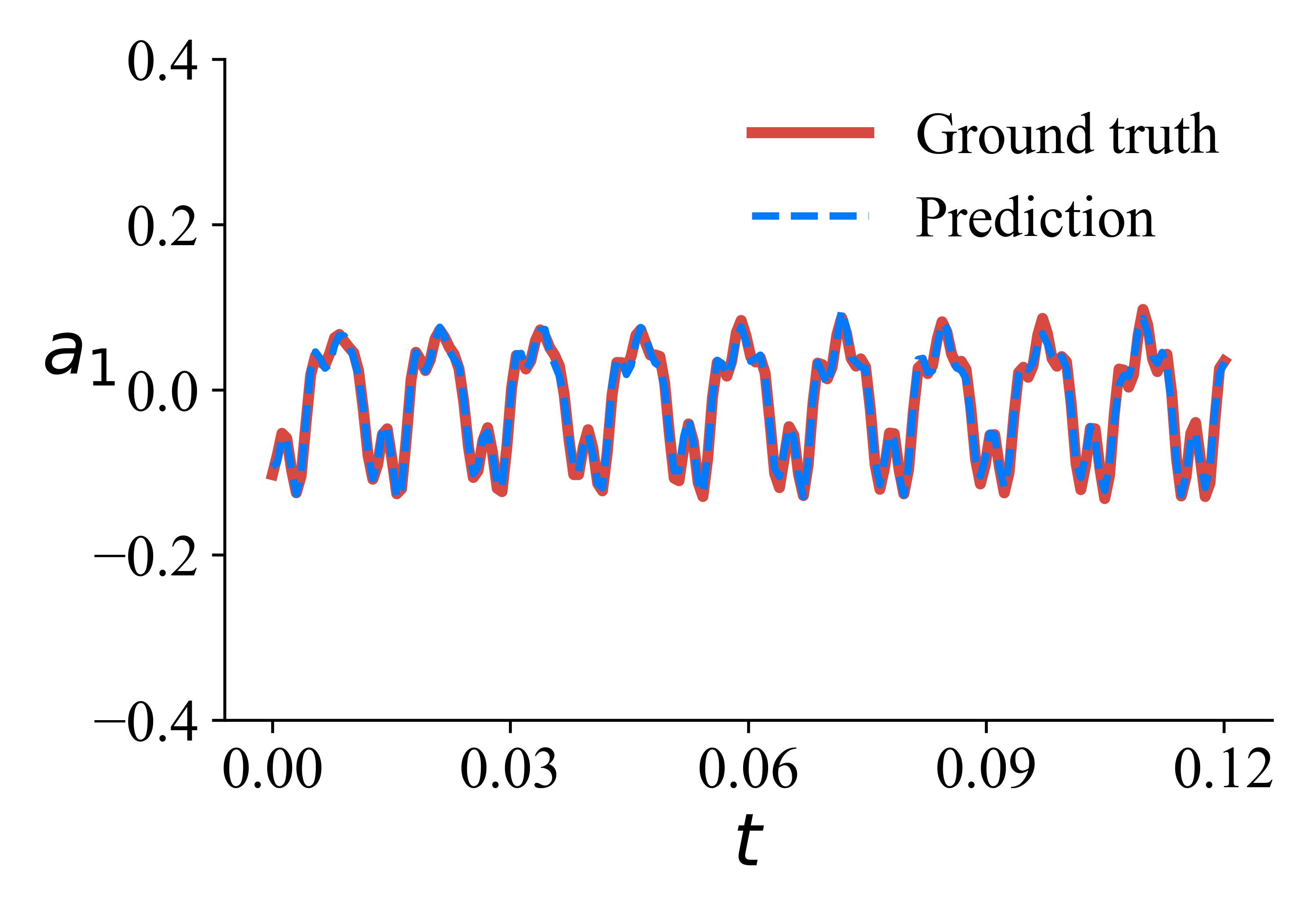}
        \caption{}
        \label{fig:pe_2}
    \end{subfigure}
    \begin{subfigure}{.33\linewidth}
        \includegraphics[width=\linewidth]{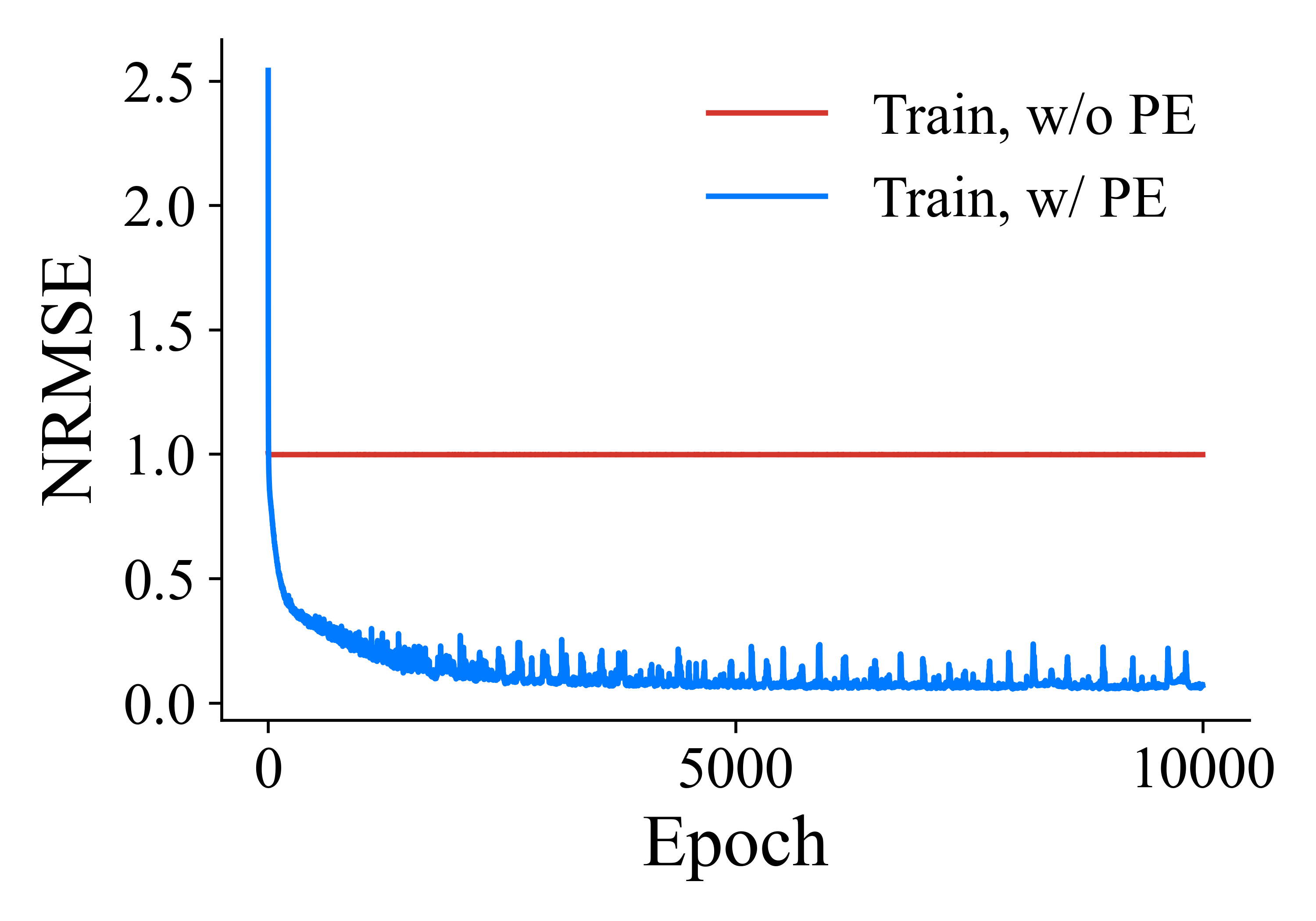}
        \caption{}
    \end{subfigure}
    \caption{DeepONet trained on SDOF response in Case 2: (a) without positional encoding (w/o PE); (b)with positional encoding (w/ PE); (c) NRMSE during the training stage.} %
    \label{fig:pe}
\end{figure*}

Following the brief introduction in Section~\ref{sec:para_deeponet}, coordinated-based MLPs that take low-dimensional location coordinates (typically points in $\mathbb{R}^1$, $\mathbb{R}^2$ or $\mathbb{R}^3$) as input and generate output at each input location, have been shown to struggle with approximating high frequency data in computer vision~\cite{tancik2020fourier}, physics-informed machine learning~\cite{wang2021eigenvector}, etc. This limitation is known as \textit{spectral bias}~\cite{rahaman2019spectral}.
The trunk net in DeepONet can be interpreted as coordinate-MLP since they take output coordinates as input (in our case, time $t$) and generate output corresponding to those coordinates.
As shown in Figure~\ref{fig:pe_1}, vanilla DeepONet fails to approximate the SDOF response (accelerometer $a_1$) in Case 2, where the sampling frequency of the response reaches $1666$ Hz.

To address this limitation, we incorporate positional encoding (PE) into the trunk net of Parametric DeepONet. PE has shown effective for learning high frequency data for MLPs~\cite{wang2021eigenvector} and neural operators~\cite{bunker2024autoencoders,seidman2023variational}.
In our implementation, for the PE in Eq.\eqref{eq:pdon_trunk}, we adopt integer-periodic Fourier feature mapping~\cite{seidman2023variational}, defined as:
\begin{equation}
    \gamma(t)=[1, \cos (\omega t), \sin (\omega t), \ldots, \cos (k \omega t), \sin (k \omega t)],
\end{equation}
with $\omega = \frac{2\pi}{L}$, $k$ is a non-negative integer controlling the number of basis spectrum features, and $L$ is the length of the coordinate domain. 
This PE maps each scalar time coordinate $t$ into a $2k$-dimensional vector of spectrum features, $\gamma(t) \in \mathbb{R}^{2k}$.
The input dimension of the trunk net with PE becomes $2k$.
PE can be viewed as introducing frequency spectrum prior to coordinates input, enabling neural networks to better tune the frequencies of output~\cite{tancik2020fourier}.
As shown in Figure~\ref{fig:pe_2}, the inclusion of PE significantly helps vanilla DeepONet in learning high frequency response.

\subsection{Decoder}
\label{sec:decoder}
 \begin{figure*}[!h]
    \centering
    \begin{subfigure}{0.5\linewidth}
        \centering
        \includegraphics[height=4.5cm]{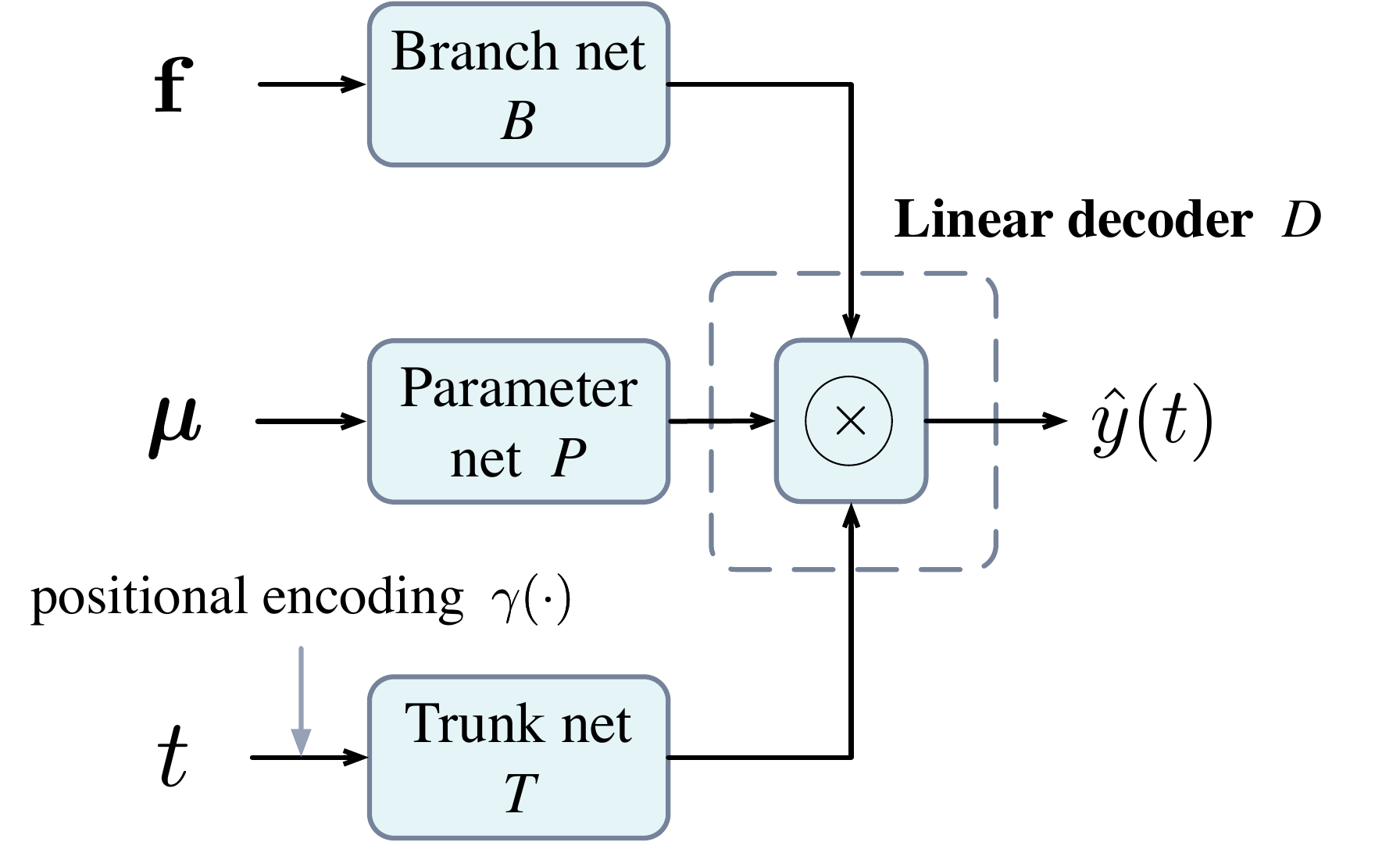}
        \caption{Parametric DeepONet (LD)}
        \label{fig:linear-decoder}
    \end{subfigure}%
    \begin{subfigure}{0.5\linewidth}
        \centering
        \includegraphics[height=4.5cm]{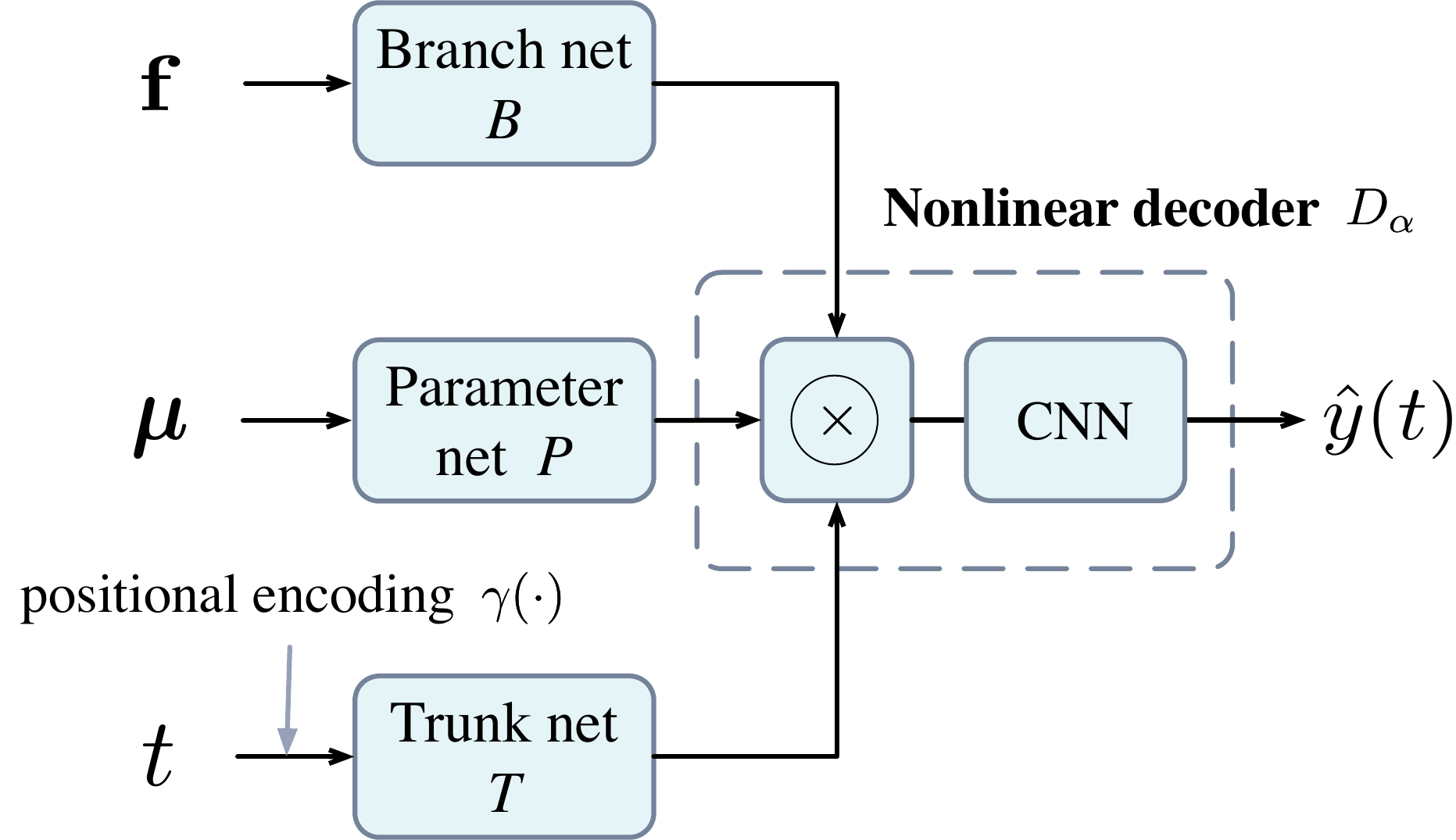}
        \caption{Parametric DeepONet (ND)}
        \label{fig:nonlinear-decoder}
    \end{subfigure}
    \caption{Illustrations of different decoding process.}
    \label{fig:decoder}
\end{figure*}
Vanilla DeepONet generates output through a dot product of the outputs of the branch net and trunk net (as computed in Eq.\eqref{eq:forward_deeponet}), which can be interpreted as a linear decoding process.
In Parametric DeepONet, we consider this dot product operation as a linear decoder (LD). 
Optionally, we introduce the nonlinear decoder (ND), a neural network that further processes the LD's output. In summary, the decoding process in Eq.~\eqref{eq:parametric_deeponet} can be categorized broadly as linear and nonlinear:
\begin{itemize}
    \item Linear decoder (LD): As shown in Figure~\ref{fig:linear-decoder}, the linear decoder $D$ contains no trainable parameters and operates similarly to the vanilla DeepONet. It computes the dot product between the outputs of the branch net, parameter net, and trunk net. 
    For SDOF systems, the output response at a time coordinate $t$ is:
        \begin{align}
            \hat{y}(t) &= D\left(B(\mathbf{f}), P(\boldsymbol{\mu}), T(\gamma(t))\right) \\
            &=\sum_{k=1}^{n} b_{k}(\mathbf{f}) p_{k}(\boldsymbol{\mu})\tau_{k}(t),
            \label{eq:pdon_ld}
        \end{align}
    where $\mathbf{f}$ is the discretized excitation force, $\boldsymbol{\mu}$ is the system parameters.
    The SDOF response results in $\hat{\mathbf{y}}_{\text{LD}} = [\hat{y}(t_1), \ldots, \hat{y}(t_r)]$, $r$ is the resolution of the output response.
    For MDOF systems, the time coordinates of the trunk net are extended $c$-fold to account for the $c$ output channels, with linearly increasing coordinate values.
    The MDOF response results in $\hat{\mathbf{y}}_{\text{LD}} = [\hat{y}(t_1), \ldots, \hat{y}(t_{c\cdot r})]$.
    The output response of length $c\cdot r$ is reshaped to shape $[c,r]$, where $c$ corresponds to the number of DOFs.
    \item Nonlinear decoder (ND): As illustrated in Figure~\ref{fig:nonlinear-decoder}, nonlinear decoder $D_{\alpha}$ is a neural network (parameterized by $\alpha$) that processes the output from LD.
    Specifically, ND takes the fully decoded response vector as input and generates the refined prediction:
    \begin{align}
        \hat{\mathbf{y}} = D_{\alpha}(\hat{\mathbf{y}}_{\text{LD}})
    \end{align}
   in which, for SDOF systems, $\hat{\mathbf{y}}_{\text{LD}} = [\hat{y}(t_1), \ldots, \hat{y}(t_r)]$, and $D_{\alpha}$ is implemented as a MLP; for MDOF systems, $\hat{\mathbf{y}}_{\text{LD}} = [\hat{y}(t_1), \ldots, \hat{y}(t_{c\cdot r})]$ is reshaped to $[c, r]$,  and $D_{\alpha}$ is implemented as a convolutional neural network (CNN), aiming to improve the learning of local patterns in dynamic response.
\end{itemize}
\begin{figure}[!h]
    \centering
    \begin{subfigure}{\textwidth}
        \includegraphics[width=\linewidth]{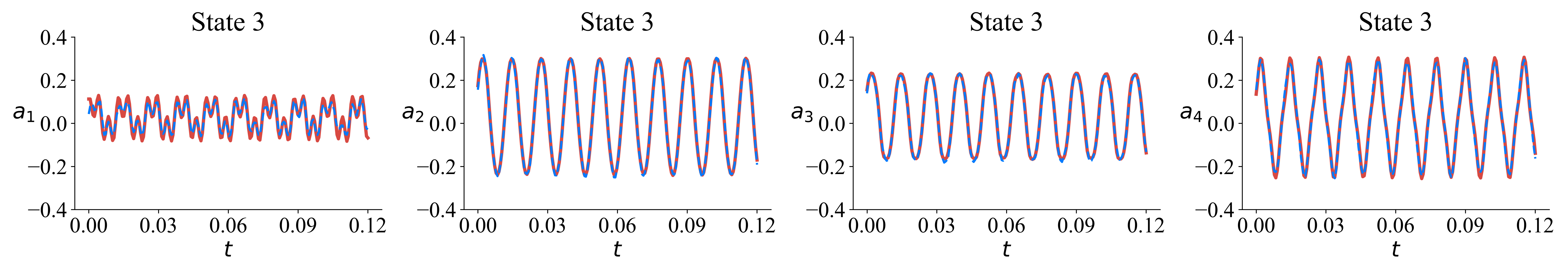}
        \includegraphics[width=\linewidth]{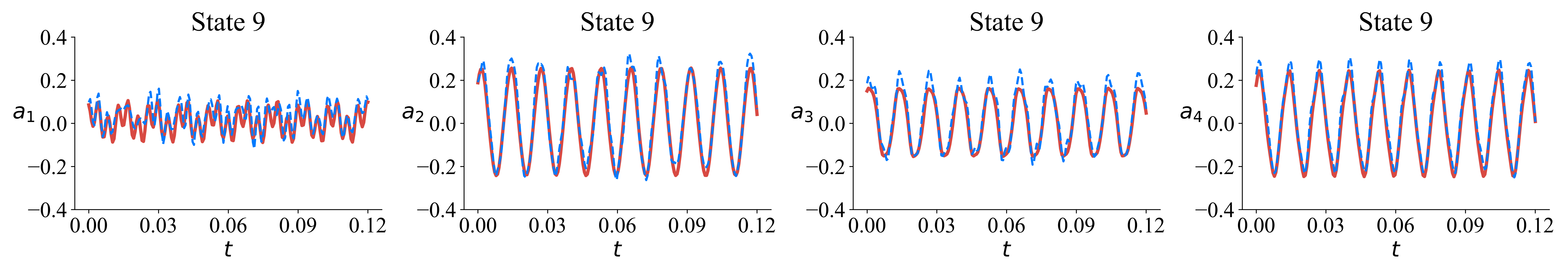}
        \caption{Parametric DeepONet (LD)}
    \end{subfigure}

    \begin{subfigure}{\textwidth}
        \includegraphics[width=\linewidth]{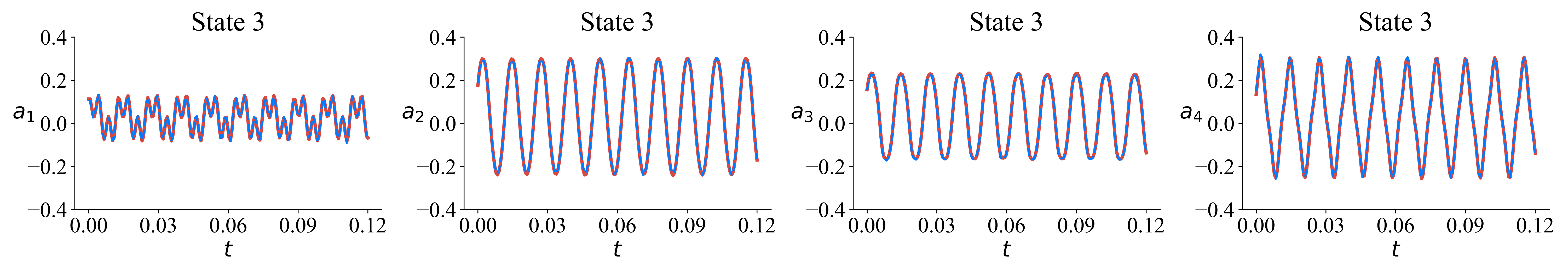}
        \includegraphics[width=\linewidth]{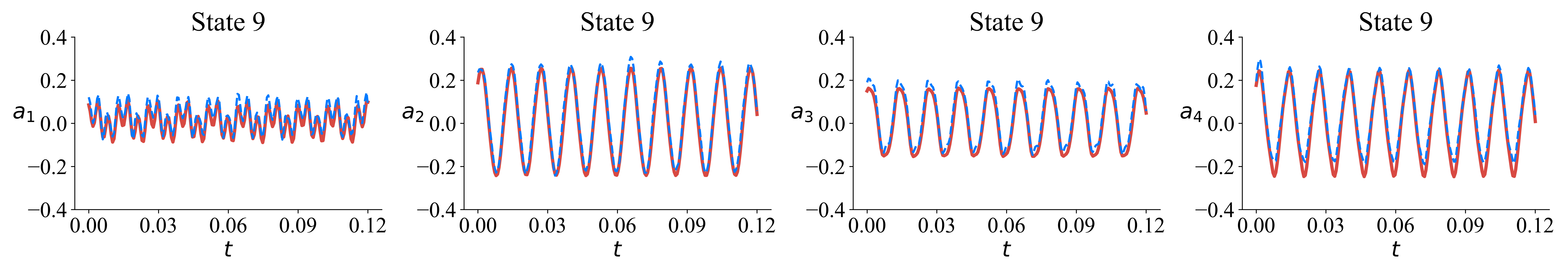}
        \caption{Parametric DeepONet (ND)}
    \end{subfigure}
    \caption{Response prediction results of Parametric DeepONet in Case 2 ($a_1 - a_4$ denote normalized acceleration).}
    \label{fig:compare-decoder}
\end{figure}
Parametric DeepONet (ND) achieves better quantitative results in response prediction, as presented in Section~\ref{sec:case1-forward} and Section~\ref{sec:case2_forward}. Figure~\ref{fig:compare-decoder} shows the qualitative results of response prediction in Case 2, where Parametric DeepONet (ND) better captures local patterns of the dynamics than Parametric DeepONet (LD).

\subsection{Zero-shot super-resolution}
\label{sec:super-resolution}
\begin{figure*}[!h]
    \centering
    \includegraphics[width=0.6\linewidth]{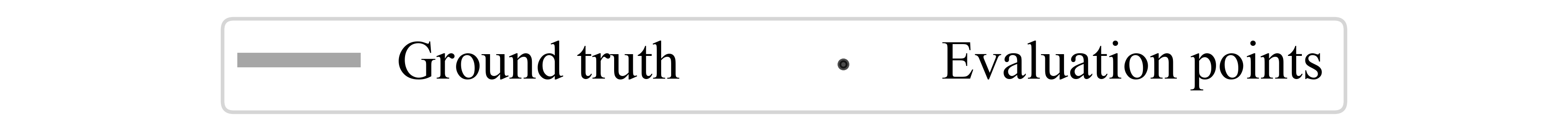}
    \begin{subfigure}{0.5\textwidth}
        \includegraphics[width=\linewidth]{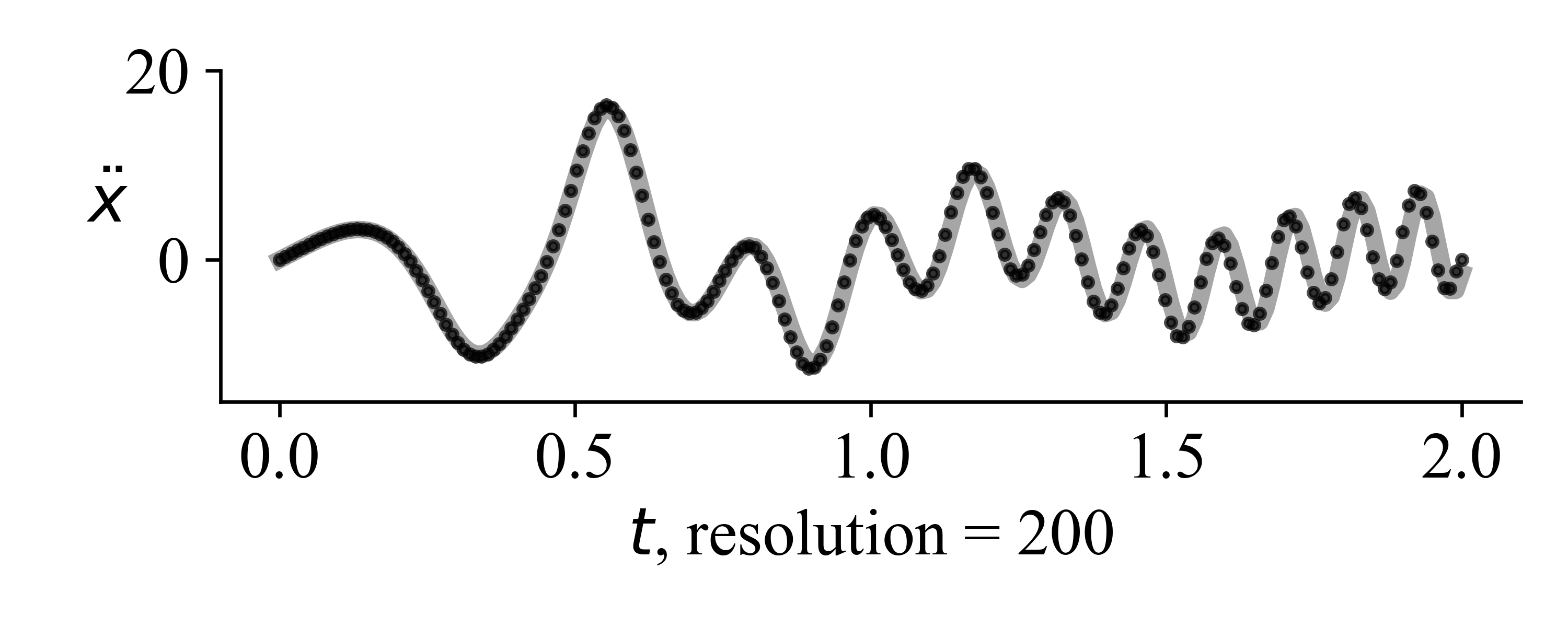}
        \includegraphics[width=\linewidth]{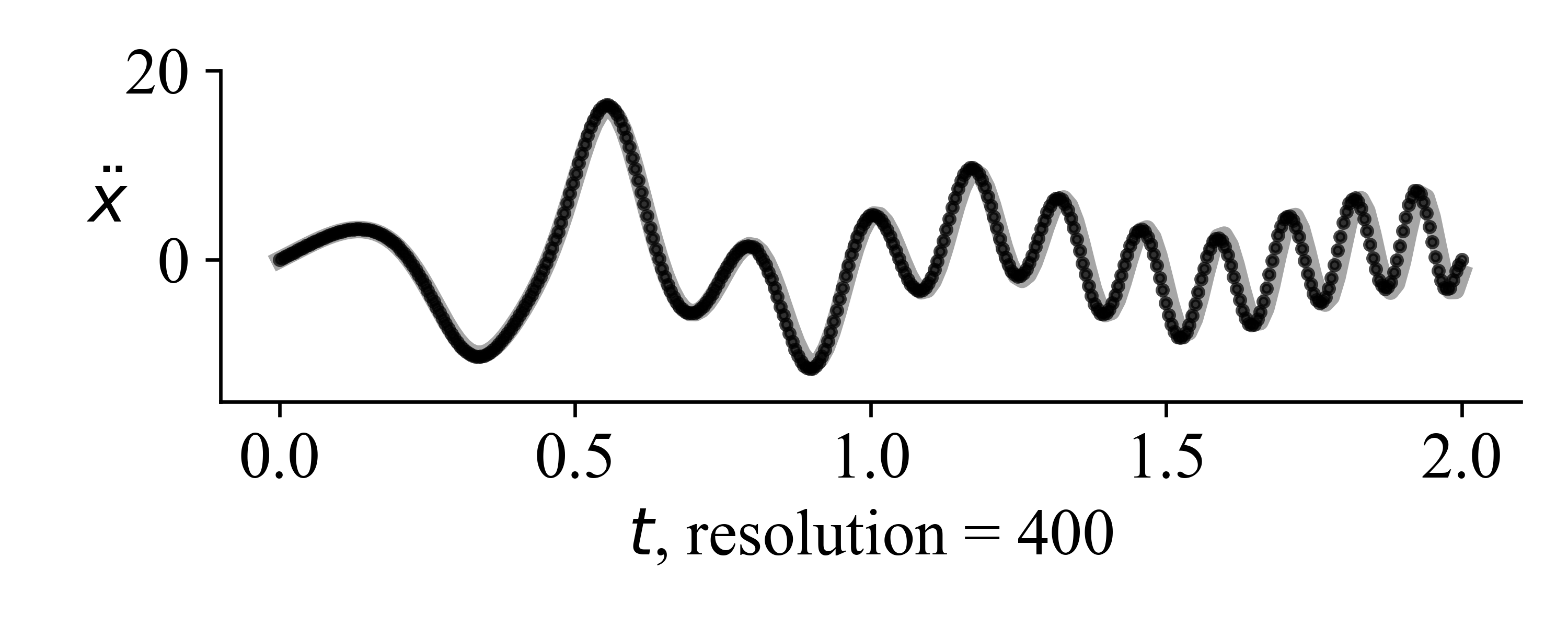}
        \includegraphics[width=\linewidth]{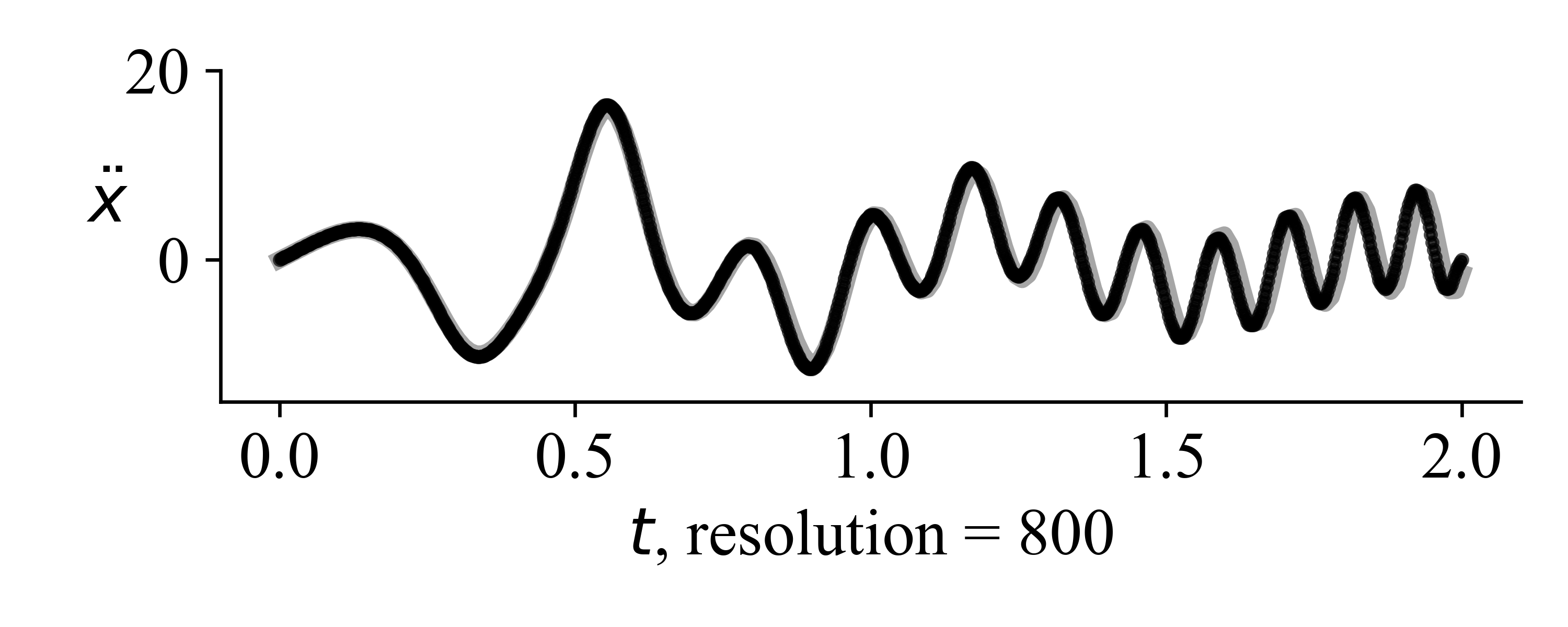}
        \caption{Case 1}
    \end{subfigure}%
    \begin{subfigure}{0.5\textwidth}
        \includegraphics[width=\linewidth]{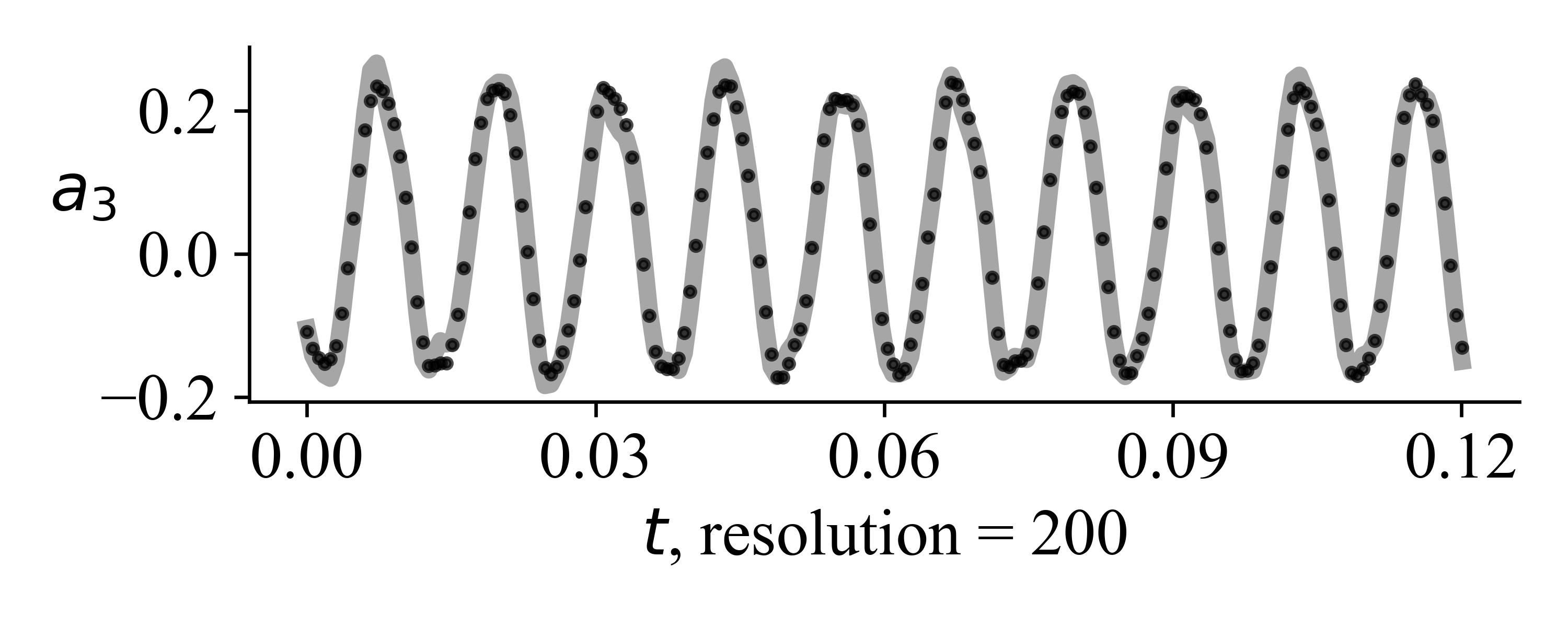}
        \includegraphics[width=\linewidth]{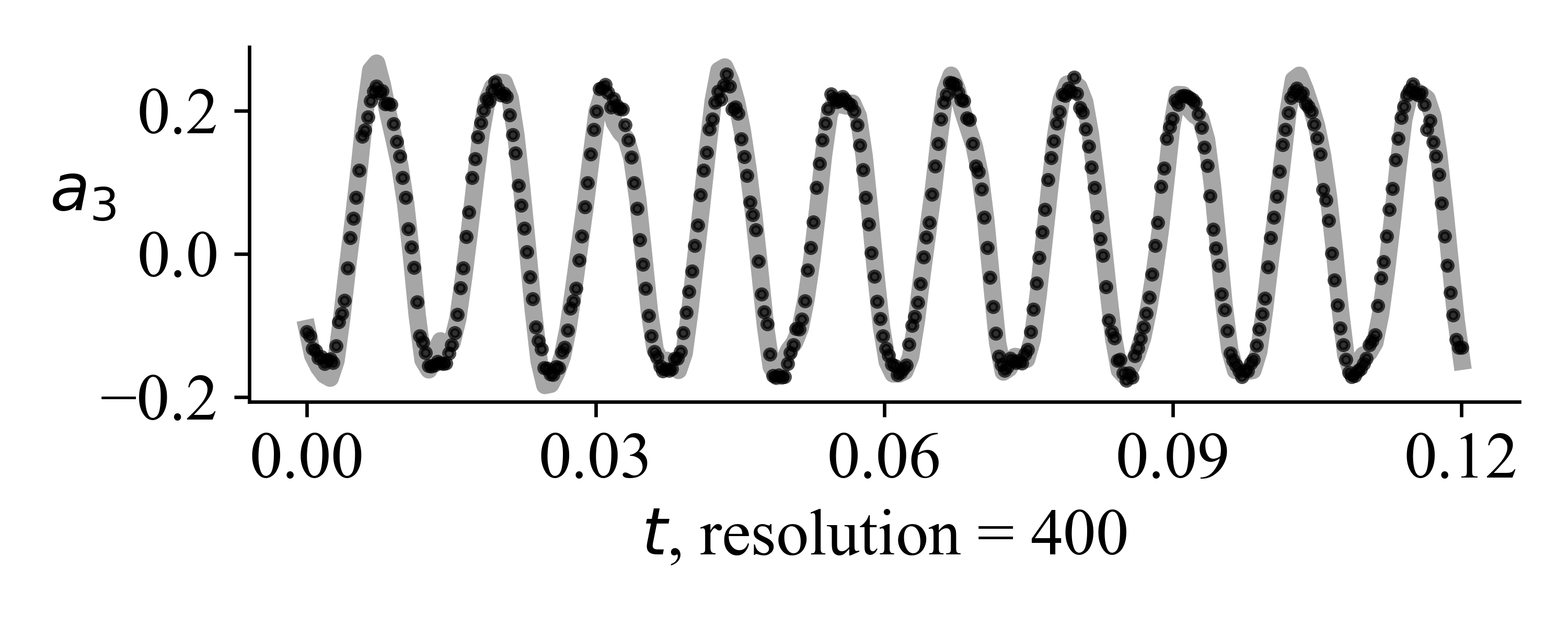}
        \includegraphics[width=\linewidth]{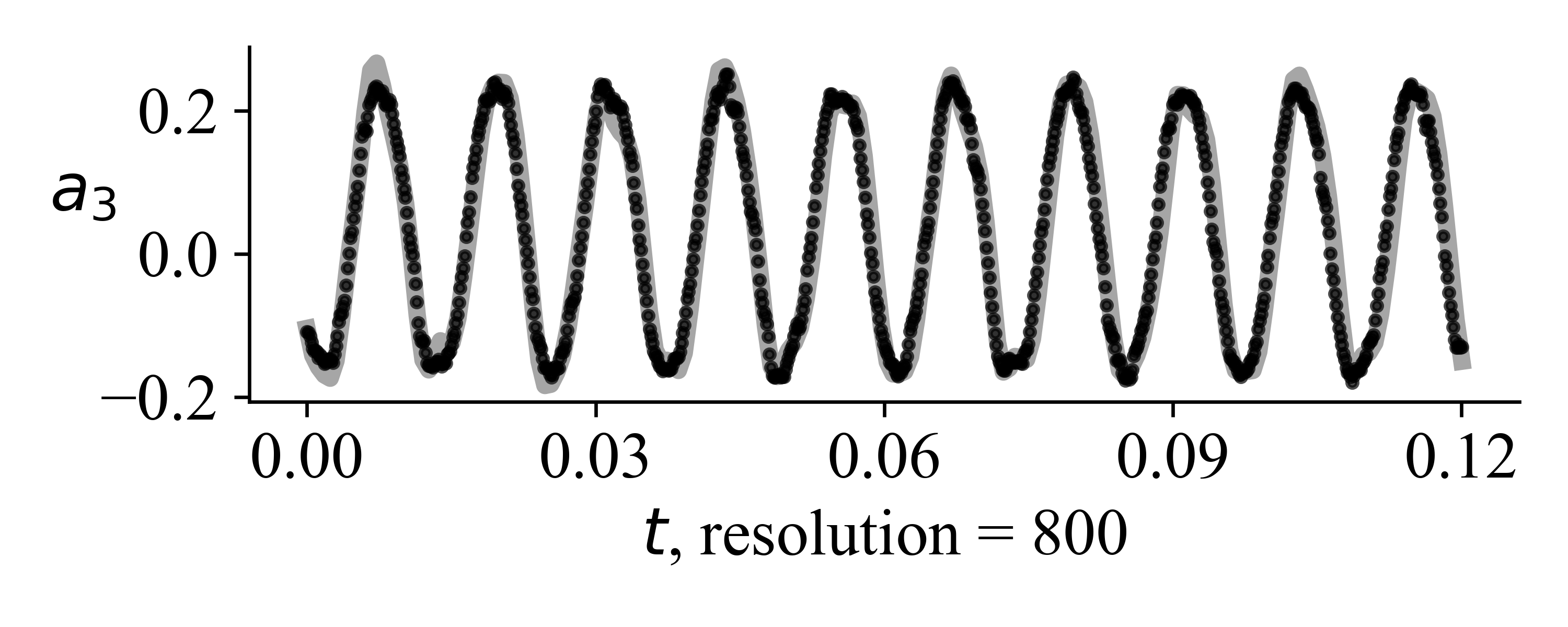}
        \caption{Case 2}
    \end{subfigure}
    \caption{Zero-shot super-resolution. Parametric DeepONet trained on 200 resolutions, and evaluated on 200, 400, and 800 resolutions.}
    \label{fig:super}
\end{figure*}

Parametric DeepONet is capable of learning a resolution-invariant mapping, enabling it to be trained on lower resolution data and evaluated at higher resolutions without seeing any higher resolution training data (zero-shot super-resolution)~\cite{li2020fourier}.
As presented in Eq.~\eqref{eq:pdon_ld}, given an arbitrary query point of $t$, Parametric DeepONet (LD) can generate the response at that query point.
Figure~\ref{fig:super} shows an example where the Parametric DeepONet is trained on responses with a resolution of 200 and evaluated at resolutions of 400 and 800, demonstrating its super-resolution capability in the time space.
The nonlinear decoder in Parametric DeepONet (ND) currently requires a fixed resolution between training and testing since the input shape to the decoder needs to be consistent.
Nevertheless, nonlinear decoders that support super-resolution are also feasible through other functional training schemes~\cite {bunker2024autoencoders,seidman2023variational}.
In contrast, MLP and CNN trained in conventional settings are only able to generate fixed solution responses, which are less flexible than Parametric DeepONet.

\section{Analysis on network's architecture}
\label{sec:optimal_architecture}
To ensure a fair comparison across baseline models, we systematically explore architectural configurations using a grid search over network depth and width.
This approach is inspired by the hyperparameter tuning strategy reported in vanilla DeepONet~\cite{lu2021learning}.
For consistency and fairness, all baseline models are designed to have similar depth and width configurations, which in turn results in comparable numbers of trainable parameters.

The learning rate is fixed at 0.001 based on prior evidence of its effectiveness.
For the depth, we consider values from [3, 4, 5]; for the width, we search over [50, 100, 200, 300, 400, 500].
Each configuration is trained and tested three times, and the mean and standard deviation of the NRMSE are reported.

\begin{figure}[!h]
    \centering
    \begin{subfigure}{0.33\textwidth}
        \includegraphics[width=\linewidth]{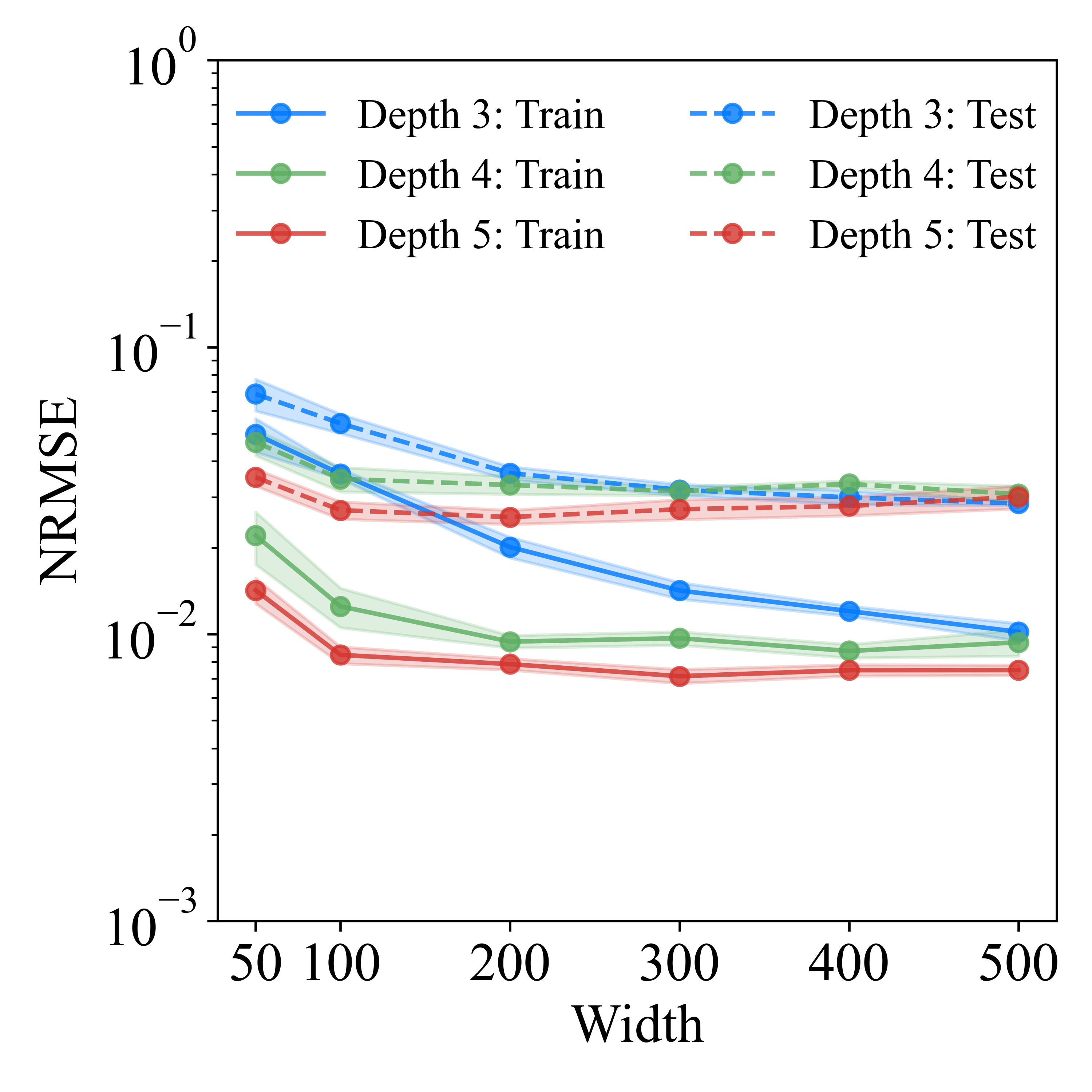}
        \caption{Parametric DeepONet (LD)}
        \label{fig:sup_case1_p_ld_deeponet}
    \end{subfigure}%
    \begin{subfigure}{0.33\textwidth}
        \includegraphics[width=\linewidth]{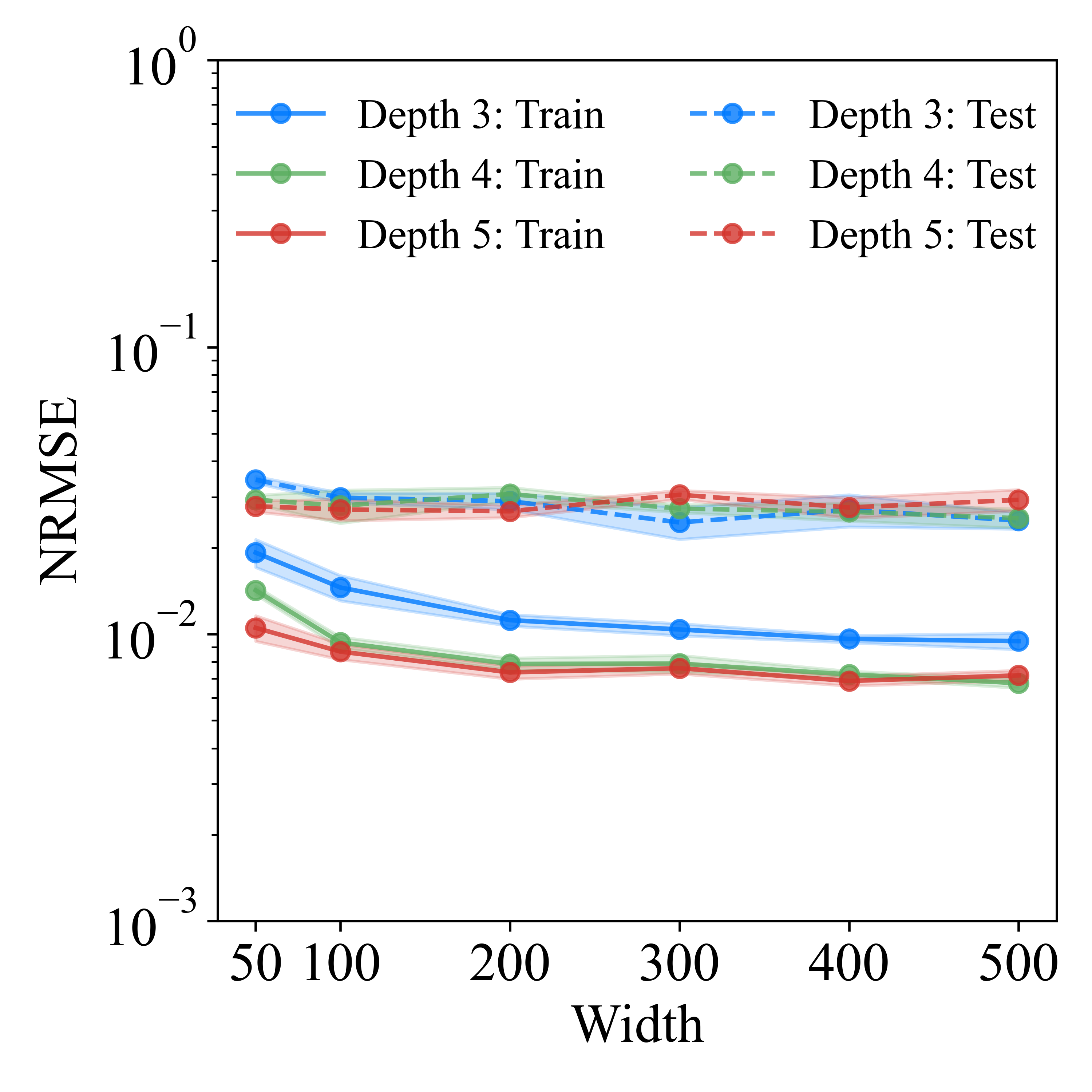}
        \caption{Parametric DeepONet (ND)}
        \label{fig:sup_case1_p_nd_deeponet}
    \end{subfigure}%
    \begin{subfigure}{0.33\textwidth}
        \centering
        \includegraphics[width=\linewidth]{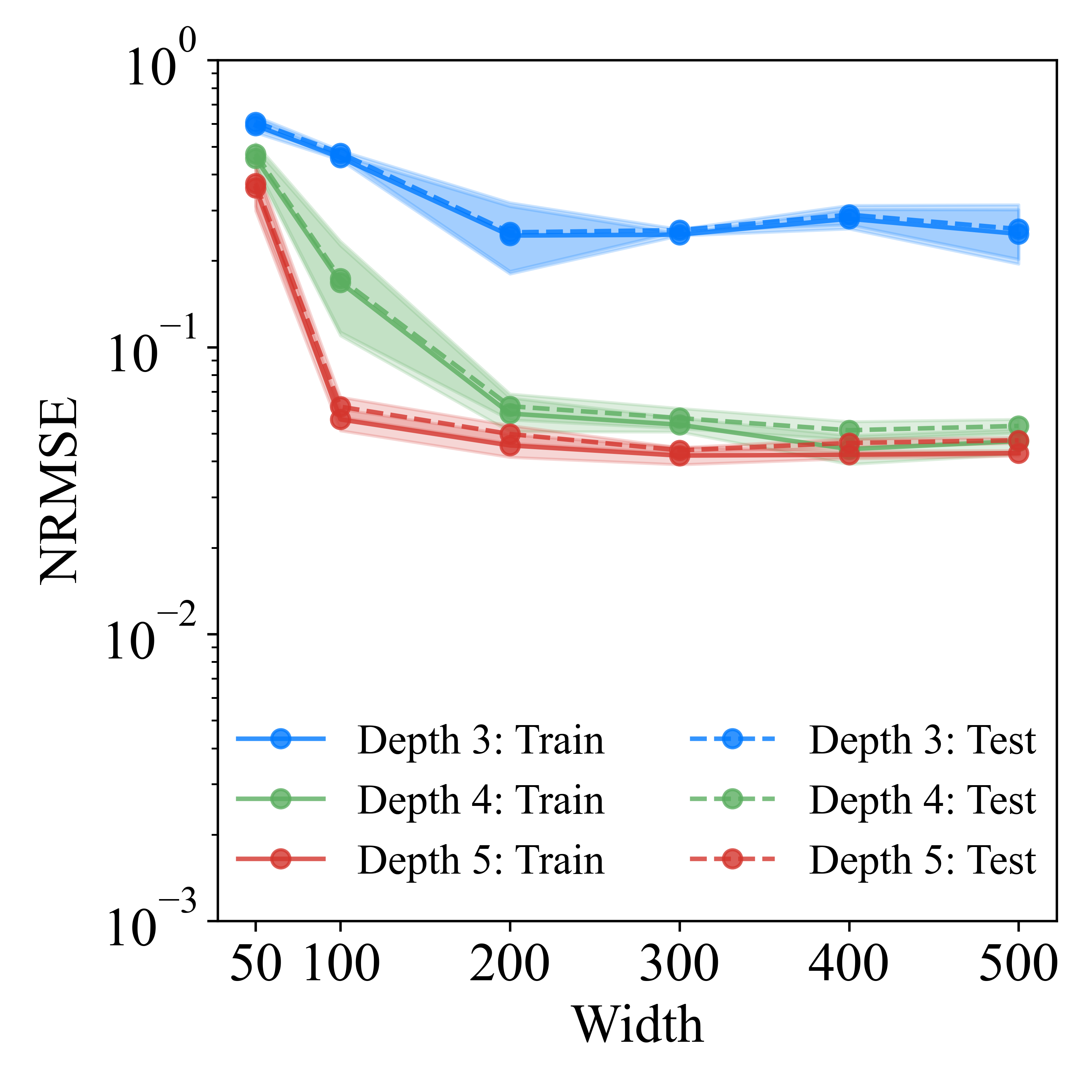}
         \caption{Vanilla DeepONet}
         \label{fig:sup_case1_deeponet}
    \end{subfigure}%

    \begin{subfigure}{0.33\textwidth}
        \includegraphics[width=\linewidth]{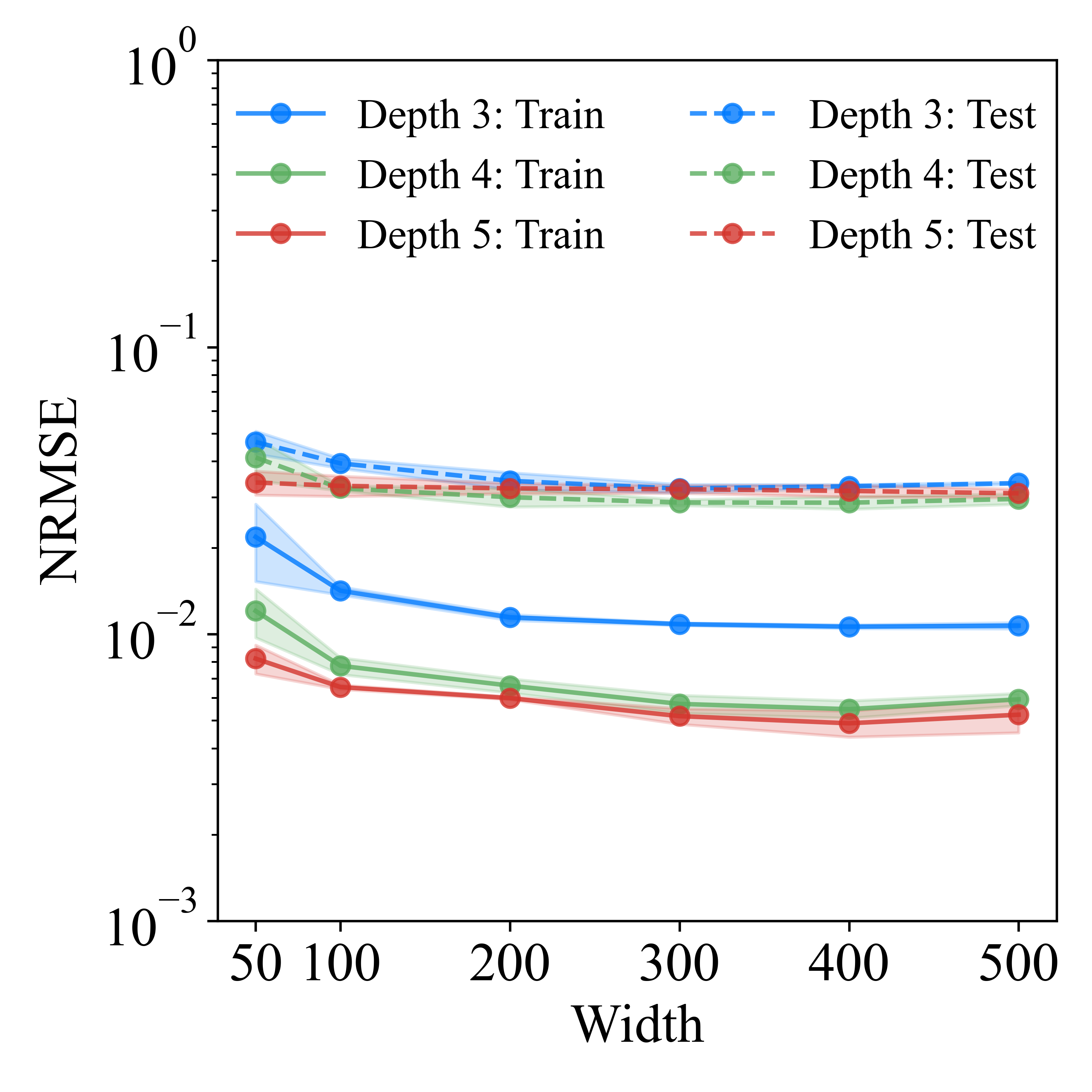}
        \caption{MLP}
        \label{fig:sup_case1_mlp}
    \end{subfigure}%
    \begin{subfigure}{0.33\textwidth}
        \includegraphics[width=\linewidth]{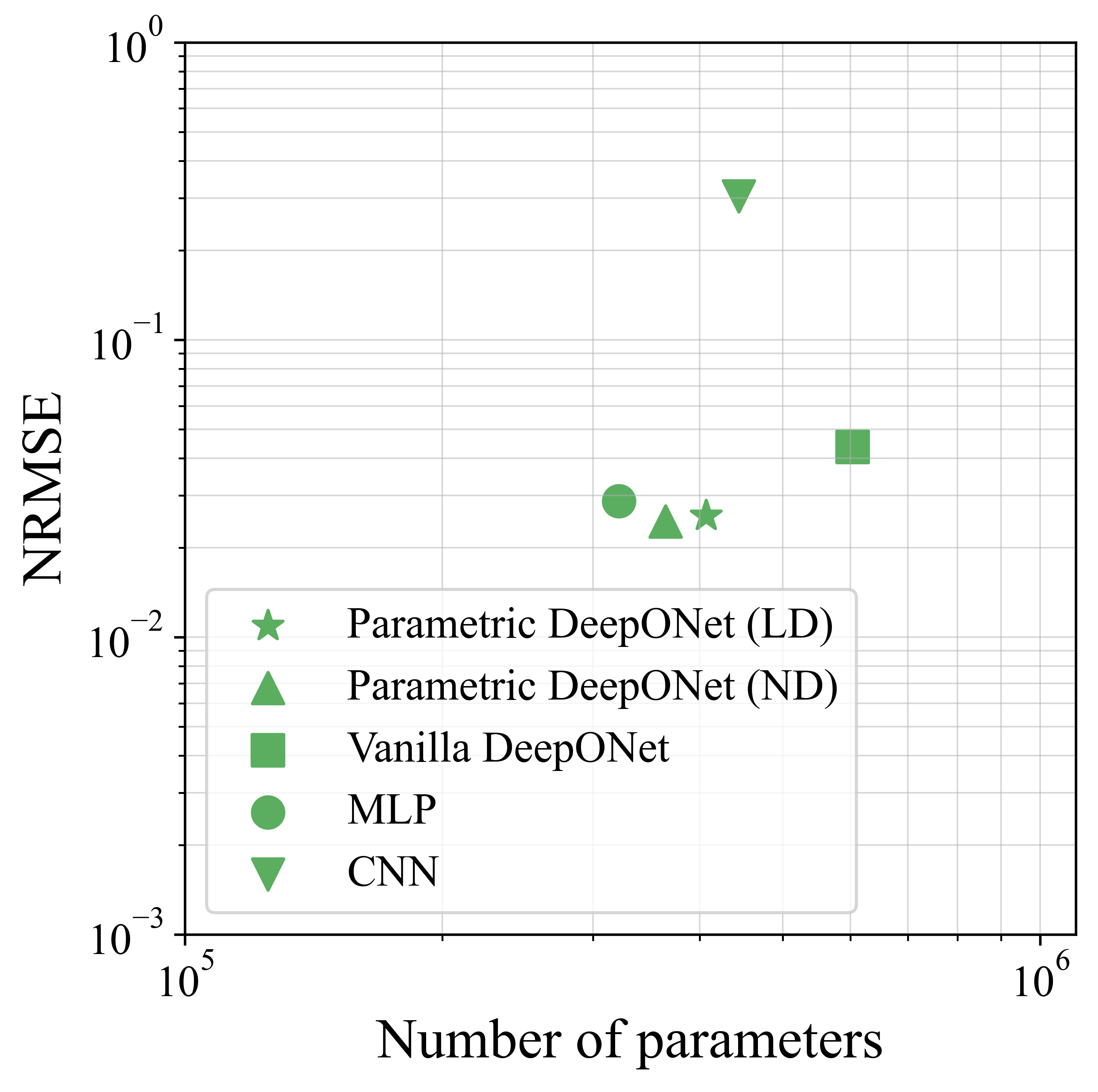}
        \caption{Performance and number of parameters.}
        \label{fig:sup_params_case1}
    \end{subfigure}
    \caption{Performance comparison of response prediction in Case 1b, the shaded regions represent the standard deviation.}
    \label{fig:sup_baselines_case1}
\end{figure}
\subsection{Parametric DeepONet}
In Case 1b, Parametric DeepONet (LD) achieves the lowest test NRMSE with a depth of 5 and width of 200 (Figure~\ref{fig:sup_case1_p_ld_deeponet}).
Parametric DeepONet (ND) performs best with a depth of 3 and width of 300 (Figure~\ref{fig:sup_case1_p_nd_deeponet}).
In Case 2, the optimal test NRMSE for Parametric DeepONet (LD) is obtained at depth 5 and width of 300 (Figure~\ref{fig:sup_case2_p_ld_deeponet}).
Parametric DeepONet (ND) has the smallest test NRMSE with a depth of 3 and width of 300 (Figure~\ref{fig:sup_case2_p_nd_deeponet}).
The branch, parameter, and trunk nets share the same depth and width in each configuration, which are all implemented as MLPs with ReLU activation in hidden layers.
The MLPs consists of input, hidden, and output layers.
The corresponding MLP architectures are as follows:
\begin{itemize}
    \item Case 1b, Parametric DeepONet (LD): branch net-[200, 200, 200, 200, 200, 200], parameter net-[2, 200, 200, 200, 200], trunk net-[20, 200, 200, 200, 200] (with PE spectrum features dimension $k = 10$).
    \item Case 1b: Parametric DeepONet (ND): branch net-[200, 300, 300], parameter net-[2, 300, 300], trunk net-[20, 300, 300] (with PE spectrum features dimension $k = 10$).
    \item Case 2, Parametric DeepONet (LD): branch net-[200, 300, 300, 300, 300], parameter net-[6, 300, 300, 300, 300], trunk net-[100, 300, 300, 300, 300] (with PE spectrum features dimension $k = 50$).
    \item Case 2: Parametric DeepONet (ND): branch net-[200, 300, 300], parameter net-[6, 300, 300], trunk net-[100, 300, 300] (with PE spectrum features dimension $k = 50$).
\end{itemize}

Overall, grid search results suggest that a width of $200-300$ yields the best performance.
For Parametric DeepONet (LD), generally improves with increased depth.
For Parametric DeepONet (ND), depth has limited impact, and performance depends more strongly on network width.

\subsection{Vanilla DeepONet}
In Case 1b (Figure~\ref{fig:sup_case1_deeponet}) and Case 2 (Figure~\ref{fig:sup_case2_deeponet}), the lowest test NRMSE is obtained with a depth of 5 and width of 300.
The branch network is a 5-layer MLP with dimensions [202, 300, 300, 300, 200] and [206, 300, 300, 300, 200] in Case 2.
The trunk network is an MLP of [1, 300, 300, 300, 300] in Case 1b, and [100, 300, 300, 300, 300] (with PE spectrum features dimension $k = 50$).
All MLPs are with ReLU activation in hidden layers.
Overall, grid search results suggest that a depth of $5$ and a width of 300 consistently yield the best performance for vanilla DeepONet in both cases.

\subsection{Multilayer perceptron (MLP)}
\begin{figure}[!h]
    \centering
    \begin{subfigure}{0.33\textwidth}
        \centering
        \includegraphics[width=\linewidth]{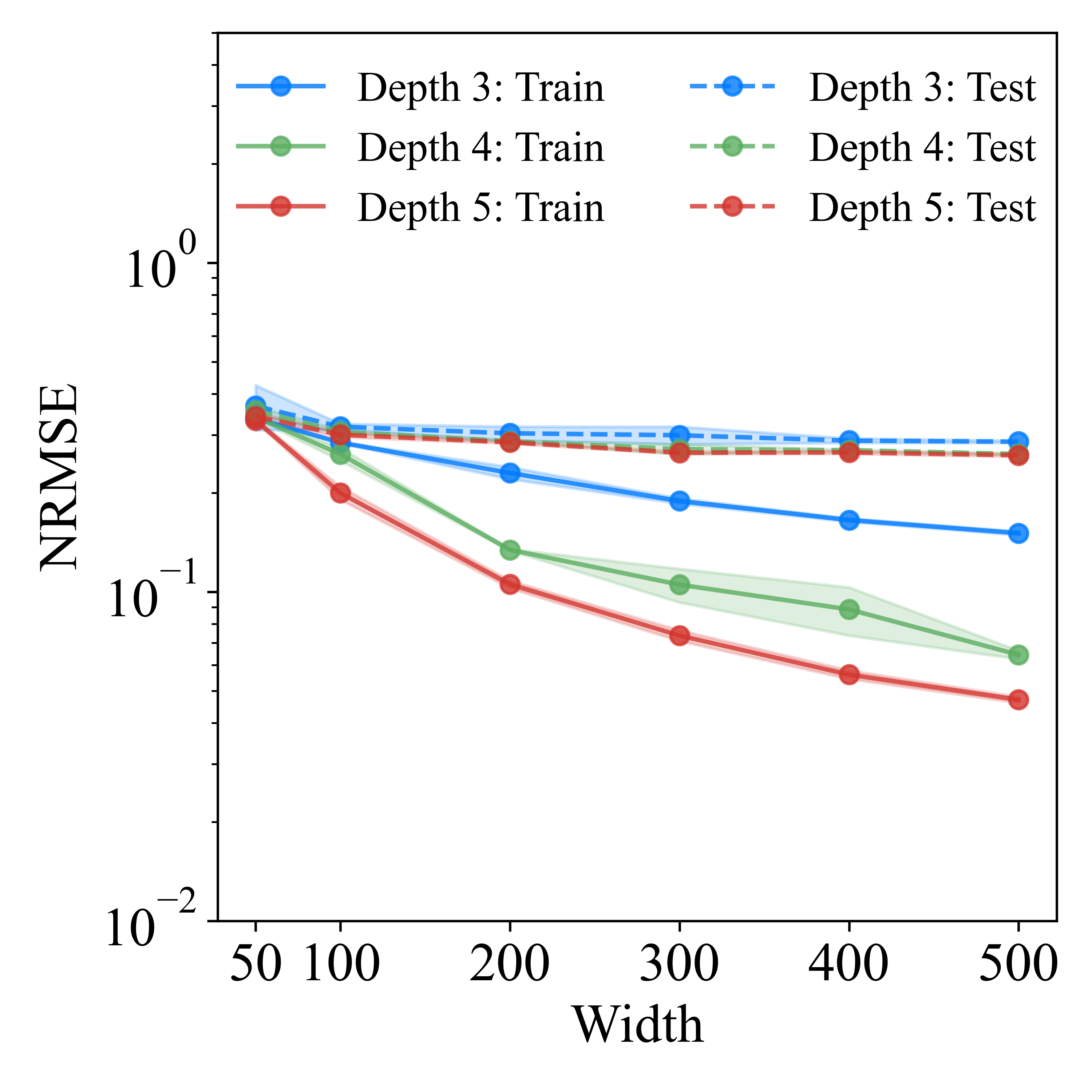}
         \caption{Parametric DeepONet (LD)}
         \label{fig:sup_case2_p_ld_deeponet}
    \end{subfigure}%
    \begin{subfigure}{0.33\textwidth}
        \includegraphics[width=\linewidth]{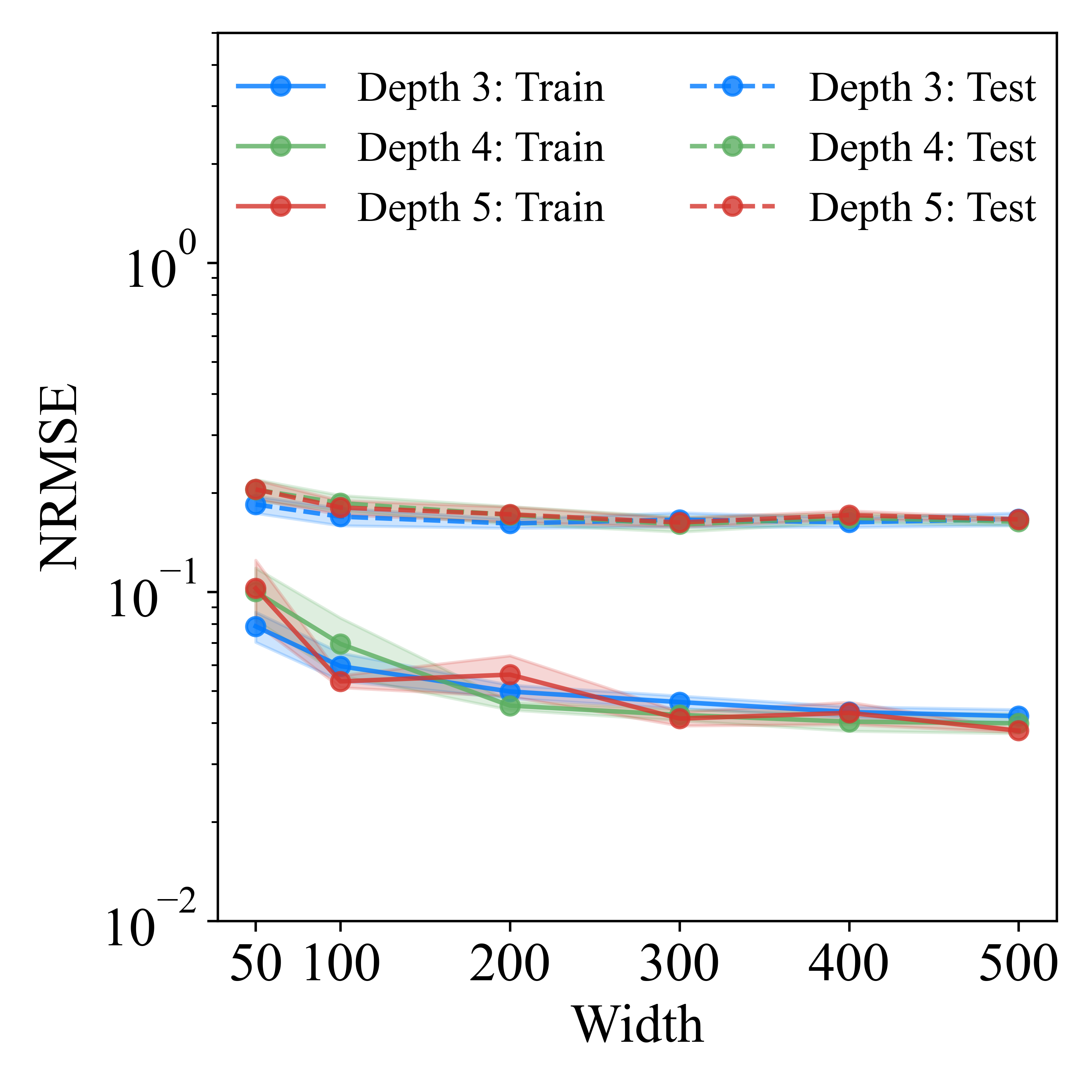}
        \caption{Parametric DeepONet (ND)}
        \label{fig:sup_case2_p_nd_deeponet}
    \end{subfigure}%
    \begin{subfigure}{0.33\textwidth}
        \includegraphics[width=\linewidth]{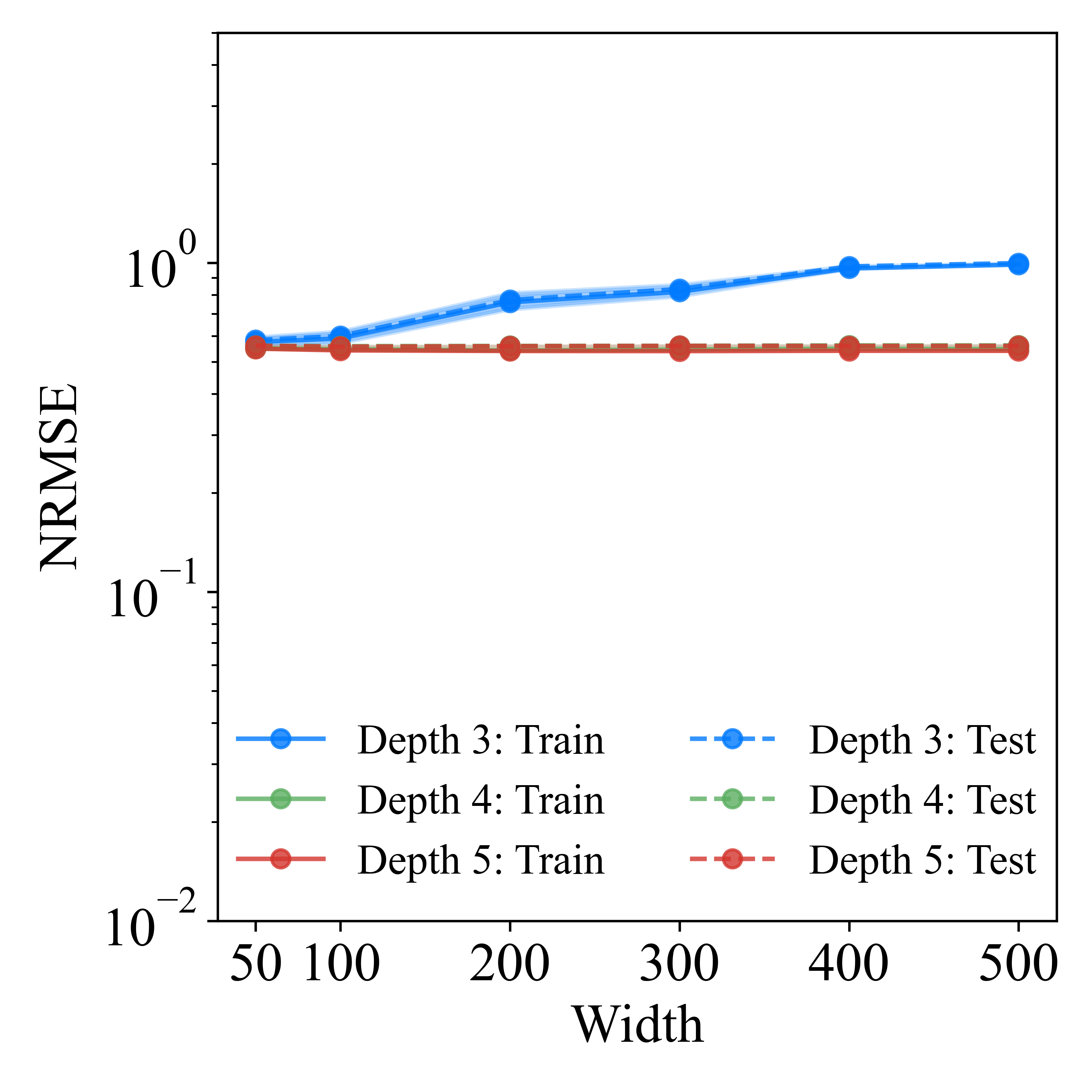}
        \caption{Vanilla DeepONet}
        \label{fig:sup_case2_deeponet}
    \end{subfigure}

    \begin{subfigure}{0.33\textwidth}
        \includegraphics[width=\linewidth]{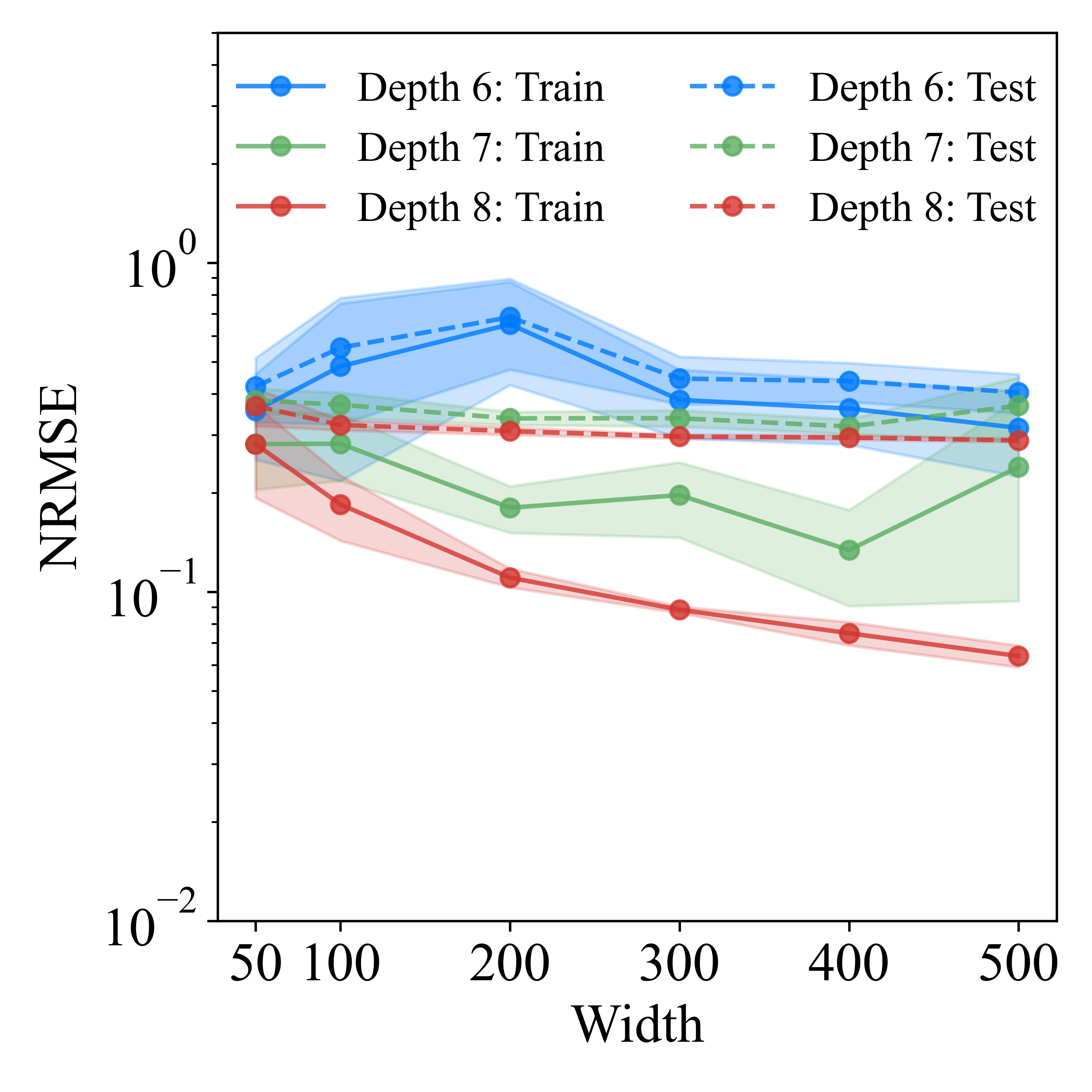}
        \caption{MLP}
        \label{fig:sup_case2_mlp}
    \end{subfigure}%
    \begin{subfigure}{0.33\textwidth}
        \includegraphics[width=\linewidth]{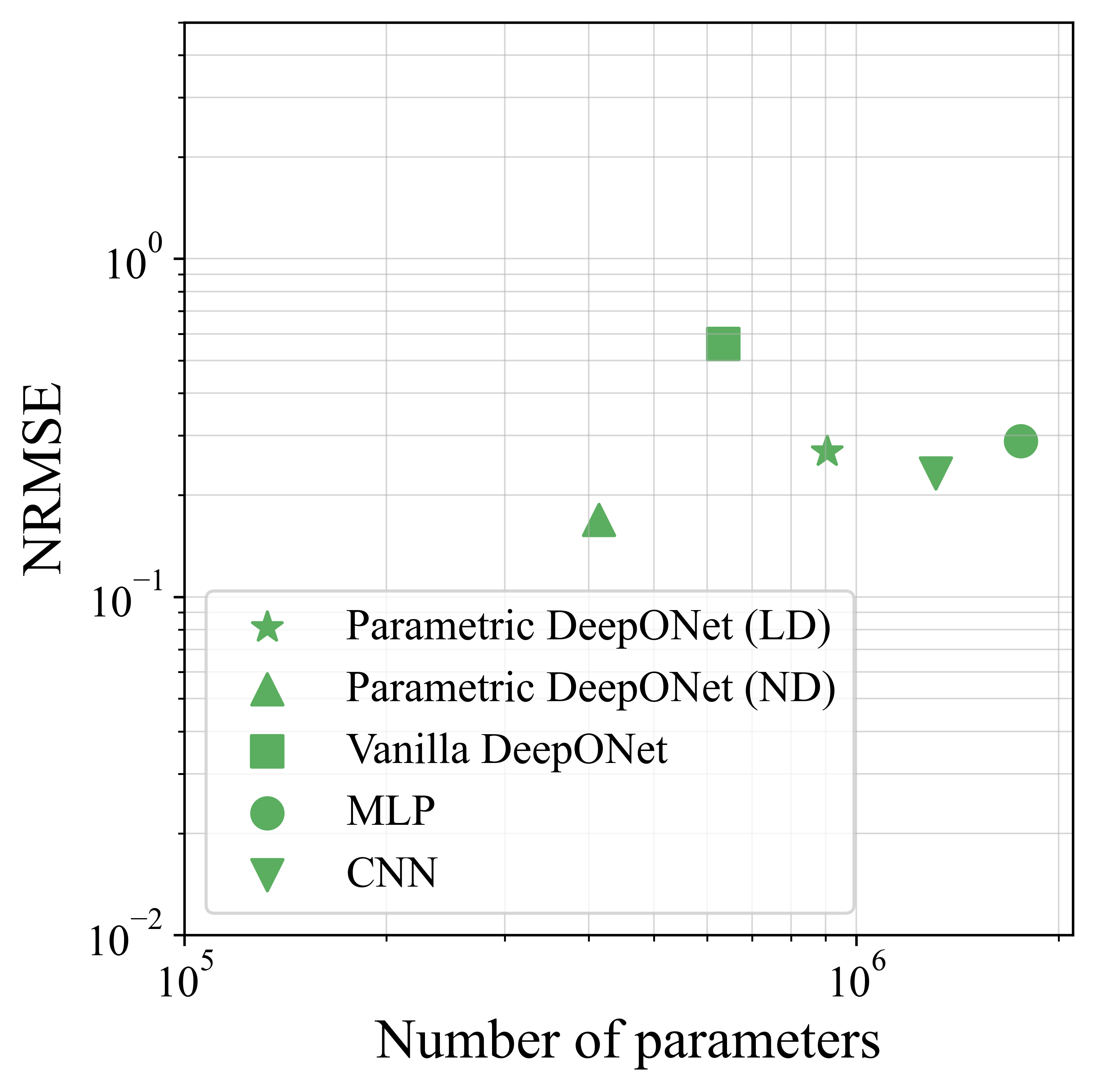}
        \caption{Performance and number of parameters}
        \label{fig:sup_params_case2}
    \end{subfigure}
    \caption{Performance of response prediction of baseline models in Case 2, the shaded regions represent the standard deviation.}
    \label{fig:sup_baselines_case2}
\end{figure}

In Case 1b (Figure~\ref{fig:sup_case1_mlp}), the smallest test NRMSE is obtained with a depth of 4 and a width of 400. 
For Case 2 (Figure~\ref{fig:sup_case2_mlp}), depth is chosen from [6, 7, 8], and width is in [50, 100, 200, 300, 400, 500] since the dynamic responses are with MDOF and increased complexity.
The optimal test NRMSE is obtained with a depth of 8 and a width of 500 (Figure~\ref{fig:sup_case2_mlp}). 
Similarly, MLPs are with ReLU activation in hidden layers.
In Case 1, the MLP consists of 4 layers with input, hidden, and output dimensions of [202, 400, 400, 200].
In Case 2, the MLP consists of 8 layers with input, hidden, and output dimensions of [206, 500, 500, 500, 500, 500, 500, 800]. Overall, the results indicate that MLP performance generally improves with increased depth and width, particularly for more complex problems such as the MDOF system in Case 2.

\subsection{Convolutional neural network (CNN)}
\label{sec:cnn}
As discussed in Section~\ref{sec:case1-forward}, CNNs with concatenated inputs may not be an effective and stable method for parametric modeling of structural dynamics, particularly when the number of system parameters is small.
CNNs operate based on local connectivity and are designed to capture local patterns.
Therefore, concatenating the system parameters at the input boundary is suboptimal, which could hinder parametric information globally encoded into the output.
For this reason, we do not perform hyperparameter tuning for CNNs, and the reported architecture and results serve as a baseline for reference, illustrating the limitations of input concatenation.

The CNN has an encoder-decoder architecture consisting of convolutional and deconvolutional layers.
Each layer includes convolution or deconvolution operations, batch normalization (BN), and leaky ReLU activation.
In Case 1, the encoder has 3 convolutional layers with output channels $[64, 128, 512]$, kernel sizes 3 and strides 2;
the decoder has 3 deconvolutional layers with output channels $[128, 64, 1]$, kernel sizes 3 and strides 2.
In Case 2, the encoder consists of 6 convolutional layers with output channels $[16, 32, 64, 128, 256, 512]$, kernel sizes $[3, 3, 3, 3, 3, 4]$ and strides $[2, 2, 2, 2, 3, 1]$; 
the decoder consists of 6 deconvolutional layers with output channels $[256, 128, 64, 32, 16, 4]$, kernel sizes $[4, 3, 3, 3, 3, 3]$ and strides $[1, 3, 2, 2, 2, 2]$.

A more systematic investigation of parameter encoding strategies in CNNs would be an interesting direction for future work. One potential solution is to first encode system parameters in a latent space (as in the Parametric DeepONet (ND), see~\ref{sec:decoder}), where parametric information is uniformly integrated into the CNN's input. This avoids placing parametric information at the boundary of the input, enabling CNN to better capture the major dynamics for response prediction.

\subsection{Summary}
Table~\ref{table:num_models} summarizes the number of trainable parameters for different models in Case 1 and Case 2.
Figure~\ref{fig:sup_params_case1} and Figure~\ref{fig:sup_params_case2} present the best test performance and the corresponding number of trainable parameters for each baseline model.
In Case 1b, Parametric DeepONet (LD) and Parametric DeepONet (ND) achieve the lowest and second-lowest NRMSEs, respectively, with fewer parameters than vanilla DeepONet and CNN models. Notably, MLP also performs competitively with the fewest trainable parameters among all models.
In Case 2, Parametric DeepONet (ND) maintains the lowest NRMSE among baseline models and with the fewest trainable parameters.
Additionally, Parametric DeepONet (LD) attains better performance than MLP and vanilla DeepONet.

Overall, Parametric DeepONet has proven to be an effective variant of vanilla DeepONet, which can significantly improve the performance of response prediction while maintaining a compact model architecture. Compared to other baseline models such as MLP and CNN, Parametric DeepONet performs more consistently and flexibly in both Case 1b and Case 2, making it suitable for parametric modeling of both SDOF and MDOF structural systems.
\begin{table*}[!h]
    \centering
    \caption{Number of trainable parameters, for different validation cases and models.}
    \label{table:num_models}
    \begin{tabular}{cccccc}
    \hline
    Validation Case & \makecell{Parameteric \\ DeepONet (LD)} & \makecell{Parameteric \\ DeepONet (ND)}& DeepONet & MLP & CNN \\
    \hline
    Case 1 & 406,800  & 364,833 &  603,301 & 321,800 & 444,417 \\
    Case 2 & 905,400 & 414,188 & 634,201 & 1,756,800 & 1,313,428 \\
    \hline
    \end{tabular}
\end{table*}

 \section{Implementation details}
All experiments are conducted via the Pytorch~\cite{paszke2019pytorch} library on an NVIDIA Geforce RTX 3080 graphic card.
For the training process of forward modeling, we set the number of training epochs to 10,000 with a batch size of 64 and a learning rate of $1 \times 10^{-3}$.
The Adam~\cite{kingma2014adam} optimizer is used in Case 1, while the SGD optimizer is used in Case 2.
For inverse modeling in Case 1, the training process consists of 5000 epochs for gradient-based initialization and 300 epochs for neural refinement with 1 iteration step.
In Case 2, the inverse training process involves 200 epochs for gradient-based initialization and 100 epochs for neural refinement with 3 iteration steps.
During gradient-based initialization, system parameters are estimated within a bounded domain. If the predicted values fall outside the defined range, they are clipped to the boundaries of the parameter space to ensure physically meaningful estimates.
The neural refinement network in Case 1 is an MLP consisting of 4 layers with input, hidden, and output dimensions of $[4, 64, 64, 2]$, with ReLU activation in hidden layers.
The neural refinement network in Case 2 is an MLP consisting of 4 layers with input, hidden, and output dimensions of $[12, 256, 256, 6]$, with ReLU activation in hidden layers.

\bibliographystyle{elsarticle-num} 
\bibliography{ref}

\end{document}